\documentclass[12pt]{article}
\pdfoutput=1
\usepackage{putex}
\usepackage{graphicx}
\usepackage{caption}
\usepackage{amsmath}
\usepackage{amssymb}
\usepackage{array}
\usepackage{bm}
\usepackage{multirow}
\usepackage{mathtools}
\usepackage{comment} 
\usepackage{subcaption}
\usepackage{epstopdf}
\usepackage{enumerate}
\usepackage{cite}
\usepackage{youngtab}
\usepackage{tensor}
\usepackage{slashed}
\usepackage[aligntableaux=center]{ytableau}
\usepackage[utf8]{inputenc}
\usepackage{rotating}
\usepackage{bigfoot}
\usepackage[
      colorlinks=true,
      linkcolor=blue,
      urlcolor=blue,
      filecolor=black,
      citecolor=red,
      linktocpage=true
      ]{hyperref}
\usepackage{braket}
\usepackage{tikz}
\usepackage{mathrsfs}

\newcommand {\be} {\begin {equation}}
\newcommand {\ee} {\end {equation}}

\newcommand {\bes} {\begin {equation*}}
\newcommand {\ees} {\end {equation*}}

\newcommand{\es}[2] {\begin{equation} \label{#1} \begin{split} #2 \end{split} \end{equation}}

\newcommand{\cO}{{\mathcal O}}

\newcommand{\rmd}{\mathrm{d}}

\newcommand{\bea}{\begin{equation}\begin{aligned}}
\newcommand{\eea}[1]{\label{#1}\end{aligned}\end{equation}}

\newcommand{\beq}{\begin{equation}}
\newcommand{\eeq}{\end{equation}}

\def\ie{\begin{equation}\begin{aligned}}
\def\fe{\end{aligned}\end{equation}}

\numberwithin{equation}{section}

\def\<{\langle}
\def\>{\rangle}

\begin{document}

\preprint{}

\institution{oxford}{Mathematical Institute, University of Oxford,
Woodstock Road, Oxford, OX2 6GG, UK }
\institution{imperial}{Abdus Salam Centre for Theoretical Physics, Imperial College London, London SW7 2AZ, UK}
\institution{china}{Institute of Theoretical Physics, Chinese Academy of Sciences, Beijing 100190, China}
\institution{chinaa}{School of Physical Sciences, University of Chinese Academy of Sciences, Beijing 100049, China}

\title{Strong coupling spectrum of $\mathrm{AdS}_3\times \mathrm{S}^3$ from string field theory}

\authors{Luis F. Alday,\worksat{\oxford} Shai M. Chester,\worksat{\imperial} Kiarash Naderi,\worksat{\imperial} and De-liang Zhong\worksat{\china,\chinaa}  }

\abstract{
We consider type IIB string theory on $\mathrm{AdS}_3\times \mathrm{S}^3\times \mathbb{T}^4$ in a large AdS radius expansion at arbitrary values of the relative RR and NS--NS fluxes. We develop a method for computing scaling dimensions in the dual CFT in this limit by combining a string field theory calculation, valid at finite quantized NS--NS flux but perturbative in RR flux, with a finite mixed-flux ansatz, which is constrained by worldsheet parity and motivated by classical string solutions. We apply the method to three families of states. In the pure-RR limit, our results agree both with a family of states computed using the worldsheet bootstrap and with another family computed using integrability.
}
\date{}

\maketitle

\tableofcontents
\newpage

\section{Introduction}\label{sec:intro}

An outstanding question in string theory is how to compute observables in the presence of Ramond-Ramond (RR) flux. Since most compactifications of string theory involve RR flux, this question is crucial to understand string theory on curved spacetimes, or to construct realistic models of cosmology or particle physics from string theory. Unfortunately, the textbook Neveu-Schwarz-Ramond (NSR) worldsheet formulation of string theory cannot handle backgrounds with RR flux. The pure spinor formalism was invented to tackle this problem \cite{Berkovits:2000fe}, but it has been difficult to compute observables with it in practice.

This problem arises even in the background where we think we understand string theory the best: AdS/CFT. Almost all examples of the AdS/CFT correspondence require RR flux \cite{Maldacena:1997re}, so most calculations in the bulk are either performed in the leading large $N$ and large 't Hooft coupling $\lambda$ limit where supergravity is valid, or by considering the flat space limit where the NSR formalism applies (see e.g. \cite{Chester:2018aca}). A notable exception is the planar limit of certain examples of AdS/CFT. The Green-Schwarz worldsheet formalism is valid for RR flux \cite{Metsaev:1998it}, but is only understood at tree level, i.e. in the planar limit. Starting from integrability of the classical Green-Schwarz string \cite{Bena:2003wd}, this program culminated in the quantum spectral curve (QSC) \cite{Gromov:2013pga}, which allows for efficient numerical evaluation of the planar spectrum at finite $\lambda$. This program has been carried to completion for the cases of type IIB on $\text{AdS}_5\times \text{S}^5$ dual to $\mathcal{N}=4$ super-Yang-Mills (SYM) \cite{Gromov:2013pga,Gromov:2023hzc}, and type IIA on $\text{AdS}_4\times \mathbb{CP}^3$ dual to $\mathcal{N}=6$ ABJM \cite{Aharony:2008ug,Bombardelli:2018bqz,Cavaglia:2014exa}.\footnote{For the ABJM case, there are still certain families of states that the QSC has not yet been applied to, as pointed out in \cite{Chester:2024esn}.}

Integrability has not yet been able to compute general higher point correlators, such as OPE coefficients.\footnote{In four-dimensional $\mathcal{N}=4$ SYM theory, three-point functions can, in principle, be computed using the hexagon formalism \cite{Basso:2015zoa}, while higher-point functions can be studied by hexagonalization \cite{Fleury:2016ykk,Fleury:2017eph,Coronado:2018ypq}; see \cite{Komatsu:2017buu} for a review. For three-point functions involving two BPS operators and one non-BPS operator, quantum corrections can be resummed to obtain finite-coupling results; see \cite{Basso:2022nny,Basso:2025mca,Basso:2026lnr}. However, the hexagon formalism has not yet been developed for ABJM theory or the AdS${}_3$ case. An alternative approach based on separation of variables has been successfully applied to certain subsectors of $\mathcal{N}=4$ SYM; see, e.g. \cite{Bercini:2022jxo,Bargheer:2025kli,Bargheer:2026kon}. However, this approach has not yet been generalized to the full correlator. A complementary approach is to input the spectrum from integrability and then determine the four-point functions by combining numerical bootstrap with localization \cite{Caron-Huot:2022sdy,Caron-Huot:2024tzr}, which in practice has allowed certain OPE coefficients to be numerically determined for all $\lambda$.
} It also has not yet been applied to many examples of AdS/CFT with less than maximal supersymmetry, such as those with flavor branes \cite{Aharony:1998xz}. These problems have been overcome by the worldsheet bootstrap \cite{Alday:2023mvu}. This method computes four point functions in holographic CFTs in the planar limit in a large AdS radius expansion by demanding consistency between the flat space limit \cite{Penedones:2010ue} of the conformal block expansion, with an ansatz for the worldsheet integral describing the correlator. Both the scaling dimensions and OPE coefficients of certain single trace operators can then be read off from this correlator in a large $\lambda$ expansion. This program was initiated for $\text{AdS}_5\times \text{S}^5$ \cite{Alday:2022uxp,Alday:2022xwz,Alday:2023jdk,Alday:2023mvu}, and has since been generalized not only to $\text{AdS}_4\times \mathbb{CP}^3$ \cite{Chester:2024esn}, but also to holographic CFTs with flavor branes where the QSC is not yet known \cite{Alday:2024yax}. In all these cases, the first two curvature corrections to flat space were computed,\footnote{To get the second curvature correction, it was important to input extra constraints from localization as derived in \cite{Binder:2019jwn,Chester:2020dja,Binder:2019mpb,Behan:2023fqq,Chester:2022sqb} for the various theories.} allowing one to read off the first two large $\lambda$ corrections to CFT data.

The case of type IIB on $\text{AdS}_3\times \text{S}^3\times \mathbb{T}^4$ is very different from its higher dimensional counterparts. In particular, the background may be supported by both RR and NSNS three-form flux threading the $\text{S}^3$, allowing for a continuous family of mixed-flux backgrounds. The AdS radius $R$ divided by string length $\ell_s$ is related to these fluxes as
\es{radius}{
\sqrt{\lambda}\equiv R^2/\ell^2_s\,, \quad k = q \sqrt{\lambda} \,,
}
where $k$ is the quantized NSNS flux, and $q$ parameterizes the amount of RR flux such that $q=0$ is pure RR flux, and $q=1$ is pure NSNS flux. In the case of pure NSNS flux, there is an exact NSR worldsheet description via a supersymmetric $\text{SL}(2,\mathbb{R})\times \text{SU}(2)$ Wess-Zumino-Witten (WZW) model together with four free bosons and fermionic superpartners \cite{Maldacena:2000hw,Maldacena:2000kv,Maldacena:2001km}. So one can in principle compute observables in the pure NSNS case for any $k$, including the strong coupling large AdS radius limit at large $k$. But once any amount of RR flux is turned on, there is no exactly solvable worldsheet description just as in higher dimensions.\footnote{The hybrid formalism \cite{Berkovits:1999im} was introduced to study $\text{AdS}_3$ in the presence of RR flux, but as with the pure spinor formalism for AdS higher dimensions, it is hard to use it to compute observables in practice.} In fact, there is no useful Lagrangian description of the CFT for general RR flux,\footnote{Two explicit descriptions of the dual CFT are known in special limits. In the limit where the volume of $\mathbb{T}^4$ becomes infinite, the CFT can be described as the Higgs branch of a 2d $\mathcal{N}=(4,4)$ supersymmetric gauge theory \cite{Aharony:1997th,Witten:1997yu}. For $k=1$ and $q=1$, the string theory is instead dual to a grand canonical ensemble of free symmetric orbifold CFTs \cite{Eberhardt:2018ouy,Eberhardt:2019ywk,Eberhardt:2020bgq,Aharony:2024fid}.} which makes the theory even harder to study than higher dimensions that had explicit gauge theory duals.

The first observable with RR flux at strong coupling was computed using the worldsheet bootstrap in \cite{Chester:2024wnb}. In particular, the first AdS correction to four point functions of closed string modes dual to half-BPS operators was computed in \cite{Chester:2024wnb,Jiang:2025oar}, which could be used to read off the large $\lambda$ expansion of the lowest Regge trajectory CFT data, including the scaling dimension:\footnote{Scaling dimensions of other states that appear in that correlator cannot be read off, because they are degenerate. The story is similar for $\text{AdS}_5\times \text{S}^5$ and $\text{AdS}_4\times \mathbb{CP}^3$.}
\es{LRTintro}{
\Delta_\text{LRT}=\lambda^{\frac14}\sqrt{2S}-1+\lambda^{-\frac14}\Big[\frac{3S^{\frac32}}{4\sqrt{2}}-\frac{\sqrt{S}}{2\sqrt{2}}+\frac{J^2}{2\sqrt{2S}}\Big]+\mathcal{O}(\lambda^{-\frac34})\,.
}
Here, $S$ denotes the even spin of the superprimary, while $J$ labels the irrep $(\tfrac{J-1}{2},\tfrac{J-1}{2})$ under the $\text{SU}(2)\times \text{SU}(2)$ R-symmetry. Note that it is not obvious from the worldsheet bootstrap that one is considering pure RR, and this can be identified mostly by comparing the scaling dimension to a classical solution. In particular, one can consider classical solutions to the Polyakov action on $\text{AdS}_3\times \text{S}^3$ for finite $q$, where all the quantum numbers $\Delta, S, J$ are assumed to scale as $\sqrt{\lambda}$, so that a saddle point solution is valid. In this paper, we construct a specific classical solution called the drift-closed folded string,\footnote{Our solution corrects an error in the previously considered folded string solution \cite{David:2014qta,Loewy:2002gf}, which did not match the $q=1$ result known from the WZW model \cite{Ferreira:2017pgt}, and also was not a valid solution for $0<q<1$.} whose energy takes the form
\es{LRTintroClass}{
\Delta^\text{class}_\text{LRT}&=\lambda^{\frac14}\sqrt{2S}+\lambda^{-\frac14}\Big[\frac{3S^{\frac32}}{4\sqrt{2}}(1-q^2)+\frac{J^2}{2\sqrt{2S}}\Big]\\
&+\lambda^{-\frac34}\Bigg[
\frac{3(1-q^2)(-7+39q^2)}{64\sqrt{2}}S^{\frac52}
+\frac{5(1-q^2)J^2}{16\sqrt{2}}\sqrt{S}
-\frac{J^4}{16\sqrt{2}S^{\frac32}}
\Bigg]+\mathcal{O}(\lambda^{-\frac54})\,.
}
This classical solution matches the classical terms in \eqref{LRTintro} for the pure RR case $q=0$,\footnote{Again up to the shift by $-1$, that the classical solution is not sensitive to.} while the $\sqrt{S}/\lambda^{1/4}$ term in \eqref{LRTintro} is a 1-loop correction.

It is much harder to derive the QSC in $3$d than in higher dimensions, because the CFT dual is less understood. Nonetheless, a QSC was conjectured in \cite{Cavaglia:2021eqr,Ekhammar:2021pys} for the pure RR case. It is also much harder to extract scaling dimensions from the QSC than in higher dimensions, so originally only some weak coupling data was obtained \cite{Cavaglia:2022xld}. Recently, a numerical prediction at strong coupling was obtained for a family of states whose superprimary has odd spin $S-1$, and transforms in the $(\tfrac{J-2}{2},\tfrac{J}{2})$ irrep \cite{Ekhammar:2026ykk}:\footnote{We have exchanged the left- and right-moving sectors with respect to \cite{Ekhammar:2026ykk}.}
\es{Intintro}{
&\Delta_\text{QSC}=\lambda^{\frac14}\sqrt{2S}-1+\lambda^{-\frac14}\Big[\frac{3S^{\frac32}}{4\sqrt{2}}-\frac{\sqrt{S}}{\sqrt{2}}+\frac{J^2}{2\sqrt{2S}}\Big]+\frac{J}{2}\lambda^{-\frac12}\\
&+\lambda^{-\frac34}\Big[
-\frac{21S^{\frac{5}{2}}}{64\sqrt{2}}
+\frac{23-24\zeta(3)}{16\sqrt{2}}S^{\frac32}
+\frac{5J^2\sqrt{S}}{16\sqrt{2}}
-\frac{\sqrt{S}}{32\sqrt{2}}
-\frac{J^2}{4\sqrt{2S}}
-\frac{J^4}{16\sqrt{2}S^{\frac32}}
\Big]
+\mathcal{O}(\tfrac{1}{\lambda})\,.
}
Curiously, the classical terms match the same classical solution \eqref{LRTintroClass} for $q=0$ that matched the lowest Regge trajectory. It has not yet been possible to use the QSC to compute the lowest Regge trajectory state in \eqref{LRTintro}, so the conjectured QSC cannot yet be compared to the worldsheet bootstrap. 

In this paper, we will introduce a method that can be used to compute the large $\lambda$ expansion of the scaling dimension $\Delta$ of any state for string theory on $\text{AdS}_3\times \text{S}^3\times \mathbb{T}^4$, and applies for general mixed flux $q$. Our main input is the string field theory (SFT) calculation of \cite{Cho:2018nfn}. We consider the marginal operator $\cO_\text{RR}$ that corresponds to deforming the WZW model by RR flux, and is uniquely fixed by $\text{AdS}_3\times \text{S}^3$ superisometries \cite{Friedan:1985ge,Kutasov:1998zh,Kutasov:1999xu}.\footnote{As discussed in \cite{Cho:2018nfn}, the overall $k$-dependent normalization of $\mathcal{O}_{\text{RR}}$ is not totally fixed, however, it is fixed up to the order $1/k^{5/2}$.} There is no linear perturbative correction $\langle \mathcal{V} \cO_\text{RR}\mathcal{V}^\dagger\rangle$ to the pure NSNS value $\Delta_\text{WZW}$, where $\mathcal{V}$ is the vertex operator for the state, because it is forbidden from picture number selection rules. The first nonzero correction to $\Delta_\text{WZW}$ is at quadratic order, and takes the schematic form
\es{IntroSFT}{
\Delta-\Delta_\text{WZW}\sim \mu^2\int_{\mathbb{C}}\text{d}^2 z \frac{\langle \mathcal{V}(0,0) \cO_\text{RR}(1,1)\cO_\text{RR} (z,\bar{z})\mathcal{V}^\dagger(\infty,\infty) \rangle}{\langle \mathcal{V}(0,0)\mathcal{V}^\dagger(\infty,\infty)\rangle}\,,
}
where $\mu$ is the coefficient of the deformation $\cO_\text{RR}$, and is related to $q$ as
\es{qmu}{
q=1-\mu^2/2+O(\mu^4)\,.
} 
The integral in \eqref{IntroSFT} is power law divergent when $z\sim1$, as is typical for conformal perturbation theory, and requires an off-shell formalism to obtain a finite answer. String field theory gives a consistent prescription to obtain the unique physical answer for the scaling dimension \cite{Cho:2018nfn,Sen:2019jpm}, which in practice amounts to simply throwing out the divergent terms. The result is a correction at small RR flux $\mu^2$ to the scaling dimension at finite $k$. This correction was in fact computed in \cite{Cho:2018nfn} for a family of states labelled by an integer $n\geq 0$, whose superprimary is a scalar transforming in the $(\tfrac{J-1}{2},\tfrac{J-1}{2})$ irrep.

We can then take the large $k$ limit of \eqref{IntroSFT} and use the relation \eqref{radius} to get the large $\lambda$ expansion to any order, except we only know each coefficient as a function of $q$ expanded around one to quadratic order as in \eqref{qmu}. On the other hand, we can compare to classical solutions for each state, which as discussed are known for any $|q|\leq 1$. For instance, for the state in \cite{Cho:2018nfn}, the corresponding classical solution is the so-called pulsating string \cite{Hernandez:2018gcd}:
\es{pulseIntro}{
\Delta_\text{CCY}^\text{class} = 2 \lambda^{1/4}\sqrt{n} - 2 q\, n + \frac{1}{\lambda^{1/4}}\!\left[\frac{5(1-q^2)}{2}\, n^{3/2} + \frac{J^2}{4\sqrt{n}}\right] + \frac{6 q\, (1-q^2)\, n^2}{\sqrt{\lambda}}+\mathcal{O}(\lambda^{-\frac34})\,.
}
We observe that each term in both classical solutions \eqref{pulseIntro} and \eqref{LRTintroClass} at order $\lambda^{\frac{1-m}{4}}J^{\ell}$ are polynomials in $q$ with degree at most $m-\ell$, which we can check to many higher orders than shown here. Our ansatz assumes this is true for loop corrections as well. As we will discuss in detail in Section~\ref{sec:SFT-comparison}, we can further constrain the ansatz using world-sheet parity which we denote as $\Omega$: this maps $q\mapsto -q$ and a physical state $\ket{\Psi}$ to another physical state $\ket{\Omega(\Psi)}$. We explicitly construct candidate parity partners for the states considered in Section~\ref{sec:NSNS} and impose, supported by string field theory computations, that\footnote{As discussed in more detail in Section~\ref{sec:NSNS}, the world-sheet parity $\Omega$ is different from the combined transformation considered in \cite{Witten:1983ar} that also sends a group element $g\mapsto g^{-1}$ and leaves the action invariant. Moreover, the spectral-parameter transformation $u\mapsto -u$ seems to be the world-sheet parity that we have considered, see \cite{Ekhammar:2026ykk}, and it be good to understand the precise relation in more detail.}
\begin{equation} \label{eq:mirror-map-spectrum-symmetry}
 	\Delta(\ket{\Omega(\Psi)},-q) = \Delta(\ket{\Psi},q) \ .
\end{equation}
We can then use the SFT result to fix the unknown coefficients of the loop corrections to order $1/\sqrt{\lambda}$.\footnote{At higher orders, the single input from SFT as computed to order $\mu^2$ is not enough to fix all the coefficients of $q$ that may appear.} For the state in \cite{Cho:2018nfn}, this gives the complete prediction for arbitrary $q$:
\es{pulseIntro2}{
\Delta_\text{CCY} &= 2 \lambda^{1/4}\sqrt{n} -1- 2 q\, n + \frac{1}{\lambda^{1/4}}\!\left[\frac{5(1-q^2)}{2}\, n^{3/2} -{\color{blue} \frac{1-q^2}{2}\sqrt{n}}+ \frac{J^2}{4\sqrt{n}}\right] \\
&\qquad+ \frac{1}{\sqrt{\lambda}}\Big[{6 \, n^2}q\, (1-q^2){\color{blue}-\frac52n\,q\, (1-q^2)}\Big]+\mathcal{O}(\lambda^{-\frac34})\,.
}
The odd coefficients in $q$ are because world-sheet parity maps the state in \cite{Cho:2018nfn} to another state, whose coefficients at order $\lambda^{-m/2}$ for $m=0,1,2,\dots$ differ by a sign, as we checked explicitly by applying SFT to that other state. We highlight in blue the terms that were not given by the classical solution \eqref{pulseIntro}, which we thus learned from our method.\footnote{Note that the classical solution does not have the $-1$ shift that we wrote in \eqref{pulseIntro2}, which can already be seen from the pure NSNS scaling dimension. This classical solution is not sensitive to this shift. } 

We can similarly compute the scaling dimensions for the lowest Regge trajectory to get
\es{LRTintroClass2}{
\Delta_\text{LRT}=\lambda^{\frac14}\sqrt{2S}-1+\lambda^{-\frac14}\Big[\frac{3S^{\frac32}}{4\sqrt{2}}(1-q^2)-{\color{blue}\frac{\sqrt{S}}{2\sqrt{2}}(1-q^2)}+\frac{J^2}{2\sqrt{2S}}\Big]+\mathcal{O}(\lambda^{-\frac34})\,.
}
Here, world-sheet parity maps the lowest regge trajectory state to itself, which is why all polynomials in $q$ are even. We again highlight in blue the term not available from the classical solution \eqref{LRTintroClass}. After setting $q=0$, this result matches the pure RR prediction from the worldsheet bootstrap in \eqref{LRTintro}, which is a non-trivial check of our method.

Finally, we similarly compute the scaling dimension for the states considered by the QSC:
 \es{Intintro2}{
 &\Delta_\text{QSC}=\lambda^{\frac14}\sqrt{2S}-1+\lambda^{-\frac14}\Big[\frac{3S^{\frac32}}{4\sqrt{2}}(1-q^2)-{\color{blue}\frac{\sqrt{S}}{\sqrt{2}}(1-q^2)}+\frac{J^2}{2\sqrt{2S}}\Big]+{\color{blue}(1-q^2)\frac{J}{2}\lambda^{-\frac12}}\\
 &+\lambda^{-\frac34}\Bigg[
 \frac{3(1-q^2)(-7+39q^2)}{64\sqrt{2}}S^{\frac52}
 +{\color{purple}\frac{(1-q^2)\left(23-24\zeta(3)+q^2\left(-79+72\zeta(3)\right)\right)}{16\sqrt{2}}S^{\frac32}}\\
 &+{\color{purple}\frac{(2-98q^2)(-1+q^2)}{64\sqrt{2}}\sqrt{S}}
 +\frac{5(1-q^2)J^2}{16\sqrt{2}}\sqrt{S}
 +{\color{blue}\frac{(-1+q^2)J^2}{4\sqrt{2S}}}
 -\frac{J^4}{16\sqrt{2}S^{\frac32}}
 \Bigg]+\mathcal{O}(\lambda^{-1})\,.
 }
This state is also mapped to itself under world-sheet parity, which explains why the answer is even in $q$. The black terms continue to match the classical solution \eqref{LRTintroClass} that we also identified with the lowest Regge trajectory, now at arbitrary $q$, where for the $S^\frac{5}{2}$ term we used the classical solution to fix the full polynomial in $q$. The blue terms are not captured by the classical solution, and for $q=0$ they match the pure RR prediction from the QSC in \eqref{Intintro}, which is another nontrivial check of our method, as well as a check of the conjectured QSC. For the purple terms, SFT alone was not enough to fix the full $q$ dependence, so we used the pure RR QSC prediction to fully fix it.

The rest of this paper is organized as follows. In Section \ref{sec:NSNS} we discuss the pure NSNS theory, where we explicitly construct the states we will consider from the WZW model, and assemble the ingredients we will need later for the SFT calculation. In Section \ref{sec:semiclassics}, we review the pulsating string solution, and derive the drift-closed folded string solution, which are valid for arbitrary $q$. In Section \ref{sec:SFT}, we use SFT to compute the scaling dimensions of the three families of states we consider at finite $k$ and to quadratic order in $\mu$, which we then use to fix the large $\lambda$ and arbitrary $q$ expansion as described above. We conclude in Section \ref{sec:conclusion} with a review of our results and a discussion of future directions. Technical details of the calculations are given in the various Appendices.

\section{Pure NS-NS string theory}\label{sec:NSNS}
In this section, we review and spell out our conventions for Type IIB superstring theory with pure NS-NS flux on the background $\text{AdS}_3 \times \text{S}^3 \times \mathbb{T}^4$. We will discuss the field contents of the world-sheet CFT and the physical spectrum by closely following the methods and conventions of \cite{Ferreira:2017pgt}. We will also explicitly construct the various states whose deformed energies we will compute in this work: the QSC state, the LRT \cite{Ferreira:2017pgt}, and the CCY state from \cite{Ferreira:2017pgt,Cho:2018nfn}, as well as the states these map to under world-sheet parity.

\subsection{The world-sheet CFT} \label{sec:ws-cft}
The $\text{AdS}_3 \times \text{S}^3$ world-sheet theory is described as a WZW model on $\text{SL}(2,\mathbb{R})\times \text{SU}(2)$. The $\mathbb{T}^4$ theory possesses a small $\mathcal{N}=(4,4)$ superconformal algebra with $c=6$. As most of the following analysis is independent of $\mathbb{T}^4$, we will mainly focus on $\text{AdS}_3 \times \text{S}^3$ in this section.

The WZW theory on $\text{SL}(2,\mathbb{R})\times \text{SU}(2)$ has both left- and right-moving sectors. We will primarily discuss the left-moving sector, as the right-moving sector is completely analogous. Throughout the text, given a left-moving field $F$, we will denote the corresponding right-moving field as $\bar F$.

The left-moving part of the $\text{AdS}_3 \times \text{S}^3$ theory is described by an $\mathfrak{sl}(2,\mathbb{R})^{(1)}_k \oplus \mathfrak{su}(2)^{(1)}_{k^{\prime}}$ affine super Lie algebra. Here the superscript `$(1)$' denotes an $\mathcal{N}=1$ world-sheet supersymmetry, while the subscripts $k$ and $k^{\prime}$ denote the corresponding supersymmetric levels. We will discuss these two affine algebras in more detail now.

The $\mathfrak{sl}(2,\mathbb{R})^{(1)}_k$ affine algebra consists of bosonic currents $J^a$, together with real fermions $\psi^a$ where $a\in\{+,-,3\}$ denotes the adjoint index. The currents $J^a$ form an $\mathfrak{sl}(2,\mathbb{R})_k$ while $\psi^a$ transform in the adjoint representation of the global bosonic subalgebra. The non-zero (anti-)commutation relations, in our conventions, are
\begin{subequations} \label{eq:decoupled-sl2-commutators}
      \begin{equation}
            [J^3_n,J^{\pm}_m] = \pm J^{\pm}_{n+m} \ , \quad [J^3_n,J^3_m]=-\frac{k}{2} n \delta_{n+m,0} \ , \quad [J^+_n,J^-_m] = k n \delta_{n+m,0} - 2 J^3_{n+m} \ ,
      \end{equation}
      \begin{equation}
            [J^{\pm}_n,\psi^3_r] = \mp \psi^{\pm}_{n+r} \ , \quad [J^3_n,\psi^{\pm}_r]= \pm \psi^{\pm}_{n+r} \ , \quad [J^{\pm}_n,\psi^{\mp}_r] = \mp 2 \psi^3_{n+r} \ ,
      \end{equation}
      \begin{equation}
            \{\psi^+_r,\psi^-_s\} = k \delta_{r+s,0} \ , \quad \{\psi^3_r,\psi^3_s\} = -\frac{k}{2} \delta_{r+s,0} \ ,
      \end{equation}
\end{subequations}
where the subscripts refer to the corresponding modes. In particular, for the bosonic fields $J^a_m$, we have that $m\in \mathbb{Z}$ and for the fermions $\psi^a_r$, we have that either $r\in \mathbb{Z}+\frac{1}{2}$ in the NS-sector or $r\in\mathbb{Z}$ in the R-sector. Upon defining,
\begin{equation}
      \mathcal{J}^{\pm} = J^{\pm} \pm \frac{2}{k} (\psi^3 \psi^{\pm}) \ , \quad \mathcal{J}^3 = J^3 - \frac{1}{k} (\psi^+ \psi^-) \ , 
\end{equation}
the currents $\mathcal{J}^a$ commute with the fermions $\psi^b$. Moreover, one can check that $\mathcal{J}^a$ form an $\mathfrak{sl}(2,\mathbb{R})_{k+2}$, i.e.\
\begin{equation}
      \begin{split}
            [\mathcal{J}^3_n,\mathcal{J}^{\pm}_m] &= \pm \mathcal{J}^{\pm}_{n+m} \ , \quad [\mathcal{J}^3_n,\mathcal{J}^3_m]=-\frac{k+2}{2} n \delta_{n+m,0} \ , \\ &[\mathcal{J}^+_n,\mathcal{J}^-_m] = (k+2) n \delta_{n+m,0} - 2 \mathcal{J}^3_{n+m} \ .
      \end{split}
\end{equation}
We will refer to $J^a$ and $\mathcal{J}^a$ as $\mathfrak{sl}(2,\mathbb{R})$ coupled and decoupled currents, respectively.

The $\mathfrak{su}(2)^{(1)}_{k^{\prime}}$ theory is described similarly: we will denote the bosonic currents as $K^a$ and the fermions as $\chi^a$, where $a\in\{+,-,3\}$ is the adjoint index. The currents $K^a$ form an $\mathfrak{su}(2)_{k^{\prime}}$ and $\chi^a$ transform in the adjoint representation of the global bosonic subalgebra. In our conventions, the non-zero (anti-)commutators are
\begin{subequations} \label{eq:decoupled-su2-commutators}
      \begin{equation}
            [K^3_n,K^{\pm}_m] = \pm K^{\pm}_{n+m} \ , \quad [K^3_n,K^3_m]=\frac{k^{\prime}}{2} n \delta_{n+m,0} \ , \quad [K^+_n,K^-_m] = k^{\prime} n \delta_{n+m,0} + 2 K^3_{n+m} \ ,
      \end{equation}
      \begin{equation}
            [K^{\pm}_n,\chi^3_r] = \mp \chi^{\pm}_{n+r} \ , \quad [K^3_n,\chi^{\pm}_r]= \pm \chi^{\pm}_{n+r} \ , \quad [K^{\pm}_n,\chi^{\mp}_r] = \pm 2 \chi^3_{n+r} \ ,
      \end{equation}
      \begin{equation}
            \{\chi^+_r,\chi^-_s\} = k^{\prime} \delta_{r+s,0} \ , \quad \{\chi^3_r,\chi^3_s\} = \frac{k^{\prime}}{2} \delta_{r+s,0} \ .
      \end{equation}
\end{subequations}
Similar to AdS$_3$, if we define
\begin{equation}
      \mathcal{K}^{\pm} = K^{\pm} \mp \frac{2}{k^{\prime}} (\chi^3 \chi^{\pm}) \ , \quad \mathcal{K}^3 = K^3 - \frac{1}{k^\prime} (\chi^+ \chi^-) \ , 
\end{equation}
the currents $\mathcal{K}^a$ commute with the fermions $\chi^b$ and form an $\mathfrak{su}(2)_{k^\prime-2}$:
\begin{equation}
      \begin{split}
            [\mathcal{K}^3_n,\mathcal{K}^{\pm}_m] &= \pm \mathcal{K}^{\pm}_{n+m} \ , \quad [\mathcal{K}^3_n,\mathcal{K}^3_m]=\frac{k^\prime-2}{2} n \delta_{n+m,0} \ , \\ &[\mathcal{K}^+_n,\mathcal{K}^-_m] = (k^\prime-2) n \delta_{n+m,0} + 2 \mathcal{K}^3_{n+m} \ .
      \end{split}
\end{equation}
We will refer to $K^a$ and $\mathcal{K}^a$ as $\mathfrak{su}(2)$ coupled and decoupled currents, respectively.

The relation between $k$ and $k^\prime$ is imposed by the criticality of the world-sheet theory to be $k^\prime=k$. In order to see this, we note that the central charges of $\mathfrak{sl}(2,\mathbb{R})^{(1)}_k$, $\mathfrak{su}^{(1)}(2)_{k^{\prime}}$ and $\mathbb{T}^4$ are
\begin{equation} \label{eq:ads3-central-charge}
      c_{\mathfrak{sl}(2,\mathbb{R})^{(1)}_k} = 3 \left(\frac{k+2}{k} + \frac{1}{2}\right) \ , \quad c_{\mathfrak{su}(2)_{k^\prime}^{(1)}} = 3 \left(\frac{k^\prime-2}{k^\prime} + \frac{1}{2}\right) \ , \quad c_{\mathbb{T}^4}=6 \ .
\end{equation}
The criticality of the world-sheet string requires that
\begin{equation}
      c_{\mathfrak{sl}(2,\mathbb{R})^{(1)}_k} + c_{\mathfrak{su}(2)_{k^\prime}^{(1)}} + c_{\mathbb{T}^4} = 15 \ ,
\end{equation}
which imposes $k^\prime=k$. For this reason, from now on we assume that $k^\prime=k$. 

In the following, we will need the representations of the algebras discussed so far. For the compact manifold $\text{S}^3$, the finite-dimensional representations of $\text{SU}(2)$ appear on the world-sheet spectrum, which we denote as $\mathcal{D}^\prime_{j^\prime}$: $j^{\prime}$ is the decoupled $\text{SU}(2)$ spin defined via eqs.~\eqref{eq:su2-quadratic-casimir} and \eqref{eq:su2-def-spin}. In our conventions, $\mathcal{K}^a_0$ act as
\begin{equation} \label{eq:su2-rep-action}
	\mathcal{K}^\pm_0 \ket{j^\prime,m^\prime} = \mp (m^\prime\mp j^\prime) \ket{j^\prime,m^\prime\pm1} \ , \quad \mathcal{K}^3_0 \ket{j^\prime,m^\prime} = m^\prime \ket{j^\prime,m^\prime} \ .
\end{equation}
The unitarity of the affine Lie algebra $\mathfrak{su}(2)_{k-2}$ imposes that $0 \leq j^\prime \leq \frac{k-2}{2}$. For the non-compact manifold $\text{AdS}_3$, a similar analysis implies that the following representations appear on the world-sheet spectrum \cite{Maldacena:2000hw}: $\mathcal{C}^j_\alpha \times \mathcal{C}^j_\alpha$ and $\mathcal{D}^\pm_j \times \mathcal{D}^\pm_j$. Here $\mathcal{C}^j_\alpha$ is the continuous representation of the decoupled $\mathfrak{sl}(2,\mathbb{R})$ algebra while $\mathcal{D}^\pm_j$ are discrete representations. In these representations, $\mathcal{J}^a_0$ act as
\begin{equation} \label{eq:sl2-continuous-action}
	\mathcal{J}^\pm_0 \ket{j,m} = (m\pm j) \ket{j,m\pm1} \ , \quad \mathcal{J}^3_0 \ket{j,m} = m \ket{j,m} \ ,
\end{equation}
where $m\in\mathbb{Z}+j$, and the spin $j$ is specified by the value of the quadratic Casimir, see eqs.~\eqref{eq:sl2-quadratic-casimir} and \eqref{eq:sl2-def-spin}. We have defined and spelled out our conventions for all these representations, as well their fermionic counterparts, in more detail in Appendix~\ref{app:ws}. For $\mathbb{T}^4$, the representations are labelled by $\mathfrak{u}(1)$ left-moving momenta $\vec{p}_j$ with $j\in\{1,2,3,4\}$. For the purposes needed in this paper, we actually do not need to discuss these representations in any more detail.

\subsection{The physical fields and correlation functions} \label{sec:physical-fields}
The superstring theory discussed in the previous section possesses an $\mathcal{N}=1$ superconformal algebra. In the NS-sector, the following fields for $\text{AdS}_3$\footnote{$(AB)$ denotes the radial normal-ordering of $A$ and $B$, see e.g.\ \cite{DiFrancesco:1997nk}.}
\begin{subequations} \label{eq:ads3-n=1}
	\begin{equation}
		T^{\text{AdS}_3} = \frac{1}{2k}\left( \mathcal{J}^+ \mathcal{J}^- + \mathcal{J}^- \mathcal{J}^+ - 2 \mathcal{J}^3 \mathcal{J}^3 \right) + \frac{1}{2k} \left( -\psi^+\partial \psi^- - \psi^- \partial \psi^+ + 2\psi^3 \partial \psi^3 \right) \ ,
	\end{equation}
	\begin{equation}
		G^{\text{AdS}_3} = \frac{1}{k} \left( \mathcal{J}^+ \psi^- + \mathcal{J}^- \psi^+ - 2 \mathcal{J}^3 \psi^3 \right) - \frac{2}{k^2} ((\psi^+ \psi^-)\psi^3) \ ,
	\end{equation}
\end{subequations}
form an $\mathcal{N}=1$ superconformal algebra with the central charge given in \eqref{eq:ads3-central-charge}. S$^3$ fields, in the NS-sector, form an $\mathcal{N}=1$ superconformal algebra with the central charge given in \eqref{eq:ads3-central-charge} with $k^\prime=k$:
\begin{subequations} \label{eq:s3-n=1}
	\begin{equation}
		T^{\text{S}^3} = \frac{1}{2k}\left( \mathcal{K}^+ \mathcal{K}^- + \mathcal{K}^- \mathcal{K}^+ + 2 \mathcal{K}^3 \mathcal{K}^3 \right) + \frac{1}{2k} \left( -\chi^+\partial \chi^- - \chi^- \partial \chi^+ -2\chi^3 \partial \chi^3 \right) \ ,
	\end{equation}
	\begin{equation}
		G^{\text{S}^3} = \frac{1}{k} \left( \mathcal{K}^+ \chi^- + \mathcal{K}^- \chi^+ + 2 \mathcal{K}^3 \chi^3 \right) + \frac{2}{k^2} ((\chi^+ \chi^-)\chi^3) \ .
	\end{equation}
\end{subequations}
Similarly, $\mathbb{T}^4$ fields form an $\mathcal{N}=1$ superconformal algebra with $c_{\mathbb{T}^4}=6$, which we denote the generators as $T^{\mathbb{T}^4}$ and $G^{\mathbb{T}^4}$. In Appendix~\ref{app:ws}, we have spelled out our conventions for an $\mathcal{N}=1$ superconformal algebra with central charge $c$, and we have discussed in more detail the generators in the R-sector, as well as for $\mathbb{T}^4$. Having these, in either NS- or R-sectors, we define
\begin{equation} \label{eq:tot-t-g}
	T^{\text{tot}}_m = T^{\text{AdS}_3}_m + T^{\text{S}^3}_m + T^{\mathbb{T}^4}_m \ , \quad G^{\text{tot}}_r = G^{\text{AdS}_3}_r + G^{\text{S}^3}_r + G^{\mathbb{T}^4}_r \ .
\end{equation}
These generators form an $\mathcal{N}=1$ superconformal algebra with $c=15$.

The physical fields are then defined via super-Virasoro constraints, which depend on NS- and R-sectors. The world-sheet theory can be formulated in two equivalent ways: the old covariant quantization and the BRST formalism. As it will be useful later on, we will fix our conventions by briefly reviewing both formalisms. In the old covariant quantization, the physical state conditions are \cite{Green:1987sp,Polchinski:1998rq,Blumenhagen:2013fgp}\footnote{The right-moving physical state conditions, either in the old covariant quantization or BRST formalism, must be imposed as well. Here $L^{\text{tot}}_0$ denotes the weight with the conformal dimensions of Ramond ground states subtracted, see eqs.~\eqref{eq:ramond-L0-weight-shift} and \eqref{eq:t4-spin-fields}.}
\begin{equation} \label{eq:old-physical}
	(L^{\text{tot}}_m - \nu \delta_{m,0}) \Psi = G^{\text{tot}}_r \Psi = 0 \ , \quad (m\geq 0) \ , \quad (r\geq \nu) \ .
\end{equation}
where $\nu$ is a number that depends on the sector:
\begin{equation} \label{eq:def-nu}
	\nu =
	\begin{cases}
	\frac{1}{2} \ , & \text{NS sector} \\
	0 \ , & \text{R sector}
	\end{cases} \ .
\end{equation}
In addition to \eqref{eq:old-physical}, one must impose Gliozzi–Scherk–Olive (GSO) projection \cite{Gliozzi:1976qd} to ensure that the space-time theory is supersymmetric and tachyon-free. We will briefly discuss this in the next section. The no-ghost theorem \cite{Goddard:1972iy,Hwang:1990aq,Evans:1998qu,Maldacena:2000hw,Pakman:2003cu} shows that the conditions \eqref{eq:old-physical} indeed remove the negative-normed states provided that for the discrete representations $\mathcal{D}^{\pm}_j$,
\begin{equation} \label{eq:unitarity-bound}
	\frac{1}{2} < j < \frac{k+1}{2} \ ,
\end{equation}
where $j$ is the spin of the $\mathfrak{sl}(2,\mathbb{R})_{k+2}$ representations.

Alternatively, following \cite{Friedan:1985ge}, one can form a BRST operator $Q$ where the physical fields are described as a cohomology:
\begin{equation} \label{eq:brst-physical}
	Q V = 0 \ , \quad V \sim V + Q W \ .
\end{equation}
In order to discuss the BRST charge in more detail, recall that the superstring theory includes the diffeomorphism ghosts $(b,c)$ as well as the superdiffeomorphism ghosts $(\beta,\gamma)$. We have spelled out our conventions for these ghosts in Appendix~\ref{app:ws}. The BRST charge can then be written as \cite{Blumenhagen:2013fgp}
\begin{subequations} \label{eq:q-brst-charge}
\begin{equation}
	Q = Q^{(0)} + Q^{(1)} + Q^{(2)} \ ,
\end{equation}
where in our conventions
\begin{equation} \label{eq:brst-q0}
	Q^{(0)} = \oint \text{d}z \big[ c (T^{\text{tot}} + T_{\beta\gamma}) - b ((\partial c) c) \big]\ ,
\end{equation}
\begin{equation} \label{eq:brst-q1}
	Q^{(1)} = \oint \text{d}z \, \gamma \, G^{\text{tot}} \ ,
\end{equation}
\begin{equation} \label{eq:brst-q2}
	Q^{(2)} = - \oint \text{d}z \, \gamma^2 \, b \ .
\end{equation}
\end{subequations}
In superstring theory there is a notion of picture ambiguity, so that a physical field admits equivalent representations in different pictures. The picture number is defined as
\begin{equation} \label{eq:picture-number-def}
	P = -(\eta\xi) - (\beta\gamma) \ ,
\end{equation}
where $(\eta,\xi)$ are defined through the bosonization of $(\beta,\gamma)$, see eq.~\eqref{eq:beta-gamma-bosonization}. In particular, one requires that the physical fields lie in the \textit{small Hilbert space}, i.e.\ they do not depend on $\xi_0$, by imposing
\begin{equation} \label{eq:small-hilbert-space}
	\eta_0 V = 0 \ .
\end{equation}
The picture-raising operator defined via
\begin{equation} \label{eq:picture-raising}
	P_+ V = Q \xi_0 V \ ,
\end{equation}
then maps a physical field with the picture number $p$ to a physical field with picture number $p+1$.

As discussed in \cite{Goddard:1972iy,Kato:1982im,Friedan:1985ge,Polchinski:1998rq,Polchinski:1998rr,Asano:2003qb}, the old covariant quantization is equivalent to BRST formalism. Given a physical field in the old covariant quantization, how would one represent it in the BRST formalism? In the NS-sector, the canonical form of the physical fields are expressed in the picture $P=-1$ as
\begin{equation} \label{eq:ns-canonical}
	V_{\text{NS}} = c \, \bar c \, e^{\phi+\bar \phi} \, \Psi \ .
\end{equation}
One can check that $V_{\text{NS}}$ is BRST closed provided that $\Psi$ satisfies \eqref{eq:old-physical} both in the left- and right-moving sectors. Similarly, in the R-sector, the following field with picture $P=-\frac{1}{2}$,
\begin{equation} \label{eq:r-canonical}
	V_{\text{R}} = c \, \bar c \, e^{\frac{\phi}{2}+\frac{\bar \phi}{2}} \, \Psi \ ,
\end{equation}
is BRST closed provided that $\Psi$ satisfies \eqref{eq:old-physical} both in the left- and right-moving sectors.

Given a collection of $n$ physical fields $V_j(z_j,\bar z_j)$ that satisfy \eqref{eq:brst-physical} for both left- and right-moving sectors, the tree-level $n$-pt function is defined as\footnote{There is no integration in \eqref{eq:n-pt-function} if $n=2,3$. As explained in \cite{Witten:2012bh}, for the case of $2$-pt function one inserts an additional $0$-mode of the $c$-ghost, see eq.~\eqref{eq:2-pt-function-rule}.\label{footnote:2pt-function-special}}
\begin{equation} \label{eq:n-pt-function}
	\int_{\mathbb{C}^{n-3}} \text{d}^2 z_4 \cdots \text{d}^2 z_n \, \langle [\xi_0 \bar{\xi}_0 V_1](0,0) V_2(1,1) V_3(\infty,\infty) \prod_{j=4}^{n} b_{-1} \bar{b}_{-1} V_j(z_j,\bar z_j) \rangle \ .
\end{equation}
In Appendix~\ref{app:ws}, we have discussed in more detail various elements entering this correlator. Let us consider the case of $4$-pt function by denoting $z_4=z$ and $\bar{z}_4=\bar{z}$, and assuming that the two physical fields inserted at $(1,1)$ and $(z,\bar z)$ are identical fields in the R-sector, i.e.\
\begin{equation} \label{eq:special-RR-choice}
	V_2 = c \, \bar c \, e^{\frac{\phi}{2}+\frac{\bar \phi}{2}} \, \Psi_2 \ , \quad V_4 = P_+ \bar{P}_+ V_2 \ ,
\end{equation}
and the other two vertex operators are both either in NS-sector or both in R-sector. The reason for the picture-raised vertex operator $V_4$ will become clear momentarily. If both other fields are in the NS-sector, one has
\begin{equation} \label{eq:ns-sector-canonical-form}
	V_j = c \, \bar c \, e^{\phi+\bar \phi} \, \Psi_j \ , \quad j\in\{1,3\} \ .
\end{equation}
In fact, choosing $V_4$ to be picture-raised as in \eqref{eq:special-RR-choice}, the sum of the picture numbers is then correctly $(-2)$, see eq.~\eqref{eq:picture-sum-condition}. For the case where both fields are in R-sector, we choose\footnote{As discussed in Appendix~\ref{app:ws}, it does not matter how the picture number is chosen as long as the condition \eqref{eq:picture-sum-condition} is satisfied. The reason for the particular choice \eqref{eq:RR-picture-choice} is that then both NS and R-sectors have the same field insertions at $(1,1)$ and $(z,\bar z)$, see Section~\ref{sec:SFT}.}
\begin{equation} \label{eq:RR-picture-choice}
	V_1 = c \, \bar c \, e^{\frac{\phi}{2}+\frac{\bar \phi}{2}} \, \Psi_1 \ , \quad V_3 = c \, \bar c \, e^{\frac{3\phi}{2}+\frac{3\bar \phi}{2}} \, \Psi_3 \ .
\end{equation}
$V_3$ at picture $P=-\frac{3}{2}$ is chosen such that under $P_+ \bar P_+ V_3$ it is mapped to the field in the canonical picture $P=-\frac{1}{2}$. Having specified these choices, the 4-pt function of interest is 
\begin{equation} \label{eq:4-pt-function}
	\int_{\mathbb{C}} \text{d}^2 z \, \langle [\xi_0 \bar{\xi}_0 V_1](0,0) V_2(1,1) V_3(\infty,\infty) [b_{-1}\bar{b}_{-1} P_+ \bar{P}_+ V_2](z,\bar z) \rangle \ .
\end{equation}

\subsection{CCY, LRT and the world-sheet parity} \label{sec:leading-regge}
As discussed in the Introduction, the superstring theory on $\text{AdS}_3 \times \text{S}^3 \times \mathbb{T}^4$ is expected to be holographically dual to a $2$d CFT that possesses a small space-time $\mathcal{N}=(4,4)$ superconformal algebra with $c=6kp$, see Appendix~\ref{app:n4}.\footnote{Here $p$ is the number of fundamental strings F1.} One would expect that the string spectrum is organized into representations of this superconformal algebra. In fact, the world-sheet generators of the dual superconformal algebra have been explicitly found in \cite{Giveon:1998ns,Kutasov:1998zh,Kutasov:1999xu,Eberhardt:2019qcl}. In particular, M\"{o}bius generators of the dual CFT are represented on the world-sheet via
\begin{equation} \label{eq:st-mobius}
	L_1^{\text{ST}} = J^-_0 \ , \quad L_0^{\text{ST}} = J^3_0 \ , \quad L_{-1}^{\text{ST}} = J^+_0 \ ,
\end{equation}
where it is important to note that the \textit{coupled} $\mathfrak{sl}(2,\mathbb{R})_k$ currents $J^a$ have been used.\footnote{One way of seeing this is by noting that $L_m^{\text{ST}}$ must map physical fields to physical fields on the world-sheet. While $J^a_0$ commute with $L^{\text{tot}}_m$ and $G^{\text{tot}}_r$, $\mathcal{J}^a_0$ do \textit{not} commute with $G^{\text{tot}}_r$.} Similarly, $K^a_0$ represent the global part of the space-time $\mathfrak{su}(2)$ R-symmetry generators, see Appendix~\ref{app:n4}. Having \eqref{eq:st-mobius}, given a physical state $\ket{\Psi}$, its space-time energy $\Delta$ and spin $S$ are given by
\begin{equation} \label{eq:energy-spin-def}
	(J^3_0+\bar{J}^3_0) \ket{\Psi} = \Delta \ket{\Psi} \ , \quad (J^3_0-\bar{J}^3_0) \ket{\Psi} = S \ket{\Psi} \ .
\end{equation}
In \cite{Ferreira:2017pgt}, the superstring spectrum was studied and in particular, states on the leading Regge trajectory (LRT) were identified. In this subsection, we will provide an explicit formula for the states on the leading Regge trajectory in our conventions. We will also discuss the world-sheet parity action, to be studied in more detail below, on the CCY state considered \cite{Cho:2018nfn}. In the next subsection, we will construct the state studied using the QSC in \cite{Ekhammar:2026ykk}.

In order to discuss the state mentioned above, we will focus on discrete representations $\mathcal{D}^+_j$. The reason for this particular choice is that we are eventually interested in a large $k$ limit (while the unitarity bound \eqref{eq:unitarity-bound} is respected) where the energy of states scales as $\sqrt{k}$. Following \cite{Maldacena:2000hw}, one then expects to find such states in discrete representations.

Let us first consider the GSO projection for a physical field $\ket{\Psi}$ in the discrete representation $\mathcal{D}^+_j$. This is discussed in detail in \cite{Ferreira:2017pgt} and here we merely state the final result: let us assume that the total left- and right-moving excitation numbers of $\ket{\Psi}$ are $N$ and $\bar N$, respectively. Upon defining
\begin{equation} \label{eq:n-def}
	n = N - \nu \ ,
\end{equation}
see \eqref{eq:def-nu}, the GSO projection enforces $n\in\mathbb{Z}_{\geq 0}$. Moreover, in the R-sector, it imposes additional constraints on the spin fields, see \eqref{eq:gso-projection-r-sector}. The mass-shell condition in \eqref{eq:old-physical} implies
\begin{equation} \label{eq:delta-def}
	-\frac{j(j-1)}{k} + \frac{j^\prime(j^\prime+1)}{k} + h^{\mathbb{T}^4} + n = 0 \ ,
\end{equation}
where we have used eqs.~\eqref{eq:l0-ads3-def} and \eqref{eq:l0-s3-def} together with \eqref{eq:n-def}.\footnote{In particular, $h^{\mathbb{T}^4}$ is the weight of the $\mathbb{T}^4$ state minus the weight of the NS- or R-sector ground state $h^{\mathbb{T}^4}_0$, see \eqref{eq:l0-m4-def}.} Solving for $j$, we get
\begin{equation} \label{eq:sol-j}
	j = \frac{1}{2} \left( 1 + \sqrt{(2j^\prime+1)^2+ 4 k (n+h^{\mathbb{T}^4})} \right) \ .
\end{equation}
We now review the classification of physical states in discrete representations studied in detail in \cite{Ferreira:2017pgt}, see in particular eq.~\eqref{eq:label-states}. As each field on the world-sheet increases or decreases the $J^3_0$ eigenvalue at most by $1$, see eqs.~\eqref{eq:decoupled-sl2-commutators}, for the space-time lowest weight states one has that $j-n-1 \leq m_3 \leq j+n+1$ where $m_3$ labels the coupled $J^3_0$ eigenvalue. Suppressing other internal quantum numbers, one can then label physical states as
\begin{equation} \label{eq:label-states}
	\ket{(n,r,\bar r)} = \ket{j+r-n-1} \overline{\ket{j+\bar r - n - 1}} \ , \qquad 0 \leq r,\bar r \leq 2n+2 \ ,
\end{equation}
where
\begin{equation} \label{eq:energy-spin-r-rbar}
	\Delta = 2j + r + \bar r - 2 n - 2 \ , \quad S = r - \bar r \ ,
\end{equation}
see eq.~\eqref{eq:energy-spin-def}. Moreover, given $(n,r,\bar r)$, there might be multiple such states and with a slight abuse of notation, we will show them collectively as in \eqref{eq:label-states}. In this section, we specialize to $h^{\mathbb{T}^4}=0$ and suppress the resulting trivial representation of $\mathbb{T}^4$. For a given $j^\prime$, this specialization gives the least space-time energy at a fixed spin because $h^{\mathbb{T}^4} \geq 0$ enters \eqref{eq:sol-j}; see also the discussions below \eqref{eq:su2-rep-action} and \eqref{eq:l0-m4-def}.

\begin{figure}[htbp]
\centering
\begin{tikzpicture}
  \node (bb) at (0,-3) {$\ket{\mathcal{V}^{\text{CCY}}_n}$};
  \node (tb) at (3,0) {$\ket{\mathcal{V}^{\text{LRT}}_n}$};
  \node (tt) at (0,3) {$\ket{\Omega(\mathcal{V}^{\text{CCY}}_n)}_{\text{top}}$};
  \node (bt) at (-3,0) {$\ket{\rho(\mathcal{V}^{\text{LRT}}_n)}$};
  \draw[blue] (bb) -- node[below right] {} (tb);
  \draw[blue] (tb) -- node[above right] {} (tt);
  \draw[blue] (bb) -- node[below left] {} (bt);
  \draw[blue] (bt) -- node[above left] {} (tt);
  \draw[purple,<->,densely dashed] (bb) -- node[pos=0.2,right] {} (tt);
  \draw[purple,<->,densely dashed] (bt) -- node[pos=0.2,above] {} (tb);
\end{tikzpicture}

\caption{The relation between CCY, its image under world-sheet parity and LRT. Here by an abuse of notation we have denoted $\ket{\rho(\mathcal{V}^{\text{LRT}}_n)}=\ket{j-n-1}\overline{\ket{j+n+1}}$. As shown in \cite{Ferreira:2017pgt}, the number of physical states are invariant under $r \mapsto 2n+2-r$ and they form a `diamond', as we have also depicted in this figure.}
\label{fig:reflection-diamond}
\end{figure}

In \cite{Ferreira:2017pgt}, it is shown that states on the leading Regge trajectory with the top component's spin being $(2n+2)$ are the unique states with $(r,\bar r)=(2n+2,0)$.\footnote{The energy of the leading Regge trajectory (LRT) states with nonzero \(j^\prime\) was computed in \cite{Jiang:2025oar} using the AdS Virasoro–Shapiro method for the pure RR case. We emphasize that the notion of LRT used in this paper differs from that of \cite{Ferreira:2017pgt}, where the LRT is defined as the physical state with the lowest energy at fixed space-time spin. In the present work, LRT instead refers to the leading operator appearing in the \(T\)-channel expansion of the \(\langle JJ11\rangle\) correlator where $J=2j^\prime+1$, see \cite{Jiang:2025oar} for more details. In other words, given a fixed $j^\prime$, the LRT state has the least space-time energy.} Similarly, it was shown that for a given $n$, there is a unique state with $(r,\bar r)=(0,0)$. The uniqueness of the state with $r=2n+2$ is indeed not a coincidence, as noted, following a character analysis in that paper. More specifically, it was explained that the degeneracy of states is invariant under sending $r \mapsto 2n+2-r$, or equivalently, the degeneracy of $\ket{j+m}$ and $\ket{j-m}$ is the same. In order to reformulate this symmetry in a language suitable for our purposes, recall that the Weyl group of the Lie algebra $\mathfrak{sl}(2,\mathbb{R})$ is $\mathbb{Z}_2$. We will refer to its non-trivial element as the (reduced) left-moving reflection and denote it as $\rho_L$. The action of $\rho_L$ on the left-moving currents and fermions of $\mathfrak{sl}(2,\mathbb{R})^{(1)}_k$ is defined as
\begin{subequations} \label{eq:reflection-map}
	\begin{equation}
	\rho_L(J^{\pm}_p) = J^{\mp}_p \ , \quad \rho_L(J^3_p) = - J^3_p \ ,
	\end{equation}
	\begin{equation}
		\rho_L(\psi^{\pm}_s) = \psi^{\mp}_s \ , \quad \rho_L(\psi^3_s) = - \psi^3_s \ .
	\end{equation}
\end{subequations}
The right-moving reflection $\rho_R$ is defined analogously. It can be checked that $\rho_L$ preserves the (anti-)commutation relations in eqs.~\eqref{eq:decoupled-sl2-commutators}, as well as $L^{\text{tot}}_p$ and $G^{\text{tot}}_s$, see eqs.~\eqref{eq:ads3-n=1}. Given a level $N$, set $V_{N,m}$ to be the vector space spanned by products of oscillators that have $\text{ad}(J^3_0)(\cdot)=[J^3_0,\cdot]$ eigenvalue equal $m$. $\rho_L$ maps $V_{N,m}$ to $V_{N,-m}$ bijectively, and therefore, $\dim V_{N,m} = \dim V_{N,-m}$. The reason for referring to $\rho_L$ as the \textit{reduced} reflection is as follows: as discussed in \cite{Ferreira:2017pgt}, the total character is \textit{not} invariant under $\rho_L$ if one also acts on $\mathcal{D}^+_j$ which is mapped to $\mathcal{D}^-_{j}$. In other words, while the argument reviewed above shows the equality of $\dim V_{N,m}$ and $\dim V_{N,-m}$, it does not by itself explain how a physical state with label $r$ is mapped to another physical state with $r^\prime=2n+2-r$. Despite this, we will show below that there is a relatively simple way of answering this question for the extremal cases.

Now we would like to explicitly write the states mentioned in the title of this section. Let us start with the state found in \cite{Ferreira:2017pgt} and considered in \cite{Cho:2018nfn}: substituting $(r,\bar r)=(0,0)$ in \eqref{eq:label-states}, one can see that $J^3_0=j-n-1$ and $\bar{J}^3_0=j-n-1$. As mentioned above, each field can decrease the $J^3_0$ eigenvalue at most by $1$. Therefore, the only possibility is
\begin{equation} \label{eq:xi-states}
	\ket{\mathcal{V}^{\text{CCY}}_n} = \ket{j-n-1} \overline{\ket{j-n-1}} \ ,
\end{equation}
where we have defined
\begin{equation} \label{eq:leading-regge-right}
	\ket{j-n-1} = (\mathcal{J}^-_{-1})^n \psi^-_{-\frac{1}{2}} \ket{j,j} \ket{j^\prime,j^\prime} \ , \quad \overline{\ket{j-n-1}} = (\bar{\mathcal{J}}^-_{-1})^n \bar{\psi}^-_{-\frac{1}{2}} \overline{\ket{j,j} \ket{j^\prime,j^\prime}} \ .
\end{equation}
It is straightforward to check that $\ket{\mathcal{V}^{\text{CCY}}_n}$ is physical, i.e.\ it satisfies \eqref{eq:old-physical} for any $n \in \mathbb{Z}_{\geq 0}$.\footnote{Moreover, using \eqref{eq:super-charges-holo}, it can be verified that $\ket{\mathcal{V}^{\text{CCY}}_n}$ is a bottom component of a space-time $\mathcal{N}=4$ multiplet.} Note that the space-time energy and spin of this state are
\begin{equation} \label{eq:ccy-exact-spectrum-ns-ns}
	\Delta_{\text{CCY}} = 2(j-n-1) \ , \quad S_{\text{CCY}} = 0 \ ,
\end{equation}
where
\begin{equation} \label{eq:regge-sol-j}
	j = \frac{1}{2} \left( 1 + \sqrt{(1+2j^\prime)^2+4 k n} \right) \ .
\end{equation}
In particular, at large $k$ one has
\begin{equation} \label{eq:CCY-NS-NS-large-k}
	\Delta_{\text{CCY}} = 2\sqrt{n \, k} - 2n - 1 + \frac{J^2}{4 \sqrt{n} \sqrt{k}} + \mathcal{O}(k^{-\frac{3}{2}}) \ .
\end{equation}
A general procedure for finding the leading Regge trajectory states explicitly is discussed in \cite{Ferreira:2017pgt}. In particular, it was argued that the state $\ket{(n,2n+2,0)}$, see eq.~\eqref{eq:label-states}, is the top component of a long multiplet in the space-time $\mathcal{N}=(4,4)$ superconformal algebra.\footnote{For $n=0$, the long multiplet is shortened, see Appendix~B of \cite{Ferreira:2017pgt}.} The right-moving part has $\bar{J}^3_0=j-n-1$ and therefore is given by the same expression as in \eqref{eq:leading-regge-right}. The left-moving part is more involved, since it has $J^3_0=j+n+1$ and an explicit expression was not written in \cite{Ferreira:2017pgt}. As mentioned in that paper, these states can be found by directly imposing the physical state conditions in \eqref{eq:old-physical} on the most general linear combination of states, $\ket{\Psi}$, that has a fixed excitation level $N=n+\frac{1}{2}$, and furthermore satisfies
\begin{equation} \label{eq:left-st-conditions}
	J^3_0 \ket{\Psi} = (j+n+1) \ket{\Psi} \ , \quad J^-_0 \ket{\Psi} = 0 \ .
\end{equation}
As explained in Appendix~\ref{app:n4}, the $J^-_0=L^{\text{ST}}_{1}=0$ condition, see eqs.~\eqref{eq:st-mobius} and \eqref{eq:top-component-unique-condition}, is imposed in order to uniquely select the top component of a long multiplet. We have explicitly checked in \texttt{Mathematica} \cite{Headrick:Virasoro} that for $n\in\{0,1,2\}$, the physical state conditions \eqref{eq:old-physical} together with \eqref{eq:left-st-conditions}, have only one solution up to an overall normalization. In our conventions, these states can be represented as
\begin{equation} \label{eq:leading-regge-left}
	\ket{j+n+1} = \frac{(\partial_y^{2n+2})|_{y=0}}{(2n+2)!} \left(\sum_{m=0}^{2n+2} y^m e^{y J^+_0} (\mathcal{J}^-_{-1})^n \psi^-_{-1/2} e^{-y J^+_0} \ket{j,j+m} \ket{j^\prime,j^\prime} \right) \ .
\end{equation}
Although this equation may look complicated, it can be explained using representation theory \cite{Ferreira:2017pgt}: $(\mathcal{J}^-_{-1})^n \psi^-_{-\frac{1}{2}}$ has $J^3_0=-n-1$ eigenvalue and is annihilated by the adjoint action of $J^-_0$. Therefore, it is the bottom component of a spin $n+1$ finite representation of the coupled $\mathfrak{sl}(2,\mathbb{R})$. For the actual physical state $\ket{j-n-1}$, this spin $n+1$ representation is tensored with $\mathcal{D}^+_j$. Following the discussion about the reflection above, we expect the state with $r=2n+2$ is a component in the same tensor product that has $J^3_0 = j+n+1$ and is annihilated by $J^-_0$. In particular, let us write
\begin{equation}
	e^{y J^+_0} [(\mathcal{J}^-_{-1})^n \psi^-_{-1/2}] e^{-y J^+_0} = \sum_{p=0}^{2n+2} \frac{y^p}{p!} \text{ad}^p(J^+_0) [(\mathcal{J}^-_{-1})^n \psi^-_{-1/2}] \ .
\end{equation}
Then a general state in the mentioned tensor product with $J^3_0=j+n+1$ is
\begin{equation}
	\sum_{p=0}^{2n+2} \frac{u_p}{p!} \text{ad}^p(J^+_0) [(\mathcal{J}^-_{-1})^n \psi^-_{-1/2}] \ket{j,j+2n+2-p} \ket{j^\prime,j^\prime} \ ,
\end{equation}
for some coefficients $u_p$. Imposing the lowest weight condition in \eqref{eq:left-st-conditions} implies that $u_p$ is a constant not depending on $p$ and one recovers \eqref{eq:leading-regge-left}. Moreover, we have directly confirmed that $\ket{j+n+1}$ is indeed physical: one can replace $\mathcal{J}^-_{-1}$ with $J^-_{-1}$ inside \eqref{eq:leading-regge-left} without changing the state. Then using
\begin{equation}
	[G^{\text{tot}}_r,J^-_m] = - m \, \psi^-_{r+m} \ , \quad \{ G^{\text{tot}}_r,\psi^-_s \} = J^-_{r+s} \ ,
\end{equation}
it is straightforward to confirm that $\ket{j+n+1}$ is physical. To summarize, the top component of the leading Regge trajectory multiplet is
\begin{equation} \label{eq:leading-regge-trajectory-states}
	\ket{\mathcal{V}^{\text{LRT}}_n} =  \ket{j+n+1} \overline{\ket{j-n-1}} \ ,
\end{equation}
where $\ket{j+n+1}$ and $\overline{\ket{j-n-1}}$ are given in eqs.~\eqref{eq:leading-regge-left} and \eqref{eq:leading-regge-right}, respectively. Moreover, $j$ is defined in \eqref{eq:regge-sol-j} and the space-time energy and spin of the LRT bottom component for $n\geq 1$, using \eqref{eq:energy-spin-def}, are
\begin{equation} \label{eq:reggeNSNS}
	\Delta_{\text{LRT}} = 2j - 2 = -1 + \sqrt{(1+2j^\prime)^2 +4k n} \ , \quad S_{\text{LRT}} = 2n \ .
\end{equation}
Let us finally discuss the action of the world-sheet parity $\Omega$ on the CCY state defined in \eqref{eq:leading-regge-right}. As discussed in \cite{Witten:1983ar}, since the WZ term changes sign, the WZW action is not invariant under sending $\sigma \to -\sigma$ where $\sigma$ denotes the space-like string coordinate. Throughout this paper we have referred to this transformation as world-sheet parity $\Omega$. Combining $\Omega$ with the map that sends $g\mapsto g^{-1}$ where $g$ is a group element is indeed a symmetry of the WZW theory. In fact, this combined transformation exchanges the left- and right-moving algebras. Despite this, as $\Omega$ reverses the sign of the WZ term, it relates the theories at $q$ and $-q$ while preserving the space-time energy and consequently, we expect the spectrum to satisfy \eqref{eq:mirror-map-spectrum-symmetry}. It is, however, not immediately clear how to express parity images of states in our affine currents conventions, since the currents on the left- and right are not mapped to one another under $\Omega$. In Section~\ref{sec:SFT}, we will report a first-principles string field theory calculation of the perturbative space-time energies of the proposed images of the states considered in this paper under $\Omega$. This will provide evidence for the proposed identification of the states entering \eqref{eq:mirror-map-spectrum-symmetry}. In fact, as $g\mapsto g^{-1}$ is a space-time $\mathbb{Z}_2$ action and motivated by the Weyl reflection discussed below Figure~\ref{fig:reflection-diamond}, we propose that the top component of the parity image of the CCY state is
\begin{equation} \label{eq:reflected-ccy-state}
	\ket{\Omega(\mathcal{V}^{\text{CCY}}_n)}_{\text{top}} =  \ket{j+n+1} \overline{\ket{j+n+1}} \ ,
\end{equation}
where $\ket{j+n+1}$ is defined in \eqref{eq:leading-regge-left}. We note that the space-time energy and spin of the corresponding bottom component are (for $n\geq 1$)
\begin{equation} \label{eq:reflection-ccy-exact-spectrum-ns-ns}
	\Delta_{\Omega(\text{CCY})} = 2 j + 2n - 2 \ , \quad S_{\Omega(\text{CCY})} = 0 \ .
\end{equation}
In particular, at large $k$ we have
\begin{equation} \label{eq:reflection-ccy-spectrum-large-k}
	\Delta_{\Omega(\text{CCY})} = 2\sqrt{n \, k} + 2n - 1 + \frac{J^2}{4\sqrt{n} \sqrt{k}} +  \mathcal{O}(k^{-\frac{3}{2}}) \ .
\end{equation}
Comparing eq.~\eqref{eq:reflection-ccy-exact-spectrum-ns-ns} with \eqref{eq:ccy-exact-spectrum-ns-ns}, and using eq.~\eqref{eq:regge-sol-j}, we see that the CCY state and its image under $\Omega$ both have the same coefficients at any odd power of $\sqrt{k}$ while the $k^0$ terms are equal only upon sending $n\mapsto -n$. As we will discuss in Section~\ref{sec:SFT}, in the presence of RR flux more precisely we have:\footnote{Note that we have $\Delta(\ket{\Omega(\mathcal{V}^{\text{CCY}}_n)},q)+4=\Delta(\ket{\Omega(\mathcal{V}^{\text{CCY}}_n)}_{\text{top}},q)$.}
\begin{equation} \label{eq:ccy-mirror-equation}
	\Delta(\ket{\mathcal{V}^{\text{CCY}}_n},q) = \Delta(\ket{\Omega(\mathcal{V}^{\text{CCY}}_n)},-q) \ .
\end{equation}

\subsection{The QSC state} \label{sec:extremal-r-sector}
We now explicitly construct the state considered in \cite{Ekhammar:2026ykk} at pure NS-NS point. We will denote by $\Delta_{\text{QSC}}$ and $(S-1)$ the energy and spin of the bottom component of its supermultiplet in the pure NS-NS background, respectively.\footnote{We note that compared to \cite{Ekhammar:2026ykk}, we have exchanged the left- and right-moving sectors.} Although that paper is for pure RR flux, in eq.~(15) of that paper, one does not expect the leading term in $\sqrt{k}$ to depend on the amount of fluxes. Moreover, although the $k^{0}$ term can in principle depend on fluxes, for the leading Regge trajectory it does not, see \cite{Chester:2024wnb}. Therefore, in order to find their state at pure NS-NS point, we assume
\begin{equation} \label{eq:qsc-ansatz}
	\Delta_{\text{QSC}} \sim \sqrt{2 S} \sqrt{k} - 1 + \mathcal{O}(1/\sqrt{k}) \ .
\end{equation}
In fact, as it is already clear from the previous section, it is technically easier to find the top component rather than the bottom component which will only shift $\Delta_{\text{QSC}} \mapsto \Delta_{\text{QSC}}+2$ (for $n\geq 1$, which is in a long multiplet, also see below \eqref{eq:qsc-state-form}). Assuming the labelling discussed in \eqref{eq:label-states}, we want to find $r$ and $\bar r$ corresponding to the top component of the left-moving part and the bottom of the right-moving part. Expanding \eqref{eq:energy-spin-r-rbar} at large $k$ (assuming $h^{\mathbb{T}^4}=0$) using \eqref{eq:sol-j}, and comparing with $\Delta_{\text{QSC}}+2$, see \eqref{eq:qsc-ansatz}, fixes\footnote{These are the states called `extremal R-sector states' in \cite{Ferreira:2017pgt} but an explicit expression was not provided.}
\begin{equation}
	r = 2n+\frac{3}{2} \ , \quad \bar r = \frac{1}{2} \ ,
\end{equation}
where $S=2n$. This in particular shows that the corresponding state is in the R-sector. Moreover, the R-charges, as mentioned in the Introduction, are $(\tfrac{J-2}{2},\tfrac{J}{2})$ where
\begin{equation}\label{eq:definition-J}
	J = 2j^\prime + 1 \ .
\end{equation}
We will write the top component in picture $P=-\frac{1}{2}$ as
\begin{equation} \label{eq:qsc-state-form}
	\ket{\mathcal{V}^{\text{QSC}}_n}_{P=-\frac{1}{2}} \stackrel{!}{=} \ket{j+n+\tfrac{1}{2};\varepsilon=-}_{P=-\frac{1}{2}} \overline{\ket{j-n-\tfrac{1}{2};\varepsilon=+}}_{P=-\frac{1}{2}} \, \Theta^{\alpha\beta} \bar{\Theta}^{\bar{\alpha}\bar{\beta}} \ ,
\end{equation}
and explicitly construct the left- and right-moving components. We note that the subscript $P=-\frac{1}{2}$ refers to the picture number $(-\frac{1}{2})$, see eq.~\eqref{eq:r-canonical}. $\varepsilon\in\{+,-\}$ is a doublet index where labels the $\text{SU}(2)$ representation as $j^\prime + \frac{\varepsilon}{2}$.\footnote{$\rho,\varepsilon\in\{+,-\}$ inside an expression are $\pm 1$ here and below, e.g.\ in \eqref{eq:alpha-omega-epsilon}.} $\Theta^{\alpha\beta}$ are the $\mathbb{T}^4$ spin fields, see eq.~\eqref{eq:t4-spin-fields}. Before we continue, let us mention that one expects that when $n=0$, the multiplet is protected. The reason is that the spin of the top component is ($2n+1$), so if $n=0$ the multiplet is shortened. Since we assume that $k$ is large enough, we will search for states \eqref{eq:qsc-state-form} in the discrete representations $\mathcal{D}^+_j$.

Let us start with the right-moving state. This state has $\bar{J}^3_0=j-n-\frac{1}{2}$ eigenvalue. Similar to the argument that led us to \eqref{eq:xi-states}, one can fairly simply show that the only possibility is to act by exactly $n$ operators chosen out of $\{\bar{\mathcal{J}}^-_{-1},\bar{\psi}^-_{-1}\}$ on the highest weight states. The $\text{S}^3$ part of the state is in fact uniquely fixed, since it has R-charge $j^\prime+\frac{1}{2}$, see eq.~\eqref{eq:definition-J}, and no $\text{S}^3$ oscillators. Thus, in our conventions, for $n\geq 0$ we have
\begin{gather} \label{eq:qsc-state-right}
	\begin{align}
		\overline{\ket{j-n-\tfrac{1}{2};\varepsilon=+}}_{P=-\frac{1}{2}} = &(\bar{\mathcal{J}}^-_{-1})^n \overline{\ket{j,j,-;j^\prime,j^\prime,+;-}} \\ &+  \alpha^{\rho=-,\varepsilon=+}_n (\bar{\mathcal{J}}^-_{-1})^{n-1}\bar{\psi}^-_{-1} \overline{\ket{j,j,-;j^\prime,j^\prime,+;+}} \ , \nonumber
	\end{align}
\end{gather}
see eq.~\eqref{eq:generic-spinfield-ads3-s3-states} for our conventions. In fact, $\alpha^{\rho=-,\varepsilon=+}_n$ is fixed by the physical state conditions \eqref{eq:old-physical} to be
\begin{equation} \label{eq:alpha-omega-epsilon}
	\alpha_n^{\rho,\varepsilon} = - n - \rho(j-\frac{1}{2}) - \varepsilon(j^\prime+\frac{1}{2}) \ .
\end{equation}
For later usages, we have defined the more general formula \eqref{eq:alpha-omega-epsilon}.\footnote{We note that $\alpha^{\rho=-,\varepsilon=+}_{n=0}=\alpha^{\rho=+,\varepsilon=-}_{n=0}=0$ on-shell.} The reason for the choice `$\gamma=-$' for the outer index of the spin fields in the first line of \eqref{eq:qsc-state-right}, see eq.~\eqref{eq:generic-spinfield-ads3-s3-states}, is the following: we are looking for a state that is \textit{not} in the same multiplet as $\overline{\ket{j-n-1}}$, see \eqref{eq:leading-regge-right}. The global modes of the right-moving space-time supercharges, in our conventions are \cite{Kutasov:1999xu}
\begin{equation} \label{eq:super-charges-holo}
	\oint \text{d} \bar{z} \, e^{\bar{\phi}/2} \bar{S}^{\alpha\beta}_+ \bar{\Theta}^{\eta,+} \ ,
\end{equation}
see Appendix~\ref{app:ws}. In particular, this operator anti-commutes with the BRST charge defined in eqs.~\eqref{eq:q-brst-charge}, while the other outer index does not anti-commute with $Q$. This justifies the choice of the outer index `$\gamma=-$' in \eqref{eq:qsc-state-right}. Imposing the GSO projection, see eq.~\eqref{eq:gso-projection-r-sector}, implies that both $\beta$ and $\bar{\beta}$ are `$-$' in \eqref{eq:qsc-state-form}. In passing, we also mention that $\overline{\ket{j-n-\tfrac{1}{2};\varepsilon=-}}_{P=-\frac{1}{2}}$ can be constructed for $n\geq 1$:
\begin{align} \label{eq:qsc-state-right-epsilon-}
	&\overline{\ket{j-n-\tfrac{1}{2};\varepsilon=-}}_{P=-\frac{1}{2}}  = (\bar{\mathcal{J}}^-_{-1})^n \overline{\ket{j,j,-;j^\prime,j^\prime,-;-}} - (\bar{\mathcal{J}}^-_{-1})^n \overline{\ket{j,j,-;j^\prime,j^\prime-1,+;-}} \\ &+ \alpha^{\rho=-,\varepsilon=-}_n (\bar{\mathcal{J}}^-_{-1})^{n-1}\bar{\psi}^-_{-1} \overline{\ket{j,j,-;j^\prime,j^\prime,-;+}} - \alpha^{\rho=-,\varepsilon=-}_n (\bar{\mathcal{J}}^-_{-1})^{n-1}\bar{\psi}^-_{-1} \overline{\ket{j,j,-;j^\prime,j^\prime-1,+;+}} \ , \nonumber
\end{align}
where $\alpha^{\rho=-,\varepsilon=-}_n$ is fixed by the physical state conditions \eqref{eq:old-physical}, see eq.~\eqref{eq:alpha-omega-epsilon}. Having this, similar to the argument that led us to the leading Regge trajectory state in \eqref{eq:leading-regge-left}, the left-moving states $\ket{j+n+\tfrac{1}{2};\varepsilon=\pm}$ can be obtained using $\ket{j-n-\frac{1}{2};\varepsilon=\pm}$. Via imposing
\begin{equation} \label{eq:left-qsc-st-conditions}
	J^3_0 \ket{\Psi} = (j+n+\frac{1}{2}) \ket{\Psi} \ , \quad J^-_0 \ket{\Psi} = 0 \ ,
\end{equation}
it turns out that for $n\geq 1$,
\begin{align} \label{eq:qsc-state-left-epsilon=+}
	&\ket{j+n+\tfrac12;\varepsilon=+}_{P=-\frac{1}{2}} \nonumber \\ &= \frac{\partial_y^{2n+1}|_{y=0}}{(2n+1)!} \Big(\sum_{m=0}^{2n+1} \sum_{r=0}^1 y^{m+r} e^{y J^+_0} (\mathcal{J}^-_{-1})^n e^{-y J^+_0} \ket{j,j+m,-1+2r;j^\prime,j^\prime,+;-} \Big) \\ &+ \alpha^{\rho=+,\varepsilon=+}_n \frac{\partial_y^{2n+1}|_{y=0}}{(2n+1)!} \Big(\sum_{m=0}^{2n+1} \sum_{r=0}^1 y^{m+r} e^{y J^+_0} (\mathcal{J}^-_{-1})^{n-1}\psi^-_{-1} e^{-y J^+_0} \ket{j,j+m,-1+2r;j^\prime,j^\prime,+;+} \Big) \ , \nonumber
\end{align}
where $\alpha^{\rho=+,\varepsilon=+}_n$ given by \eqref{eq:alpha-omega-epsilon} is fixed by the physical state conditions \eqref{eq:old-physical}. Finally, for $n\geq 0$ we also have
\begin{align} \label{eq:qsc-state-left}
	&\ket{j+n+\tfrac12;\varepsilon=-}_{P=-\frac{1}{2}} \nonumber \\ &= \frac{\partial_y^{2n+1}|_{y=0}}{(2n+1)!} \Big(\sum_{m=0}^{2n+1} \sum_{r=0}^1 y^{m+r} e^{y J^+_0} (\mathcal{J}^-_{-1})^n e^{-y J^+_0} \ket{j,j+m,-1+2r;j^\prime,j^\prime,-;-} \Big) \nonumber \\ &- \frac{\partial_y^{2n+1}|_{y=0}}{(2n+1)!} \Big(\sum_{m=0}^{2n+1} \sum_{r=0}^1 y^{m+r} e^{y J^+_0} (\mathcal{J}^-_{-1})^n e^{-y J^+_0} \ket{j,j+m,-1+2r;j^\prime,j^\prime-1,+;-} \Big) \\ &+ \alpha^{\rho=+,\varepsilon=-}_n \frac{\partial_y^{2n+1}|_{y=0}}{(2n+1)!} \Big(\sum_{m=0}^{2n+1} \sum_{r=0}^1 y^{m+r} e^{y J^+_0} (\mathcal{J}^-_{-1})^{n-1}\psi^-_{-1} e^{-y J^+_0} \ket{j,j+m,-1+2r;j^\prime,j^\prime,-;+} \Big) \nonumber \\ &-\alpha^{\rho=+,\varepsilon=-}_n \frac{\partial_y^{2n+1}|_{y=0}}{(2n+1)!} \Big(\sum_{m=0}^{2n+1} \sum_{r=0}^1 y^{m+r} e^{y J^+_0} (\mathcal{J}^-_{-1})^{n-1}\psi^-_{-1} e^{-y J^+_0} \ket{j,j+m,-1+2r;j^\prime,j^\prime-1,+;+} \Big) \ , \nonumber
\end{align}
where again $\alpha^{\rho=+,\varepsilon=-}_n$ is fixed by the physical state conditions \eqref{eq:old-physical}, see \eqref{eq:alpha-omega-epsilon}. To summarize, the state considered in \cite{Ekhammar:2026ykk}, at the WZW point, is given by\footnote{$\alpha,\bar{\alpha}\in\{+,-\}$ are free parameters, and so, we find $4$ such states. In fact, they can be decomposed into a singlet and a triplet. The calculations in Section~\ref{sec:SFT} is insensitive to these doublet indices, and therefore, for notational convenience we do not write them explicitly.}
\begin{equation} \label{eq:qsc-state-result}
	\ket{\mathcal{V}^{\text{QSC}}_n}_{P=-\frac{1}{2}} = \ket{j+n+\tfrac{1}{2};\varepsilon=-}_{P=-\frac{1}{2}} \overline{\ket{j-n-\tfrac{1}{2};\varepsilon=+}}_{P=-\frac{1}{2}} \, \Theta^{\alpha-} \bar{\Theta}^{\bar{\alpha}-} \ ,
\end{equation}
where $\ket{j+n+\frac{1}{2};\varepsilon=-}_{P=-\frac{1}{2}}$ and $\overline{\ket{j-n-\frac{1}{2};\varepsilon=+}}_{P=-\frac{1}{2}}$ are spelled out in eqs.~\eqref{eq:qsc-state-left} and \eqref{eq:qsc-state-right}, respectively. As discussed around eq.~\eqref{eq:RR-picture-choice}, we actually also need these states in picture $P=-\frac{3}{2}$. The answer turns out to be similar to the state at $P=-\frac{1}{2}$, except that the fermions $\psi^-_{-1}$ terms are dropped and the outer spin index $\gamma$ is sent to $\gamma \mapsto -\gamma$. For instance for $\ket{\mathcal{V}^{\text{QSC}}_n}_{P=-\frac{1}{2}}$ one has\footnote{$\ket{\mathcal{V}^{\text{QSC}}_n}_{P=-\frac{3}{2}}$ in picture $P=-\frac{3}{2}$ is not unique, as $G_0$ has a non-zero kernel. In fact, there is a one-parameter family of solutions that enters in the relative coefficient of the currents terms and $\psi^-_{-1}$ terms. We have taken advantage of this freedom to set the coefficients of the fermionic terms to zero, which then fixes the overall coefficient as in \eqref{eq:qsc-state-overall-coefficient-p=-3/2}. The final correlation function must be independent of this freedom, as the state at picture $P=-\frac{1}{2}$ is.}
\begin{equation} \label{eq:qsc-state-result-picture-p=-3/2}
	\ket{\mathcal{V}^{\text{QSC}}_n}_{P=-\frac{3}{2}} = \gamma^{\rho=+,\varepsilon=-}_n \gamma^{\rho=-,\varepsilon=+}_n \ket{j+n+\tfrac{1}{2};\varepsilon=-}_{P=-\frac{3}{2}} \overline{\ket{j-n-\tfrac{1}{2};\varepsilon=+}}_{P=-\frac{3}{2}} \, \Theta^{\alpha-} \bar{\Theta}^{\bar{\alpha}-} \ ,
\end{equation}
where
\begin{equation} \label{eq:qsc-state-right-p=-3/2}
	\overline{\ket{j-n-\tfrac{1}{2};\varepsilon=+}}_{P=-\frac{3}{2}} = (\bar{\mathcal{J}}^-_{-1})^n \overline{\ket{j,j,-;j^\prime,j^\prime,+;+}} \ ,
\end{equation}
and
\begin{align} \label{eq:qsc-state-left-p=-3/2}
	\begin{split}
		&\ket{j+n+\tfrac12;\varepsilon=-}_{P=-\frac{3}{2}} \\ & \hspace{1em} = \frac{\partial_y^{2n+1}|_{y=0}}{(2n+1)!} \Big(\sum_{m=0}^{2n+1} \sum_{r=0}^1 y^{m+r} e^{y J^+_0} (\mathcal{J}^-_{-1})^n  e^{-y J^+_0} \ket{j,j+m,-1+2r;j^\prime,j^\prime,-;+} \Big) \\ & \hspace{1em} - \frac{\partial_y^{2n+1}|_{y=0}}{(2n+1)!} \Big(\sum_{m=0}^{2n+1} \sum_{r=0}^1 y^{m+r} e^{y J^+_0} (\mathcal{J}^-_{-1})^n  e^{-y J^+_0} \ket{j,j+m,-1+2r;j^\prime,j^\prime-1,+;+} \Big) \ .
	\end{split}
\end{align}
Here
\begin{equation} \label{eq:qsc-state-overall-coefficient-p=-3/2}
	\gamma^{\rho,\varepsilon}_n = \frac{k}{\alpha_n^{\rho,-\varepsilon}} \ ,
\end{equation}
see \eqref{eq:alpha-omega-epsilon}, is fixed such that the state at $P=(-\frac{3}{2})$ is mapped to the state in picture $P=(-\frac{1}{2})$. Before we close this section, let us discuss the image of the QSC state under $\Omega$. Following the discussions above about the reflection map, we propose
\begin{equation} \label{eq:mirror-qsc}
	\ket{\Omega(\mathcal{V}^{\text{QSC}}_n)}_{P=-\frac{1}{2}} = \ket{\mathcal{V}^{\text{QSC}}_n}_{P=-\frac{1}{2}} = \ket{j+n+\tfrac{1}{2};\varepsilon=-} \overline{\ket{j-n-\tfrac{1}{2};\varepsilon=+}} \Theta^{\alpha-} \bar{\Theta}^{\bar{\alpha}-} \ ,
\end{equation}
or in other words, the QSC state is mapped to itself under $\Omega$. We note that a priori it is not clear how the index $\varepsilon$ should behave under $\Omega$. As a sanity check, for the state
\begin{equation} \label{eq:qsc-check-1}
	\ket{\mathcal{V}^{\text{RR},\varepsilon=-}_n} = \ket{j-n-\tfrac{1}{2};\varepsilon=-} \overline{\ket{j-n-\tfrac{1}{2};\varepsilon=-}} \Theta^{\alpha-} \bar{\Theta}^{\bar{\alpha}-} \ ,
\end{equation}
one has
\begin{equation} \label{eq:qsc-check-mirror-1}
	\ket{\Omega(\mathcal{V}^{\text{RR},\varepsilon=-}_n)} = \ket{j+n+\tfrac{1}{2};\varepsilon=-} \overline{\ket{j+n+\tfrac{1}{2};\varepsilon=-}} \Theta^{\alpha-} \bar{\Theta}^{\bar{\alpha}-} \ ,
\end{equation}
such that \eqref{eq:mirror-map-spectrum-symmetry} is consistent with the string field theory calculations, see Section~\ref{sec:SFT-summary}.\footnote{A similar analysis also holds for the states obtained via replacing $\varepsilon=-$ everywhere with $\varepsilon=+$ in both eqs.~\eqref{eq:qsc-check-1} and \eqref{eq:qsc-check-mirror-1}.} This explains why world-sheet parity does not act on the $\varepsilon$ indices in \eqref{eq:mirror-qsc}.

\section{Semiclassical strings with mixed flux}\label{sec:semiclassics}

The previous section was restricted to the pure NS-NS point $q=1$, where the worldsheet CFT is solvable. For general $q$, there is no such exactly solvable worldsheet theory. Instead, one can compute certain terms at each order in the large $\lambda$ expansion using a semiclassical expansion of the bosonic Polyakov worldsheet on $\text{AdS}_3\times \text{S}^3$ where all quantum numbers are large. These classical solutions do not depend on the $\mathbb{T}^4$ factor. We will start by setting up this calculation for general states on $\text{AdS}_3\times \text{S}^3$, and discuss how classical solutions change under worldsheet parity. We will then discuss a slight generalization of the pulsating string solution of \cite{Hernandez:2018gcd} to general winding number $w$, where $w\to-w$ under worldsheet parity, which we will compare to the CCY state for $w=1$.  We will then derive a new solution, that we call the drift-closed folded string, which we will compare to the LRT and QSC states. 

\subsection{Setup}
\label{ClassSetup}

The Polyakov action in the conformal gauge, with a flat world-sheet metric $\text{diag}(-1,+1)$, is
\begin{equation} \label{eq:sc-polyakov-confgauge}
	S_{\text{pol}} = \frac{\sqrt{\lambda}}{4\pi} \int \text{d}\tau\, \text{d}\sigma \left[ -G_{MN}\, \dot X^M \dot X^N + G_{MN}\, X'^M X'^N - 2 B_{MN}\, \dot X^M X'^N \right] \ ,
\end{equation}
where $\sqrt{\lambda}$ is the dimensionless string tension, $G_{MN}$ is the spacetime metric, and $B_{MN}$ is the usual 2-form field. The fields $X^M(\tau,\sigma)$ are the embedding coordinates, with $\dot X^M \equiv \partial_\tau X^M$ and $X'^M \equiv \partial_\sigma X^M$. Here the space-time indices $M,\, N$ run from $0$ to $5$, where the directions $0,\,1,\,2$ label $\text{AdS}_3$ while $3,\,4,\,5$ label $\text{S}^3$. The Euler-Lagrange equations following from \eqref{eq:sc-polyakov-confgauge} read
\begin{equation} \label{eq:sc-eom-general}
	-\ddot X^P + X''^P + \Gamma^P{}_{MN}\, \big(\!\!-\dot X^M \dot X^N + X'^M X'^N\big) + H^P{}_{MN}\, \dot X^M X'^N = 0 \ ,
\end{equation}
where $\Gamma^P{}_{MN}$ are the Christoffel symbols of $G_{MN}$ and $H^P{}_{MN} = G^{PQ} H_{QMN}$. Conformal gauge fixes the world-sheet diffeomorphisms only up to a residual Weyl rescaling; the remaining constraints are the Virasoro conditions
\begin{equation} \label{eq:sc-virasoro-1}
	G_{MN}\, \big( \dot X^M \dot X^N + X'^M X'^N \big) = 0 \ ,
\end{equation}
\begin{equation} \label{eq:sc-virasoro-2}
	G_{MN}\, \dot X^M X'^N = 0 \ .
\end{equation}
Note that the $B$-field, being an antisymmetric two-form on the two-dimensional world-sheet, does not enter the Virasoro constraints. 

The space-time metric for $\text{AdS}_3\times \text{S}^3$ with a non-zero $B$-field is \cite{David:2014qta}
\begin{equation} \label{eq:sc-metric}
	\begin{split}
		\text{d}s^2 &= \text{d}s^2_{\text{AdS}_3} + \text{d}s^2_{\text{S}^3} + \text{d}s^2_{\mathbb{T}^4} \ , \\
		\text{d}s^2_{\text{AdS}_3} &= -\cosh^2\!\rho\, \text{d}t^2 + \text{d}\rho^2 + \sinh^2\!\rho\, \text{d}\phi^2 \ , \\
		\text{d}s^2_{\text{S}^3} &= \text{d}\beta_1^2 + \cos^2\!\beta_1\, \big( \text{d}\beta_2^2 + \cos^2\!\beta_2\, \text{d}\beta_3^2 \big) \ ,
	\end{split}
\end{equation}
where the radii of $\text{AdS}_3$ and $\text{S}^3$ are set to unity, and $(t,\rho,\phi)$ and $(\beta_1,\beta_2,\beta_3)$ are the coordinates of $\text{AdS}_3$ and $\text{S}^3$, respectively. The NS-NS and RR three-form fluxes, in the convention $H^{(3)} = \text{d}B$, are parametrized by a single dimensionless coupling $q \in [0,1]$:
\begin{equation} \label{eq:sc-fluxes}
	\begin{split}
		H^{(3)}_{t\rho\phi} &= -2 q \cosh\rho \sinh\rho \ , \\
		F^{(3)}_{t\rho\phi} &= -2 \sqrt{1-q^2}\, \cosh\rho \sinh\rho \ , \\
		H^{(3)}_{\beta_1 \beta_2 \beta_3} &= +2 q \cos^2\!\beta_1 \cos\beta_2 \ , \\
		F^{(3)}_{\beta_1 \beta_2 \beta_3} &= +2 \sqrt{1-q^2}\, \cos^2\!\beta_1 \cos\beta_2 \ .
	\end{split}
\end{equation}
Tuning $q$ from $0$ to $1$ interpolates between pure RR flux and pure NS-NS flux. A local two-form potential $B_{MN}$ satisfying $H^{(3)} = \text{d}B$ exists; the only component used in the folded and pulsating ans\"atze of Sections~\ref{sec:sc-folded}--\ref{sec:sc-pulsating} is the $\text{AdS}_3$ piece, for which a convenient gauge choice is $B_{t\phi} = q \sinh^2\!\rho$. 

The conserved spacetimes charges are then defined as\footnote{The conserved charge $J$ is associated with the Killing vector $\partial_{\beta_3}$. Another Killing vector is
$
\partial_\perp
=
-\sin\beta_2\,\partial_{\beta_1}
+\tan\beta_1\cos\beta_2\,\partial_{\beta_2},
$
and we denote the corresponding conserved charge by $\widetilde J$. The relation between these $SO(4)$ charges and the Cartan charges of $\text{SU}(2)_L\times \text{SU}(2)_R$ is
$
J_L^3=\frac{J+\widetilde J}{2},
J_R^3=\frac{J-\widetilde J}{2}.
$
}
\es{consCharge}{
\Delta
    &=
    \frac{\sqrt{\lambda}}{2\pi}
    \int_0^{2\pi}d\sigma\,
    \left(
        \cosh^2\rho\,\dot t
        -q\sinh^2\rho\,\phi'
    \right)\,,\\
 S&=   \frac{\sqrt{\lambda}}{2\pi}
    \int_0^{2\pi}d\sigma\,
    \left(
        \sinh^2\rho\,\dot\phi
        -q\sinh^2\rho\,t'
    \right)\,,\\
    J
    &=
  \frac{\sqrt{\lambda}}{2\pi}
    \int_0^{2\pi} d\sigma\,
   \left[\cos^2 \beta_1 \cos^2 \beta_2 \dot\beta_3
+q\sin\beta_2\,\beta_1'
-q\sin\beta_1\cos\beta_1\cos\beta_2\,\beta_2'\right] \ ,
}
where we restrict to the equal spin sector $J_1=J_2=J/2$ of the $\text{SU}(2)_L\times \text{SU}(2)_R$ R-symmetry, which is all we will consider for our classical solutions. All these charges scale as $\sqrt{\lambda}$, so we can then look for large $\lambda$ saddle point solutions to the Polyakov action, given by solving \eqref{eq:sc-eom-general}, \eqref{eq:sc-virasoro-1}, and \eqref{eq:sc-virasoro-2}. Since all fermions are set to zero anyway for these classical solutions, they will also apply to the supersymmetric theory we consider.

As a sanity check for \eqref{eq:mirror-map-spectrum-symmetry}, let us study the behavior of these classical solutions under world-sheet parity, which sends $\sigma\to-\sigma$. Under this action the kinetic term in \eqref{eq:sc-polyakov-confgauge} is invariant, while the Wess--Zumino term $-2B_{MN}\dot X^M X'^N$ is odd. Since the $B$-field is proportional to $q$ (here $B_{t\phi}=q\sinh^2\!\rho$), we thus see that worldsheet parity maps $q\to-q$. Each $\partial_\tau$-translation charge has the form $Q_K\propto\int\!\text{d}\sigma\,[G_{KN}\dot X^N+B_{KN}X'^N]$, which is thus invariant under the combined transformations $\sigma\to-\sigma$ and $q\to-q$. On the other hand, some string solutions have winding $w\propto\int\!\text{d}\sigma\,X'^\phi$, which changes sign under worldsheet parity. 

\subsection{Pulsating string} \label{sec:sc-pulsating}

We start by discussing the pulsating-string ansatz on the mixed-flux background, which is obtained by switching on a single great-circle rotation on $\text{S}^3$ \cite{Hernandez:2018gcd}.

We assume the spacetime fields depend on the worldsheet coordinates as
\begin{equation} \label{eq:sc-puls-ansatz}
	t = t(\tau) \ , \quad r = r(\tau) \ , \quad \phi = w\sigma \ , \quad \beta_1 = \beta_2 = 0 \ , \quad \beta_3 = \omega \tau \ .
\end{equation}
The string carries a great-circle rotation on $\text{S}^3$ at angular velocity $\omega$, with $\text{S}^3$ angular momentum $J$. It also has winding $w$ around the $\text{AdS}_3$ angular direction $\phi$, where $w$ can be any non-zero integer, and negative winding denotes reversing the winding. The $\text{AdS}_3$ radial coordinate is $r \equiv \sinh\rho$, so that $1+r^2 = \cosh^2\!\rho$. The conserved charges \eqref{consCharge} for this ansatz are
\es{eq:sc-puls-E}{
\Delta = \sqrt{\lambda}\big((1+r^2)\, \dot t - qw\, r^2 \big)\,,\qquad J=\sqrt{\lambda}\omega\,,\qquad S=0\,.
}

The equations of motion along $\phi,\, \beta_1,\, \beta_2$ are satisfied identically on \eqref{eq:sc-puls-ansatz}, leaving only the $t$ and $r$ equations as non-trivial. We can substitute the ansatz \eqref{eq:sc-puls-ansatz} together with \eqref{eq:sc-puls-E} into the diagonal Virasoro constraint \eqref{eq:sc-virasoro-1} and eliminate $\dot t$ to get a first-order equation for $r(\tau)$,
\begin{equation} \label{eq:sc-puls-rad}
	\dot r^2 = C_1\, r^4 + C_2\, r^2 + C_3 \ ,
\end{equation}
with coefficients
\begin{equation} \label{eq:sc-puls-Cj}
	C_1 = w^2(q^2 - 1) \ , \quad
	C_2 = 2 qw\Delta/\sqrt{\lambda} - w^2 - {J}^2/\lambda \ , \quad
	C_3 = (\Delta^2 - {J}^2)/\lambda \ .
\end{equation}
The off-diagonal Virasoro constraint \eqref{eq:sc-virasoro-2} is automatic on the ansatz. The only effect of ${J} \ne 0$ is the constant shift ${J}^2$ in $C_2$ and $C_3$. 

Let $r_0$ denote the outer turning point at which $\dot r$ vanishes. Setting the right-hand side of \eqref{eq:sc-puls-rad} to zero gives a quadratic in $\Delta$ whose solution is
\begin{equation} \label{eq:sc-puls-Eturning}
	\Delta= -qw r_0^2\sqrt{\lambda} + \sqrt{(1+r_0^2)\,(w^2\lambda r_0^2 + {J}^2)} \ ,
\end{equation}
together with the polynomial factorization
\begin{equation} \label{eq:sc-puls-factor}
	C_1\, r^4 + C_2\, r^2 + C_3 = (r_0^2 - r^2)\left[  \frac{\Delta^2 - {J}^2}{\lambda r_0^2}  + w^2(1-q^2)\, r^2 \right] \,.
\end{equation}
At $r_0 \to 0$, eq.~\eqref{eq:sc-puls-Eturning} gives $\Delta \to {J}$, the BPS-saturated endpoint.

The radial motion is bounded, $0 \leq r \leq r_0$. Integer quantization of the radial action gives the Bohr-Sommerfeld condition:
\begin{equation} \label{eq:sc-puls-BS}
	\sqrt{\lambda} \int_0^{r_0}\! \frac{\sqrt{C_1\, r^4 + C_2\, r^2 + C_3}}{1+r^2}\, \text{d}r = \pi n \ , \quad n \in \mathbb{Z}_{\geq 0} \ ,
\end{equation}
where $n$ is the radial Bohr--Sommerfeld quantum number for the pulsating mode. In the semiclassical regime, $n\sim\sqrt{\lambda}$. We then extrapolate the semiclassical result to the regime $n,J=\mathcal{O}(\lambda^1)$ to obtain the dispersion relation for the short operator,
\es{CCYclassw}{
\Delta(w)
=
2\lambda^{1/4}\sqrt{|w|\,n}
-2q\,\operatorname{sgn}(w)\,n
+
\frac{1}{\lambda^{1/4}\sqrt{|w|}}
\left[
\frac{5}{2}(1-q^2)n^{3/2}
+\frac{J^2}{4\sqrt{n}}
\right]
\\+
\frac{6q(1-q^2)}{w\sqrt{\lambda}}\,n^2
+
\mathcal{O}\!\left(\lambda^{-3/4}\right).
}
For $w=1$, this recovers the result for the CCY state in \eqref{pulseIntro}. Note that the result is invariant under simultaneous $w\to-w,q\to-q$ as advertized. As a consistency check when $w=1$, we can set $q=1$ and compare to the pure NSNS result in \eqref{eq:CCY-NS-NS-large-k}. We find it matches for all $k$ up to a shift by $-1$, which the classical solution is not sensitive to because this is subleading at large $\lambda$, where $n,J\sim\sqrt{\lambda}$.

\subsection{Folded string} \label{sec:sc-folded}

We now discuss the folded string solution for general $q$. A putative solution was found in \cite{David:2014qta}, but this solution is not periodic in $\sigma$ as required except at $q=0,1$. For $q=0$, it becomes the pure RR solution of \cite{Frolov:2002av}, which was successfully matched to the worldsheet bootstrap in \cite{Chester:2024wnb,Jiang:2025oar}. At $q=1$, it becomes the solution of  \cite{Loewy:2002gf}, whose dependence on $S$ does not match any known state in the pure NS-NS theory. We will modify the construction of \cite{David:2014qta} to be periodic for all $q$, continue to match \cite{Frolov:2002av} at $q=0$, and now successfully match a physical pure NS-NS state for $q=1$.

We start with the ansatz
\begin{equation} \label{eq:drift-fold-ansatz}
	t=c_1\tau+f(\xi)\ , \qquad
	\phi=c_2\tau+g(\xi)\ , \qquad
	r=r(\xi)\ , \qquad
	\beta_3=\omega\tau\ , \qquad
	\xi=\sigma+v\tau \ ,
\end{equation}
where $t$ is the global $\text{AdS}_3$ time, $\phi$ the $\text{AdS}_3$ angle, $r \equiv \sinh\rho$ the radial coordinate, and $v$ is what we call the world-sheet drift velocity. This drift velocity was zero in \cite{David:2014qta}. Closure requires the embedding functions $f,e^{i g},r$ to be $2\pi$-periodic in $\sigma$ at fixed $\tau$, equivalently $2\pi$-periodic in $\xi$, so the drift describes a rigidly translating fold rather than a gauge transformation. Substituting \eqref{eq:drift-fold-ansatz} into the equations of motion \eqref{eq:sc-eom-general} along $t$ and $\phi$ and integrating once gives
\begin{equation} \label{eq:drift-fold-first-integrals}
	f'=\frac{c_1v(1+r^2)+q c_2r^2-k_1}{(1+r^2)(1-v^2)}\ , \qquad
	g'=\frac{q c_1r^2+c_2vr^2-k_2}{r^2(1-v^2)}\ ,
\end{equation}
where the prime denotes $\partial_\xi$ and $k_1,k_2$ are integration constants. The conserved charges \eqref{consCharge} then take the form
\es{consFold}{
{\Delta}
    &=
    \frac{\sqrt{\lambda}}{1-v^2} \, \Big[c_1 - v k_1 + q k_2
    + \frac{c_1(1-q^2)}{2\pi}
    \int_0^{2\pi} d\xi\, r^2\Big] \,,
    \\
    {S}
    &=
   \frac{\sqrt{\lambda}}{1-v^2} \, \Big[-v k_2 + q k_1 - q c_1 v
    + \frac{c_2(1-q^2)}{2\pi}
    \int_0^{2\pi} d\xi\, r^2 \Big]\,,\\
    J&=\sqrt{\lambda}\,{\omega}\,,
}
so we can exchange $c_1,c_2$ for $\Delta,S$.

The off-diagonal Virasoro condition \eqref{eq:sc-virasoro-2} ties the integration constants to the drift and the $\text{S}^3$ velocity,
\begin{equation} \label{eq:drift-fold-vir}
	c_1k_1-c_2k_2=\omega^2 v\ .
\end{equation}
With \eqref{eq:drift-fold-vir} imposed, the diagonal Virasoro constraint \eqref{eq:sc-virasoro-1} reduces to a cubic in $r^2$,
\begin{equation} \label{eq:drift-fold-cubic}
	(1-v^2)^2\, r'^2 r^2=A r^6+B r^4+C r^2+D\ ,
\end{equation}
where
\begin{equation} \label{eq:drift-fold-ABCD}
	\begin{split}
		A &= (1-q^2)(c_1^2-c_2^2)\ ,\\
		B &= 2c_1^2-c_2^2-c_1^2q^2-2q c_2k_1+2q c_1k_2-\omega^2(1+v^2)\ ,\\
		C &= c_1^2+k_1^2+2q c_1k_2-k_2^2-\omega^2(1+v^2)\ ,\\
		D &= -k_2^2\ .
	\end{split}
\end{equation}
The drift dresses the coefficients of the rigid cubic without changing its structure.

The three roots $R_1,R_2,R_3$ of \eqref{eq:drift-fold-cubic}, ordered $R_1<R_2<R_3$, bound the radial oscillation $R_2\leq r^2\leq R_3$; reality requires $A<0$, i.e.\ $c_1^2<c_2^2$ for $q<1$. Parametrizing
\begin{equation}
	r^2(\theta)=R_2+(R_3-R_2)\sin^2\theta\ , \qquad 0\leq \theta\leq \frac{\pi}{2}\ ,
\end{equation}
and defining for any function $F$ of $r^2$ the period integral
\begin{equation} \label{eq:drift-fold-integral}
	\mathcal{I}[F]\equiv 4\int_0^{\pi/2}\frac{(1-v^2)F(r^2(\theta))}{\sqrt{-A\,(r^2(\theta)-R_1)}}\,\rmd\theta\ ,
\end{equation}
where the factor of four counts the quarter-arcs traversed by the doubly-folded string in one spatial period, the three periodicity conditions take the compact form
\begin{equation} \label{eq:drift-fold-closure}
	\mathcal{I}[1]=2\pi\ , \qquad
	\mathcal{I}[f']=0\ , \qquad
	\mathcal{I}[g']=2\pi\ .
\end{equation}
These three periodicity conditions, together with the Virasoro relation \eqref{eq:drift-fold-vir}, determine the drift velocity and integration constants in terms of the remaining parameters of the solution. At large $\lambda$ with fixed $S$ and $J$, the coefficient $c_1$ becomes small, while $c_2 \to 1$. We can therefore solve these conditions perturbatively in $c_1$ and $c_2-1$, obtaining
\begin{equation} \label{eq:drift-fold-branch}
	v=-q c_1+\frac{q(1-q^2)}{2}c_1^3+\mathcal{O}(c_1^5)\ , \qquad
	k_1=-\frac{q}{2}c_1^2+\mathcal{O}(c_1^4)\,,
\end{equation}
with the remaining quantities determined similarly order by order. We then evaluate the conserved charges \eqref{consFold} on this solution. The resulting relations between \(S,J\) and the parameters \(c_1,c_2,\omega\) cannot be inverted in closed form. Instead, in the large-\(\lambda\) regime with \(S,J=\mathcal{O}(\lambda^0)\), we solve these relations perturbatively to express \(c_1,c_2,\omega\) order by order in terms of the physical charges. Substituting these perturbative solutions into \(\Delta\) then yields the classical dispersion relation \eqref{LRTintroClass}.

As a consistency check, at $q=1$ the classical dispersion \eqref{LRTintroClass} agrees with the exact pure-NSNS spectrum of both the LRT and QSC states, up to a finite shift that the classical solution is not sensitive to. For the LRT, \eqref{eq:reggeNSNS} gives $S=2n$ and $J=2j'+1$, while for the QSC state the corresponding identification follows from \eqref{eq:qsc-ansatz} and \eqref{eq:definition-J}. The $S_{\rm cl},J_{\rm cl}$ in the classical solution differ from those of each state by finite shifts: $(S_{\rm cl},J_{\rm cl})=(S+2,J-1)$ for the LRT and $(S+1,J-1)$ for the QSC state. These shifts are invisible in the classical limit $S,J\sim\sqrt{\lambda}$, so both states are described by the same classical dispersion.

\section{Adding RR flux with string field theory}\label{sec:SFT}
In this section, we discuss the perturbative string field theory (SFT) calculation of the space-time energy in the presence of RR flux. At $q=1$, the world-sheet theory is a WZW model, see eqs.~\eqref{eq:sc-fluxes} and below. Therefore, it provides a necessary background about which SFT can be defined. Following \cite{Cho:2018nfn}, we turn on a small amount of RR flux through a string field whose leading term is $\mu\,\mathcal{O}_{\mathrm{RR}}$, see eq.~(2.14) in that paper. In the following, we will not review the SFT derivations but rather quote the final $4$-pt function which once calculated, can be used to perturbatively read the space-time energy. We will review the definition of $\mu$ and $\mathcal{O}_{\mathrm{RR}}$ below, and then explain the calculation of the $4$-pt function and relevant quantities for the states considered in Section~\ref{sec:NSNS}. We will then summarize our results, and discuss how they can be combined with our ansatz to fix the large $\lambda$ expansion of scaling dimensions at arbitrary $q$.

\subsection{Calculation of the \texorpdfstring{$4$}{4}-pt function} \label{sec:4pt-function}
The SFT calculation is performed around a pure NS-NS point with a fixed quantized WZW level $k$, see Section~\ref{sec:ws-cft} for our conventions. We parametrize the RR deformation by
\begin{equation} \label{eq:SFT-mu-def}
	q = 1 - \frac{\mu^2}{2} \ ,
\end{equation}
where $q$ is defined in eqs.~\eqref{eq:sc-fluxes}. As mentioned, the quantized NS-NS flux $k$ is held fixed, and using eq.~\eqref{radius}, we have
\begin{equation} \label{eq:lambda-k-q-relation}
	\sqrt{\lambda} = \frac{k}{q} \ .
\end{equation}
By contrast, the saddles of Section~\ref{sec:semiclassics} are expanded in the conventional strong-coupling regime where $\sqrt{\lambda}$ is large, while $q$ and the (rescaled) physical charges are kept fixed. A comparison between the SFT and the semi-classical results in Section~\ref{sec:semiclassics} therefore requires rewriting the semi-classical results using eq.~\eqref{eq:lambda-k-q-relation}, and also expanding around $q=1$ while keeping $k$ and the physical charges fixed.

We now define the RR vertex operator that turns on a small amount of RR flux defined in eqs.~\eqref{eq:sc-fluxes}. In our conventions, this operator is \cite{Friedan:1985ge,Kutasov:1998zh,Kutasov:1999xu,Cho:2018nfn}
\begin{equation} \label{eq:RR-vertex-operator}
	\mathcal{O}_{\text{RR}} = \frac{1}{4\pi} \epsilon^{IJ} \epsilon^{\alpha m} \epsilon^{\beta p} \epsilon^{\bar \alpha \bar m} \epsilon^{\bar \beta \bar p} c \bar c e^{\frac{\phi}{2}+\frac{\bar{\phi}}{2}} S^{\alpha\beta}_+ \Theta^{I+} \bar{S}^{\bar \alpha \bar \beta}_+ \bar{\Theta}^{J+} V^{\text{AdS}_3}_{m,\bar{m}} V^{\text{S}^3}_{p,\bar p} \ ,
\end{equation}
where all the indices take value in $\{+,-\}$ and the summations over repeated indices is understood with the conventions $\epsilon^{+-}=-\epsilon^{-+}=1$. Moreover, in the conventions of Appendix~\ref{app:ws}, we have set
\begin{equation} \label{eq:spin-fields-notation}
	\left(S^{\alpha \beta}_\gamma \bar{S}^{\bar \alpha\bar \beta}_{\bar \gamma} V^{\text{AdS}_3}_{m,\bar{m}}V^{\text{S}^3}_{p,\bar p}\right)(z,\bar{z}) = V\Big(\ket{m,\alpha;p,\beta;\gamma}\overline{\ket{\bar m,\bar \alpha;\bar p,\bar \beta;\bar \gamma}};z,\bar z\Big) \ .
\end{equation}
$\Theta^{I+}$ and $\bar{\Theta}^{J+}$ are the normalized $(\textbf{2},\textbf{2})$ Ramond ground states of $\mathbb{T}^4$, see below \eqref{eq:l0-m4-def}. We also note that $\mathcal{O}_{\text{RR}}$ is physical and an R-symmetry singlet.\footnote{The coefficient in front of $\mathcal{O}_{\text{RR}}$ is the coefficient in eq.~(4.10) of \cite{Cho:2018nfn} adapted to our conventions for the OPEs of fermions in Section~\ref{sec:NSNS}, see Chapter~22 of \cite{Yin:FoundationsStringTheory}. In particular, the coefficients calculated in eq.~(4.13) of \cite{Cho:2018nfn} are equal to $\frac{2}{\sqrt{3}}+\mathcal{O}(1/k^3)$, and therefore, the coefficient in front of $\mathcal{O}_{\text{RR}}$ would potentially get corrections at $\mathcal{O}(1/k^3)$.}

For a physical field $\mathcal{V}$ satisfying \eqref{eq:brst-physical}, solving the string field theory equations reduces the leading world-sheet anomalous dimension to a regularized 4-pt function \cite{Cho:2018nfn,Sen:2019jpm}
\begin{equation} \label{eq:SFT-deltaDelta-integral}
	\delta \mathcal{E}
	= - \pi \,\mu^2 \, \delta\mathcal{E}^{\text{reg.}} +\mathcal{O}(\mu^3) \ ,
\end{equation}
where $\delta\mathcal{E}$ is the world-sheet weight of $\mathcal{V}$. $\delta\mathcal{E}^{\text{reg.}}$ is the regularized 4-pt function defined as follows:
\begin{equation} \label{eq:ws-anomalous-dimension}
	\delta\mathcal{E}^{\text{reg.}} = \text{Reg} \int_{\mathbb{C}} \text{d}^2 z \, \mathcal{I}(z,\bar{z}) \ ,
\end{equation}
where\footnote{See eq.~\eqref{eq:2-pt-function-rule}. For the states considered in Section~\ref{sec:NSNS}, there is no non-trivial mixing up to the $\mu^2$ order at large $k$ and fixed quantum numbers. We also assume that $\mathcal{V}$ is not BRST exact, so that the denominator is non-zero.}
\begin{equation} \label{eq:SFT-4-pt-function}
	\mathcal{I}(z,\bar{z}) = \frac{\langle [\xi_0 \bar{\xi}_0 \mathcal{V}](0,0) \mathcal{O}_{\text{RR}}(1,1) \mathcal{V}^\dagger(\infty,\infty) [b_{-1}\bar{b}_{-1} P_+ \bar{P}_+ \mathcal{O}_{\text{RR}}](z,\bar z) \rangle}{\langle \mathcal{V}(0,0) [\xi_0 \bar{\xi}_0 c_0 \bar{c}_0 \mathcal{V}^\dagger](\infty,\infty) \rangle} \ .
\end{equation}
As reviewed in more detail in Appendix~\ref{app:SFT-large-k-integral}, we follow the string field theory prescription of \cite{Sen:2019jpm} to regularize the integral \eqref{eq:ws-anomalous-dimension}. $\mathcal{V}^\dagger$ is the dual pair (or star conjugation) of the field $\mathcal{V}$ \cite{Belavin:1984vu,Zwiebach:1992ie,Gaberdiel:1997ia,Witten:2012bh,Sen:2016bwe}, see Appendix~\ref{app:dual-pair} for the definition, and also the discussion above \eqref{eq:4-pt-function}. $\delta\mathcal{E}$, defined in \eqref{eq:SFT-deltaDelta-integral}, in principle can be computed for any state, although in practice it is a tedious exercise. The details that enter the calculation for the states considered in Section~\ref{sec:NSNS} are spelled out in Appendix~\ref{app:4pt} and below we will review the generalities.

Assuming $\delta\mathcal{E}$ is computed, one can proceed and calculate the \textit{space-time} anomalous dimension of $\mathcal{V}$. Varying the mass-shell condition \eqref{eq:delta-def} at fixed $j^\prime$ and $h^{\mathbb{T}^4}$, and assuming the dispersion relation \eqref{eq:energy-spin-r-rbar} off-shell, up to the second order in $\mu$ we have
\begin{equation} \label{eq:SFT-deltaE-deltaDelta}
	\delta \Delta = \frac{2k}{(2j-1)} \delta \mathcal{E} \ ,
\end{equation}
where $\Delta$ is the space-time energy. Therefore, once the $4$-pt function \eqref{eq:SFT-4-pt-function} is calculated, one can read the space-time energy perturbatively in the deformation parameter $\mu$, see \eqref{eq:SFT-mu-def}.

Let us illustrate the calculation of the $4$-pt function \eqref{eq:SFT-4-pt-function} with the example of the LRT state $\ket{\mathcal{V}^{\text{LRT}}_n}$ in \eqref{eq:leading-regge-trajectory-states}, i.e.\
\begin{equation}
	\mathcal{V} = c \, \bar{c} \, e^{\phi+\bar{\phi}} \, V(\ket{\mathcal{V}^{\text{LRT}}_n};z,\bar{z}) \ .
\end{equation}
As discussed around eq.~\eqref{eq:ns-canonical}, this state is in the NS-sector and has the canonical picture number $P=-1$. Our conventions for star conjugation are spelled out in Appendix~\ref{app:dual-pair} \cite{Gaberdiel:1997ia,Sen:2016bwe}. Explicitly, $\mathcal{V}^\dagger$ is obtained by applying the transformations in eqs.~\eqref{eq:ns-dual-pair} and \eqref{eq:field-conjugation} to the state in \eqref{eq:leading-regge-trajectory-states}. In particular, these transformations ensure that $\mathcal{V}^{\dagger}$ is also physical, see eq.~\eqref{eq:old-physical}. Then the $2$-pt function
\begin{equation}
	\langle \mathcal{V}(0,0) [\xi_0 \bar{\xi}_0 c_0 \bar{c}_0\mathcal{V}^\dagger](\infty,\infty) \rangle \ ,
\end{equation}
can be computed by applying Ward identities. More specifically, since $\mathcal{V}^\dagger$ is physical, in particular it is also Virasoro primary. Therefore, we can calculate the $2$-pt function conveniently first at finite points $(v,\bar{v})$ and only then send $v,\bar{v}\to \infty$, i.e.\
\begin{equation}
	\langle \mathcal{V}(0,0) [\xi_0 \bar{\xi}_0 c_0 \bar{c}_0\mathcal{V}^\dagger](\infty,\infty) \rangle = \lim_{v,\bar{v}\to \infty} \langle \mathcal{V}(0,0) [\xi_0 \bar{\xi}_0 c_0 \bar{c}_0\mathcal{V}^\dagger](v,\bar{v}) \rangle \ ,
\end{equation}
where we have used that the total weight of $\mathcal{V}^\dagger$ is $0$. As explained in Appendix~\ref{app:ward}, one can apply Ward identities to reduce this to $2$-pt functions of $\mathfrak{sl}(2,\mathbb{R})_{k+2}$ and $\mathfrak{su}(2)_{k-2}$ primaries. Our method for calculating the $4$-pt function
\begin{equation}
	\langle [\xi_0 \bar{\xi}_0 \mathcal{V}](0,0) \mathcal{O}_{\text{RR}}(1,1) \mathcal{V}^\dagger(\infty,\infty) [b_{-1}\bar{b}_{-1} P_+ \bar{P}_+ \mathcal{O}_{\text{RR}}](z,\bar z) \rangle \ ,
\end{equation}
is the same as in the case of $2$-pt function just more tedious. Again since $\mathcal{V}^\dagger$ is primary, we consider it at finite points $(v,\bar{v})$ and only then send them to infinity. The expression for the picture-raised vertex operator $P_+ \bar{P}_+ \mathcal{O}_{\text{RR}}$ is given in Appendix~\ref{app:SFT-picture-raised}. Using this and \texttt{Mathematica}, we apply Ward identities to reduce the correlator to the $4$-pt functions of $\mathfrak{sl}(2,\mathbb{R})_{k+2}$ and $\mathfrak{su}(2)_{k-2}$ primaries together with the $4$-pt functions of fermions and spin fields, i.e.\
\begin{subequations} \label{eq:basic-corr-after-ward-identity}
\begin{equation} \label{eq:ads3-primary-4pt}
	\langle V(\ket{j,m_1};0,0) V(\ket{-\tfrac{1}{2},m_2};1,1) V(\ket{j,m_3};v,\bar{v})  V(\ket{-\tfrac{1}{2},m_4};z,\bar{z}) \rangle \ ,
\end{equation}
\begin{equation} \label{eq:su2-primary-4pt}
	\langle V(\ket{j^\prime,n_1};0,0) V(\ket{\tfrac{1}{2},n_2};1,1) V(\ket{j^\prime,n_3};v,\bar{v}) V(\ket{\tfrac{1}{2},n_4};z,\bar{z}) \rangle \ ,
\end{equation}
\begin{equation} \label{eq:fermions-spin-field-4pt}
	\langle \psi^a(0) S^{\alpha_2\beta_2}_{\mu_2}(1) \psi^b(v) S^{\alpha_4\beta_4}_{\mu_4}(z) \rangle \ ,
\end{equation}
\end{subequations}
and similarly for the right-moving fermions and spin fields. As discussed in \cite{Teschner:1997ft,Teschner:1999ug,Cho:2018nfn}, the $4$-pt functions in eqs.~\eqref{eq:ads3-primary-4pt} and \eqref{eq:su2-primary-4pt} can be computed since the representation associated with the RR field is finite dimensional. This computation is reviewed in detail in Appendix~\ref{app:kz-equations}. The $4$-pt functions \eqref{eq:fermions-spin-field-4pt} can be also computed, see Appendix~\ref{app:spin-fields-correlators}. Having these, the $4$-pt function $\mathcal{I}(z,\bar{z})$ given in \eqref{eq:SFT-4-pt-function} can be calculated exactly up to the second order $\mu^2$. We then, using the methods discussed in Appendix~\ref{app:SFT-large-k-integral}, compute the integral over $(z,\bar{z})$ in eq.~\eqref{eq:ws-anomalous-dimension} as an expansion over $1/\sqrt{k}$. Using eq.~\eqref{eq:SFT-deltaE-deltaDelta}, this will give us $\delta \Delta$ as an expansion over $1/\sqrt{k}$ which can then be compared with the semi-classical or other results. In the next section, we will summarize our findings for the states considered in Section~\ref{sec:NSNS}.

\subsection{Summary of the SFT results} \label{sec:SFT-summary}
We now summarize our final results for the space-time energy in eq.~\eqref{eq:SFT-deltaE-deltaDelta}. In particular, we will discuss the results for the CCY state in \eqref{eq:xi-states}, the leading Regge trajectory state in \eqref{eq:leading-regge-trajectory-states}, the image of the CCY state under the world-sheet parity in \eqref{eq:reflected-ccy-state}, and the QSC state in \eqref{eq:qsc-state-result}. For each of these states, we write the SFT result in the form
\begin{equation}
	\delta \Delta_{\mathcal V} = \mu^2 \, \Delta^{\mu^2}_{\mathcal{V}} +\mathcal{O}(\mu^3) \ ,
\end{equation}
and report the series expansion up to the order $1/k^2$. We often use the definition \eqref{eq:definition-J} to write the final answer in terms of $J$ instead of $j^\prime$, the $\mathfrak{su}(2)_{k-2}$ spin.

\paragraph{The CCY state \eqref{eq:xi-states}.} Recall that the state $\ket{\mathcal{V}^{\text{CCY}}_n}$ depends on the parameters $n$ and $j^\prime$. We have commented at the end of this section on the case $n=0$ as a sanity check. For $n\geq 1$ we have
\begin{align} \label{eq:SFT-CCY-result}
	\Delta^{\mu^2}_{\text{CCY}}
	={}&
	\frac{\sqrt{n k}}{2}
	+n
	+\frac{40n^2-8n-J^2}{16\sqrt n\,\sqrt{k}}
	+\frac{n(-5+12n)}{2k}
	\\
	&+
	\frac{1}{k^{\frac{3}{2}}}
	\left[
	\frac{3 J^4}{256n^{3/2}}
	+\frac{J^2}{16\sqrt n}
	+\frac{3\big(4+J^2\big)\sqrt n}{16}
	+\big(6\zeta(3)-9\big)n^{3/2}
	+14n^{5/2}
	\right] + \mathcal{O}(1/k^{2})
	\ . \nonumber
\end{align}

\paragraph{The parity pair of CCY defined in \eqref{eq:reflected-ccy-state}.} In this case, for $n\geq 1$ we get
\begin{align} \label{eq:SFT-mirror-CCY-result}
	&\Delta^{\mu^2}_{\Omega(\text{CCY})}
	=
	\frac{\sqrt{nk}}{2}
	-n
	+\frac{40n^2-8n-J^2}{16\sqrt n\,\sqrt{k}}
	-\frac{n(-5+12n)}{2k}
	\\
	&+
	\frac{1}{k^{\frac{3}{2}}}
	\left[
	\frac{3 J^4}{256n^{3/2}}
	+\frac{J^2}{16\sqrt n}
	+\frac{3\big(4+J^2\big)\sqrt n}{16}
	+\big(6\zeta(3)-9\big)n^{3/2}
	+14n^{5/2}
	\right] + \mathcal{O}(1/k^2)
	\ . \nonumber
\end{align}

\paragraph{The LRT state in \eqref{eq:leading-regge-trajectory-states}.}
For the leading Regge trajectory (LRT) state $\ket{\mathcal{V}^{\text{LRT}}_n}$ with spin $S=2n\geq 2$, see \eqref{eq:reggeNSNS}, we get
\begin{align} \label{eq:SFT-LRT-result}
	&\Delta^{\mu^2}_{\mathrm{LRT}} = \frac{\sqrt{S} \sqrt{k}}{2\sqrt{2}} +\frac{6S^2-4S-J^2}{8\sqrt{2S}\,\sqrt{k}} \\
	&+\frac{1}{k^{\frac{3}{2}}} \left[ \frac{3J^4}{64\sqrt{2}S^{3/2}} +\frac{J^2(5S+2)}{16\sqrt{2S}} +\frac{\sqrt{S}}{2\sqrt{2}} +\frac{\big(6\zeta(3)-5\big)S^{3/2}}{2\sqrt{2}} +\frac{3S^{5/2}}{2\sqrt{2}} \right] +\mathcal{O}(1/k^2) \ . \nonumber
\end{align}

\paragraph{The QSC state in \eqref{eq:qsc-state-result}.}
For the $\ket{\mathcal{V}^{\text{QSC}}_n}$ state, with the spin of the middle component being $S=2n\geq 2$, we get
\begin{align} \label{eq:SFT-QSC-result}
	\Delta^{\mu^2}_{\text{QSC}} ={}& \frac{\sqrt{S}\sqrt{k}}{2\sqrt{2}} +\frac{6S^2-8S-J^2}{8\sqrt{2S}\,\sqrt{k}} +\frac{J}{2k} \\
	&+\frac{1}{k^{\frac{3}{2}}} \left[ \frac{3J^4}{64\sqrt{2}S^{3/2}} +\frac{J^2(5S-4)}{16\sqrt{2S}} +\frac{3\sqrt{S}}{2\sqrt{2}} +\frac{\big(6\zeta(3)-7\big)S^{3/2}}{2\sqrt{2}} +\frac{3S^{5/2}}{2\sqrt{2}} \right] +\mathcal{O}(1/k^2) \ . \nonumber
\end{align}

\paragraph{Further consistency checks.} We have performed a few checks on our method, and we report them here. If $n=0$, the state \eqref{eq:xi-states} is $1/2$-BPS since the mass-shell condition implies $j=j^\prime+1$. As a consistency check, both numerically and using the methods explained in Appendix~\ref{app:SFT-large-k-integral}, we have checked that $\delta \Delta = 0$ in this case, also see \cite{Cho:2018nfn}. Similarly, we have checked that the LRT and QSC states in eqs.~\eqref{eq:leading-regge-trajectory-states} and \eqref{eq:qsc-state-result}, receive no anomalous dimension when $n=0$, see below eq.~\eqref{eq:qsc-state-form}. We have also considered a level $1/2$-descendant of the CCY state with $n\geq 1$ obtained upon actions of the supercurrents \eqref{eq:super-charges-holo} on both left- and right-sectors. Although the calculation is now performed in the R-sector of the world-sheet theory, and in particular uses the spin field correlation functions discussed in Appendix~\ref{app:spin-fields-correlators}, we have confirmed that the final result of the anomalous dimension precisely agrees with \eqref{eq:SFT-CCY-result}. Finally, for the state defined in \eqref{eq:qsc-check-1} we get
\begin{align}\label{eq:SFT-RR-check-result}
	\Delta^{\mu^2}_{\text{RR},\varepsilon=-}
	&=
	\frac{\sqrt{nk}}{2}
	+n-\frac{1}{2}
	+\frac{
	40n^2-32n-4(j^{\prime})^2-12j^{\prime}-5-\frac{2}{j^{\prime}}
	}{16\sqrt n\,\sqrt{k}}
	\\
	&\quad
	-\frac{1}{k}
	\left[
	-6n^2+\frac{13n}{2}+j^{\prime}-1
	+\frac{3j^{\prime}+2}{8n j^{\prime}}
	\right]
	+\mathcal{O}\!\left(\frac{1}{k^{\frac{3}{2}}}\right)
	\ . \nonumber
\end{align}
For the proposed action of the world-sheet parity given in \eqref{eq:qsc-check-mirror-1} we have
\begin{align}\label{eq:SFT-RR-mirror-check-result}
	\Delta^{\mu^2}_{\Omega(\text{RR},\varepsilon=-)}
	&=
	\frac{\sqrt{n k}}{2}
	-\left(n-\frac{1}{2}\right)
	+\frac{
	40n^2-32n-4(j^{\prime})^2-12j^{\prime}-5-\frac{2}{j^{\prime}}
	}{
	16\sqrt n\,\sqrt{k}
	}
	\\
	&\quad
	+\frac{1}{k}
	\left[
	-6n^2+\frac{13n}{2}+j^{\prime}-1
	+\frac{3j^{\prime}+2}{8n j^{\prime}}
	\right]
	+\mathcal{O}\!\left(\frac{1}{k^{\frac{3}{2}}}\right)
	\ , \nonumber
\end{align}
which provides further evidence for the action of $\Omega$ proposed in Section~\ref{sec:extremal-r-sector}.

\subsection{Comparison with the other calculations} \label{sec:SFT-comparison}
The goal of this section is to explain in more detail the $q$-dependency of our results discussed in the Introduction: we will combine our SFT results, semi-classical solutions, the consequences of world-sheet parity as imposed by \eqref{eq:mirror-map-spectrum-symmetry}, as well as other results in the literature.

Before we discuss each state in more detail, let us describe the overall logic: motivated by the semi-classical computations which are valid for any $|q|\in[0,1]$, we observed that the coefficient of $\lambda^{\frac{1-m}{4}}J^{\ell}$ is a polynomial in $q$ of degree at most $m-\ell\geq 0$. For a given state $\ket{\Psi}$, we will then impose \eqref{eq:mirror-map-spectrum-symmetry} which implies a relation between the coefficients of $\lambda^{\frac{1-m}{4}}J^{\ell}$ at $q$ and $-q$. Moreover, we know the exact spectrum at the pure NS-NS point $q=1$, see eqs.~\eqref{eq:energy-spin-r-rbar} and \eqref{eq:regge-sol-j}. Finally, the SFT computation of the previous section, using \eqref{eq:SFT-mu-def}, gives a constraint on the derivative of the $q$-dependent part at $q=1$. Using these rules, we will fix the $q$-dependency up to the order $\lambda^{-\frac{1}{2}}$ for the CCY and the LRT states. For the QSC state, we also have the $q=0$ result of \cite{Ekhammar:2026ykk} up to $\lambda^{-\frac{3}{4}}$, and so in this case, we are actually able to fix the $q$-dependency up to this order in $\lambda$.

\paragraph{The CCY state in \eqref{eq:xi-states}.} As mentioned above, we will assume that at the order $\lambda^{\frac{1-m}{4}}J^{\ell}$, the $q$-dependency is a polynomial in $q$ of degree at most $m-\ell\geq 0$. Upon combining this with the SFT results and the conjectured symmetry \eqref{eq:mirror-map-spectrum-symmetry}, we derive \eqref{pulseIntro2}. To begin with, we see that at the pure NS-NS point $q=1$, using \eqref{eq:regge-sol-j}, the coefficients of $\lambda^{\frac{1-m}{4}}J^{\ell}$ are zero for $m\geq 2$, unless $m=\ell$. If $m=0$, the polynomial is constant. Thus, using our assumptions, in these cases there is no $q$-dependency and therefore the coefficient is completely fixed. Moreover, as mentioned in the Introduction, we are interested in the cases where $m\leq 3$ as for higher values, we cannot fix their $q$-dependency using our method. For these reasons, we will only need to consider $\lambda^{\frac{1-m}{4}}J^{\ell}$ with $0\leq \ell < m \leq 3$.

We first consider $m$ being even, so only $m=2$ and $\ell=0,1$ remain to be discussed, which implies that the relevant $q$-dependency is at most quadratic. We now use \eqref{eq:mirror-map-spectrum-symmetry}: if $\ell=1$, the $q$-dependency is linear and must vanish at $q=1$, and so is proportional to $(1-q)$. Upon comparing with the SFT result in \eqref{eq:SFT-CCY-result}, we see that its overall coefficient vanishes. If $\ell=0$, the $q$-dependency is quadratic. Via imposing \eqref{eq:mirror-map-spectrum-symmetry} and the $q=1$ results, we see that the polynomials are proportional to $(1-q^2)$. Using eqs.~\eqref{eq:SFT-CCY-result} or \eqref{eq:SFT-mirror-CCY-result}, we fix the $m=0,2$ $q$-dependencies as in eq.~\eqref{pulseIntro2}.

It remains to discuss the cases $m=1,3$: for $m=1$, the $q$-dependency is linear. The SFT result in \eqref{eq:SFT-CCY-result} and the $q=1$ input fix the $q$-dependency. For $m=3$ and $\ell=1,2$, the polynomial in $q$ has degree $2$ or less. Imposing \eqref{eq:mirror-map-spectrum-symmetry}, the SFT results, and the fact that the $q=1$ coefficients vanish, implies that the relevant coefficients are zero. For $m=3$ and $\ell=0$, the condition \eqref{eq:mirror-map-spectrum-symmetry}, together with the SFT results in eqs.~\eqref{eq:SFT-CCY-result} and \eqref{eq:SFT-mirror-CCY-result}, implies that the polynomial in $q$ is odd, has degree $3$ and has zeroes at $q=\pm1$, and so it is as written in \eqref{pulseIntro2}.

\paragraph{The LRT state in \eqref{eq:leading-regge-trajectory-states}.} The discussion is very similar to the previous case: again we only need to discuss the coefficients of $\lambda^{\frac{1-m}{4}}J^{\ell}$ with $0\leq \ell < m \leq 3$, as the other cases with $m=\ell$ or $m=0$ are fixed by the pure NS-NS case. Similar to above, we first consider the case where $m=2$ and $\ell=0,1$. Using the fact that the leading Regge trajectory is mapped to itself under $\Omega$, and the $q=1$ result, we see that the $q$-dependency is through an even polynomial with a zero at $q=1$. Combining this with the SFT result in \eqref{eq:SFT-LRT-result}, we fix the relevant $q$-dependencies in \eqref{LRTintroClass2}. For $m=1$, the dependency in $q$ is linear: world-sheet parity implies that the coefficient of the linear term in $q$ vanishes, while the constant term is fixed by the $q=1$ input. For $m=3$, the evenness of the polynomial dependency on $q$ (together with the WZW point input) implies that the dependency is through $(1-q^2)$ for $\ell=1$. However, comparing with \eqref{eq:SFT-LRT-result} shows that its overall coefficient is zero. The cases $\ell=0,2$ are similar and one recovers the $q$-dependency spelled out in \eqref{LRTintroClass2}.

We note that there is a non-trivial check that we can perform on our answer: setting $q=0$ corresponds to the pure RR flux. In this case, in \cite{Chester:2024wnb} the space-time energy has been computed using world-sheet bootstrap techniques. Our result \eqref{LRTintroClass2} matches precisely with the result obtained there, and in particular, with the one-loop term predicted in that paper.

\paragraph{The QSC state in \eqref{eq:qsc-state-result}.} In this case, we have further inputs from the QSC computations of \cite{Ekhammar:2026ykk} for the case $q=0$ given in \eqref{Intintro}. Recall that the QSC state is mapped to itself under $\Omega$, see \eqref{eq:mirror-qsc}. The $q$-dependency for $m=0,1$ is fixed similar to the previous two cases. Let us consider $m=2$ with $\ell=0,1$: the $q$-dependency is at most quadratic. For $\ell=1$, the SFT calculation and the $q=1$ result implies that the coefficient is zero. For $\ell=0$, via imposing \eqref{eq:mirror-map-spectrum-symmetry}, we see that the polynomial is even and vanishes at $q=1$, and therefore is proportional to $(1-q^2)$. The dependency is thus completely fixed using \eqref{eq:SFT-QSC-result}. As a check, it actually agrees with \eqref{Intintro} at $q=0$. Now we consider the cases where $m=3$ and $\ell=0,1,2$. After imposing \eqref{eq:mirror-map-spectrum-symmetry} and comparing with the SFT results, we see that only $\ell=1$ has a non-zero coefficient. In this case, the $q$-dependency is through an even degree $2$ polynomial in $q$, with a zero at $q=1$. Therefore, the $q$-dependency is fixed as in \eqref{Intintro2}. Again, as a non-trivial check, setting $q=0$ we reproduce the same term in \eqref{Intintro}.

Finally, we consider the case $m=4$ and $\ell=0,1,2,3$: again, imposing \eqref{eq:mirror-map-spectrum-symmetry} and the $q=1$ result implies that the polynomial in $q$ is even and has a zero at $q=1$, and via comparing with eq.~\eqref{eq:SFT-QSC-result}, shows that only the $\ell=2,0$ have non-zero coefficients. If $\ell=2$, the $q$-dependency is quadratic in $q$ and one can fix it similar to before. If $\ell=0$, the $q$-dependency is
\begin{equation}
	(1-q^2) \times (\text{even polynomial of degree $2$ in $q$}) \ ,
\end{equation}
and therefore, with two unknown coefficients for each term in powers of $S$. Combining the SFT result in \eqref{eq:SFT-QSC-result} with the QSC result at $q=0$ fixes the $q$-dependency as in \eqref{Intintro2}.

\section{Conclusions} \label{sec:conclusion}

We used the SFT formalism of \cite{Cho:2018nfn} to compute the small RR flux $\mu^2$ correction to the finite $k$ scaling dimensions of the lowest Regge trajectory (LRT), the states considered by the quantum spectral curve (QSC) in \cite{Ekhammar:2026ykk}, and the states originally considered in \cite{Cho:2018nfn} (denoted as CCY), as well as the states related by world-sheet parity to the CCY states. We used these results along with an ansatz motivated by classical solutions and world-sheet parity to fix the large $\lambda$ expansion of these scaling dimensions at arbitrary $q$ to several orders. For the QSC and LRT states, we matched results in the pure RR $q=0$ case from the quantum spectral curve and worldsheet bootstrap, respectively, which is a nontrivial check of our method.

We envision a program of computing the CFT data of all states to any order in $1/\lambda$ at arbitrary $q$, by fixing our ansatz using the small $\mu$ expansion computed with SFT to increasingly higher orders. In particular, we can compute $\mu^{2n}$ corrections to scaling dimensions using $(2+2n)$-point functions in the WZW model, and $\mu^{2n}$ to OPE coefficients using $(3+2n)$-point functions. The WZW model is in principle exactly solvable, so it should be possible to compute these higher point correlators, even if nothing beyond 4-point correlators has yet been computed in practice. While the normalization of the deformation operator $\cO_\text{RR}$ is currently only known to order $1/k^{5/2}$ \cite{Cho:2018nfn}, it should be possible to fix this normalization to higher orders by comparing to classical solutions which are known to any order. Also, this normalization need only be fixed once, in order to then apply it to any number of states.

In the main text, we proposed an identification of the states related by the world-sheet parity and provided evidence for it from string field theory computations. It would be interesting to establish this identification non-perturbatively on the world-sheet and prove \eqref{eq:mirror-map-spectrum-symmetry} for the proposed state pairs.

Our results for mixed flux should help guide the ongoing search for a mixed flux QSC. In particular, our observation of how world-sheet parity relates the scaling dimensions of different operators should be manifest in such a QSC. It would be interesting to also see how world-sheet parity is manifest in the recently proposed Thermodynamic Bethe Ansatz (TBA) for mixed flux \cite{Frolov:2025tda}. It was very recently proposed in \cite{Cavaglia:2026scj} that the pure RR QSC can be derived from the previously discussed pure RR TBA \cite{Frolov:2021bwp}, so a similar derivation might be possible for the mixed flux QSC.

We also think our program can be generalized to the other $\text{AdS}_3\times \text{S}^3\times \text{M}_4$ backgrounds for $\text{M}_4=\text{K}3$ and $\text{M}_4=\text{S}^3\times \text{S}^1$. The $\text{K}3$ case has a similar small $\mathcal{N}=4$ algebra and WZW model, while the only difference is that the free theory corresponding to $\mathbb{T}^4$ is replaced by an interacting $c=6$ K3 sigma-model \cite{Maldacena:2000hw,Maldacena:2000kv,Maldacena:2001km}. We expect the results for the LRT and CCY states to be the same for $\mathbb{T}^4$ and K3, as these states do not depend on $\mathbb{T}^4$ excitations. The QSC state, however, explicitly required modes from the $\mathbb{T}^4$, and so may differ. The $\text{S}^3\times \text{S}^1$ theory has a large $\mathcal{N}=4$ algebra and a correspondingly different WZW model \cite{Elitzur:1998mm,deBoer:1999gea}. The deformation $\cO_\text{RR}$ will thus be different, and the calculation of all three states could differ.\footnote{Recent proposals for the pure RR QSC on $\text{AdS}_3\times\text{S}^3\times\text{S}^3\times \text{S}^1$ have been put forward in \cite{Chernikov:2025jko,Cavaglia:2025icd}, but strong-coupling data are not yet available.}

In this paper, we studied the strong coupling regime where the `t Hooft coupling $\lambda$ is large. Near the tensionless symmetric orbifold point, where $\lambda$ is small, anomalous dimensions of certain operators under RR deformation have been studied in \cite{Gaberdiel:2023lco,Frolov:2023pjw,Fabri:2025rok}, and also in \cite{Gaberdiel:2015uca,Fiset:2022erp} in the hybrid formalism \cite{Berkovits:1999im}. It would be interesting to explore the precise relation between these two regimes in more detail.

Finally, it would be interesting to generalize the pure RR AdS Virasoro-Shapiro amplitude of \cite{Chester:2024wnb} to mixed flux. Our results for the scaling dimension of the LRT should be a useful input to pin down the $q$ dependence. It would also be useful to compute the corresponding pure NSNS correlator in the same basis as \cite{Chester:2024wnb}, which should help us understand how the worldsheet ansatz of \cite{Chester:2024wnb} needs to be generalized in the case of mixed flux. A first step in this direction was taken in \cite{Alday:2024rjs}, where the correlator for a bosonic AdS$_3$ model was considered. The analytic method of expanding integrated worldsheet four-point functions at large $k$ that we introduced in this work, should be useful for evaluating this pure NSNS correlator.

\section*{Acknowledgments} 

We thank Arkady Tseytlin, Minjae Cho, Xi Yin, Lorenz Eberhardt, Ofer Aharony, Matthias Gaberdiel, Nikolay Gromov, Bogdan Stefa\'{n}ski and Simon Ekhammar for useful discussions, and Ofer Aharony and Xi Yin for reviewing the manuscript. SMC is supported by the Royal Society under the grant URF\textbackslash R1\textbackslash 221310 and the UK Engineering and Physical Sciences Research council grant number EP/Z000106/1, KN is supported by EP/Z000106/1, and DlZ is supported by URF\textbackslash R1\textbackslash 221310. LFA is supported by the STFC grant ST/T000864/1. We acknowledge the use of ChatGPT and Claude for assistance with coding and the evaluation of integrals.

\appendix

\section{Conventions for the world-sheet theory} \label{app:ws}
In this appendix, we fix our conventions on various elements entering the pure NS-NS flux string theory discussed in Section~\ref{sec:NSNS}. 

\paragraph{Representations.} Let us begin by discussing how the zero modes of the decoupled currents discussed in Section~\ref{sec:NSNS} act in both NS- and R-sectors. In both sectors, the continuous representations $\mathcal{C}^j_\alpha$ of the decoupled $\mathfrak{sl}(2,\mathbb{R})_{k+2}$ are defined as follows: the zero modes act as
\begin{subequations} \label{eq:sl2-zero-mode-actions}
	\begin{equation}
		\mathcal{J}^+_0 \ket{j,m} = (m+j) \ket{j,m+1} \ ,
	\end{equation}
	\begin{equation}
		\mathcal{J}^-_0 \ket{j,m} = (m-j) \ket{j,m-1} \ ,
	\end{equation}
	\begin{equation}
		\mathcal{J}^3_0 \ket{j,m} = m \ket{j,m} \ ,
	\end{equation}
\end{subequations}
where $m \in \mathbb{Z}+\alpha$ with $\alpha\in[0,1) \mod 1$, and the quadratic Casimir, defined by
\begin{equation} \label{eq:sl2-quadratic-casimir}
      \mathcal{C}_2 = \frac{1}{2} \big( \mathcal{J}^+_0 \mathcal{J}^-_0 + \mathcal{J}^-_0 \mathcal{J}^+_0 \big)-\mathcal{J}^3_0 \mathcal{J}^3_0 \ ,
\end{equation}
defines the spin $j$ via
\begin{equation} \label{eq:sl2-def-spin}
      \mathcal{C}_2  \ket{j,m} = -j(j-1)\ket{j,m} \ .
\end{equation}
For the continuous representations, one has $j\in\frac{1}{2} + i\mathbb{R}$.

Given the actions in eqs.~\eqref{eq:sl2-zero-mode-actions}, $\mathcal{J}^-_0 \ket{j,j}=0$ and $\mathcal{J}^+_0 \ket{j,-j}=0$. The discrete representation $\mathcal{D}^+_j$ is defined via only keeping $\ket{j,m}$ with $m\geq j$ with $\alpha = j \mod 1$. The discrete representation $\mathcal{D}^-_j$ is defined similarly by only considering the states $\ket{j,m}$ with $m\leq -j$. In these cases, $j\in\mathbb{R}$. We also note that
\begin{equation} \label{eq:l0-ads3-def}
	L^{\text{AdS}_3}_0 \ket{j,m} = -\frac{j(j-1)}{k}\ket{j,m} \ ,
\end{equation}
where $L^{\text{AdS}_3}_0$ is the zero mode of the stress-tensor given in \eqref{eq:ads3-n=1}.

The decoupled currents $\mathcal{K}^a$ that form $\mathfrak{su}(2)_{k-2}$ in the NS-sector act as
\begin{subequations} \label{eq:su2-zero-mode-actions}
	\begin{equation}
		\mathcal{K}^+_0 \ket{j^{\prime},m^{\prime}} = -(m^{\prime}-j^{\prime}) \ket{j^{\prime},m^{\prime}+1} \ ,
	\end{equation}
	\begin{equation}
		\mathcal{K}^-_0 \ket{j^{\prime},m^{\prime}} = (m^{\prime}+j^{\prime}) \ket{j^{\prime},m^{\prime}-1} \ ,
	\end{equation}
	\begin{equation}
		\mathcal{K}^3_0 \ket{j^{\prime},m^{\prime}} = m^{\prime} \ket{j^{\prime},m^{\prime}} \ .
	\end{equation}
\end{subequations}
The quadratic Casimir
\begin{equation} \label{eq:su2-quadratic-casimir}
      \mathcal{C}^\prime_2 = \frac{1}{2} \big( \mathcal{K}^+_0 \mathcal{K}^-_0 + \mathcal{K}^-_0 \mathcal{K}^+_0 \big)+\mathcal{K}^3_0 \mathcal{K}^3_0 \ ,
\end{equation}
defines the spin $j^\prime \in \frac{1}{2}\mathbb{Z}_{\geq 0}$ of the finite dimensional representation $\mathcal{D}^\prime_{j^\prime}$:
\begin{equation} \label{eq:su2-def-spin}
      \mathcal{C}^\prime_2 \ket{j^{\prime},m^{\prime}} = j^\prime(j^\prime+1) \ket{j^{\prime},m^{\prime}} \ .
\end{equation}
In particular, one has $-j^\prime\leq m^\prime \leq j^\prime$ and $m^\prime \in \mathbb{Z}+j^\prime$. It is useful to also see that
\begin{equation} \label{eq:l0-s3-def}
	L^{\text{S}^3}_0 \ket{j^{\prime},m^{\prime}} = \frac{j^\prime(j^\prime+1)}{k} \ket{j^{\prime},m^{\prime}} \ ,
\end{equation}
where $L^{\text{S}^3}_0$ is the zero mode of the stress-tensor given in \eqref{eq:s3-n=1}.

In the R-sector, one must be careful about the zero modes of the fermions, as discussed in detail in \cite{Ferreira:2017pgt}. In this paper, we work with generic AdS$_3$ and S$^3$ discrete representations and also the case where these representations form doublets. Let us start with the latter case: it is useful to combine the AdS$_3$ and S$^3$ representations together as
\begin{equation}
	\ket{m,\alpha;n,\beta;\gamma} \ , \quad m,n,\alpha,\beta,\gamma\in\{+,-\} \ .
\end{equation}
$m$ denotes decoupled $\mathfrak{sl}(2,\mathbb{R})_{k+2}$ doublet index, $\alpha$ shows doublet under the fermions $\psi^a$ of AdS$_3$, $n$ depicts the decoupled $\mathfrak{su}(2)_{k-2}$ doublet index, $\beta$ denotes the doublet index under the fermions $\chi^a$ of S$^3$, and finally $\gamma$ shows an outer automorphism doublet index. The $\mathfrak{sl}(2,\mathbb{R})_{k+2}$ currents act as\footnote{Throughout the text, when $m\in\{+,-\}$ appears in products of numbers, they are equal $\pm 1$.}
\begin{subequations}
\begin{equation}
	\mathcal{J}^+_0 \ket{-,\alpha;n,\beta;\gamma} = \ket{+,\alpha;n,\beta;\gamma} \ ,
\end{equation}
\begin{equation}
	\mathcal{J}^-_0 \ket{+,\alpha;n,\beta;\gamma} = -\ket{-,\alpha;n,\beta;\gamma} \ ,
\end{equation}
\begin{equation}
	\mathcal{J}^3_0 \ket{m,\alpha;n,\beta;\gamma} = \frac{m}{2} \ket{m,\alpha;n,\beta;\gamma} \ .
\end{equation}
\end{subequations}
and the fermions $\psi^a$ act as
\begin{equation} \label{eq:psi-spinfield-0-mode}
	\psi^a_0 \ket{m,\alpha;n,\beta;\gamma}=(t^a_\gamma)^\alpha_{\alpha^\prime} \ket{m,\alpha^\prime;n,\beta;-\gamma} \ ,
\end{equation}
where a summation over $\alpha^\prime$ is assumed and the non-zero matrix elements are
\begin{equation}
	(t^3_+)^{\pm}_{\pm} = \frac{\pm 1}{2} \ , \quad (t^3_-)^{\pm}_{\pm} = \frac{\mp k}{2} \ , \quad (t^\pm_+)^{\mp}_{\pm} = \pm 1 \ , \quad (t^\pm_-)^{\mp}_{\pm} = \mp k \ .
\end{equation}
The decoupled $\mathfrak{su}(2)_{k-2}$ currents act as
\begin{subequations}
\begin{equation}
	\mathcal{K}^+_0 \ket{m,\alpha;-,\beta;\gamma} = \ket{m,\alpha;+, \beta;\gamma} \ ,
\end{equation}
\begin{equation}
	\mathcal{K}^-_0 \ket{m,\alpha;+, \beta;\gamma} = \ket{m,\alpha;-, \beta;\gamma} \ ,
\end{equation}
\begin{equation}
	\mathcal{K}^3_0 \ket{m,\alpha;n,\beta;\gamma} = \frac{n}{2}\ket{m,\alpha;n,\beta;\gamma} \ ,
\end{equation}
\end{subequations}
and the fermions $\chi^a$ act as
\begin{equation} \label{eq:chi-spinfield-0-mode}
	\chi^a_0 \ket{m,\alpha;n,\beta;\gamma}=(T^a_\gamma)^\beta_{\beta^\prime} \ket{m,\alpha;n,\beta^\prime;-\gamma} \ ,
\end{equation}
where a summation over $\beta^\prime$ is implied with the following non-zero matrix elements:
\begin{equation}
	(T^3_+)^{\pm}_{\pm} = \frac{\pm 1}{2} \ , \quad (T^3_-)^{\pm}_{\pm} = \frac{\pm k}{2} \ , \quad (T^\pm_+)^{\mp}_{\pm} = 1 \ , \quad (T^\pm_-)^{\mp}_{\pm} = k \ .
\end{equation}
Moreover, we will also need the following representations:
\begin{equation} \label{eq:generic-spinfield-ads3-s3-states}
	\ket{j,m,\alpha;j^\prime,m^\prime,\beta;\gamma} \ , \quad \alpha,\beta,\gamma\in\{+,-\} \ .
\end{equation}
$(j,m)$ and $(j^\prime,m^\prime)$ label the representation under the decoupled $\mathfrak{sl}(2,\mathbb{R})_{k+2}$ and $\mathfrak{su}(2)_{k-2}$ respectively, see eqs.~\eqref{eq:sl2-zero-mode-actions} and \eqref{eq:su2-zero-mode-actions}. $\alpha$ and $\beta$ denote the representation under $\psi^a_0$ and $\chi^a_0$ respectively, see eqs.~\eqref{eq:psi-spinfield-0-mode} and \eqref{eq:chi-spinfield-0-mode}. Finally, $\gamma$ denotes an outer index, as before. We sometimes also write
\begin{equation} \label{eq:spin-fields-notation-appendix}
	\left(S^{\alpha \beta}_\gamma \bar{S}^{\bar \alpha\bar \beta}_{\bar \gamma} V^{\text{AdS}_3}_{j,m,\bar{m}}V^{\text{S}^3}_{j^\prime,p,\bar p}\right)(z,\bar{z}) = V\Big(\ket{j,m,\alpha;j^\prime,p,\beta;\gamma}\overline{\ket{j, \bar m,\bar \alpha;j^\prime, \bar p,\bar \beta;\bar \gamma}};z,\bar z\Big) \ .
\end{equation}

\paragraph{The $\mathbb{T}^4$ theory.} At any point in the moduli space of $\mathbb{T}^4$ and $\text{K}3$ $2$d CFTs, the CFT possesses a small $\mathcal{N}=(4,4)$ superconformal algebra with $c=6$. Let us explicitly discuss these generators for the case of $\mathbb{T}^4$. Focusing on the left-moving fields, it is convenient to introduce two pairs of complex bosons
\begin{equation} \label{eq:t4-bosons}
	X^j(z,\bar z) \underline{X}^k(w,\bar w) \sim \delta^{jk}\ln|z-w|^2 \ ,
\end{equation}
and two pairs of complex fermions
\begin{equation}
	\Lambda^{j}(z) \underline{\Lambda}^k(w) \sim \frac{\delta^{jk}}{(z-w)} \ ,
\end{equation}
where $j,k\in\{1,2\}$. The other OPEs are trivial. The generators of $\mathcal{N}=(2,2)$ superconformal subalgebra with $c=6$ are
\begin{subequations} \label{eq:t4-n=2}
	\begin{equation}
		T = \partial X^j \partial \underline{X}^j - \frac{1}{2} \Lambda^j \partial \underline{\Lambda}^j - \frac{1}{2} \underline{\Lambda}^j \partial \Lambda^j \ ,
	\end{equation}
	\begin{equation}
		G^+ = \partial \underline{X}^j \Lambda^j \ ,
	\end{equation}
	\begin{equation}
		G^- = \partial X^j \underline{\Lambda}^j \ ,
	\end{equation}
	\begin{equation}
		J = \frac{1}{2} \Lambda^j \underline{\Lambda}^j \ ,
	\end{equation}
where there is a summation over $j\in\{1,2\}$. The following fields,
\begin{equation}
	J^{++} = \Lambda^1 \Lambda^2 \ , \quad J^{--} = \underline \Lambda^2 \underline \Lambda^1 \ ,
\end{equation}
together with $J$, form an $\mathfrak{su}(2)_1$ R-symmetry algebra. The additional supercurrents are then
\begin{equation}
	\widetilde{G}^+ = -\epsilon_{jk} \partial X^j \Lambda^k \ ,
\end{equation}
\begin{equation}
	\widetilde{G}^- = -\epsilon_{jk} \partial \underline{X}^j \underline{\Lambda}^k \ ,
\end{equation}
\end{subequations}
with a summation over $j,k\in\{1,2\}$ and $\epsilon_{12}=-\epsilon_{21}=1$. Having these, the following fields
\begin{subequations} \label{eq:n=1-t4}
	\begin{equation}
		T^{\mathbb{T}^4} = \partial X^j \partial \underline{X}^j - \frac{1}{2} \Lambda^j \partial \underline{\Lambda}^j - \frac{1}{2} \underline{\Lambda}^j \partial \Lambda^j \ ,
	\end{equation}
	\begin{equation}
		G^{\mathbb{T}^4} = \partial \underline X^j \Lambda^j + \partial X^j \underline \Lambda^j \ ,
	\end{equation}
\end{subequations}
form an $\mathcal{N}=1$ superconformal algebra with $c=6$, see eqs.~\eqref{eq:n=1-conventions} below. A given $\text{M}_4$ $2$d CFT consists of states that transform under its $\mathcal{N}=(4,4)$ generators. In particular, we will label the $L_0$ and $\bar L_0$ weights as $h^{\text{M}_4}$ and $\bar{h}^{\text{M}_4}$:
\begin{equation} \label{eq:l0-m4-def}
	L^{\text{M}_4}_0 \ket{\Psi} = h^{\text{M}_4} \ket{\Psi} \ , \quad \bar L^{\text{M}_4}_0 \ket{\Psi} = \bar{h}^{\text{M}_4} \ket{\Psi} \ ,
\end{equation}
where $\ket{\Psi}$ is a state in the $\text{M}_4$ CFT. Among all such states, there is the vacuum state $\ket{0}_{\text{M}_4}$ that lies in the NS-sector of the $\mathcal{N}=(4,4)$ generators and in particular has $h^{\text{M}_4}_0=\bar{h}^{\text{M}_4}_0=0$. In fact, via the spectral flow symmetry of the $\mathcal{N}=(2,2)$ subalgebra with $\eta=\frac{1}{2}$, see Chapter~12 of \cite{Blumenhagen:2013fgp}, the chiral primaries are in $1$-to-$1$ correspondence with the Ramond ground states which have $h^{\text{M}_4}_0=\bar{h}^{\text{M}_4}_0=\frac{1}{4}$. These two observations show that $h^{\text{M}_4}_0$ and $\bar{h}^{\text{M}_4}_0$ are non-negative numbers. Let us concentrate on specific Ramond ground states that is used in Section~\ref{sec:SFT}: we apply the spectral flow with $\eta=\frac{1}{2}$ of an $\mathcal{N}=(2,2)$ subalgebra to $\ket{0}_{\text{M}_4}$ and denote the result by $\sigma^{\eta=\frac{1}{2}}(\ket{0}_{\text{M}_4})$. This state transforms in the $\textbf{2}$ representation of the R-symmetry $\mathfrak{su}(2)_1$. Therefore, applying the $\mathfrak{su}(2)_1$ raising operator, one in total obtains $2$ Ramond ground states which we denote them collectively as $\Theta_I$ with $I\in\{+,-\}$. A similar construction in the right-moving sector gives $\bar{\Theta}_I$ Ramond ground states that transform in $\textbf{2}$ of the right-moving R-symmetry. We will normalize them such that
\begin{equation} \label{eq:m4-spin-fields-normalization}
    \langle \Theta_+(1) \Theta_-(0) \rangle = \frac{1}{\sqrt{2}} \ ,
\end{equation}
and similarly for $\bar{\Theta}_I$. In the case of $\mathbb{T}^4$, one can explicitly construct them. In fact, if we bosonize
\begin{equation}
	\Lambda^j = e^{iH^j} \ , \quad \underline{\Lambda}^j = e^{-i H^j} \ ,
\end{equation}
where conventionally we choose that $j,k\in\{4,5\}$ and
\begin{equation}
	H^j(z) H^k(w) \sim -\delta^{jk} \ln(z-w) \ ,
\end{equation}
then we have the following spin fields:
\begin{equation} \label{eq:t4-spin-fields}
	\Theta^{\alpha\beta} =2^{-\frac{1}{4}} e^{\frac{\alpha}{2}(iH^4+i \beta H^5)} \ , \quad \alpha,\beta\in \{+,-\} \ ,
\end{equation}
and similarly for the right-moving spin fields $\bar{\Theta}^{\alpha\beta}$. The GSO projection enforces that for a state of the form
\begin{equation}
	V(\ket{j,m,\alpha;j^\prime,m^\prime,\beta;\gamma},z) \, \Theta^{\eta\rho}(z) \ ,
\end{equation}
see eq.~\eqref{eq:generic-spinfield-ads3-s3-states}, we have
\begin{equation} \label{eq:gso-projection-r-sector}
	\gamma = \rho \ .
\end{equation}
In this paper, we only need the following $4$-pt function of these spin fields,
\begin{equation} \label{eq:t4-spin-field-corr}
    \frac{\langle \Theta^{\alpha_1-}(0) \Theta^{\alpha_2+}(1) \Theta^{\alpha_3-}(v) \Theta^{\alpha_4+}(z) \rangle}{\langle \Theta^{\alpha_1-}(0) \Theta^{\alpha_3-}(v) \rangle}
    = \frac{\epsilon^{\alpha_2 \alpha_4}}{\sqrt{2} \, (1-z)^{\frac{1}{2}}} \ , \quad \alpha_1=-\alpha_3 \ ,
\end{equation}
which can be computed using the bosonization \eqref{eq:t4-spin-fields} similar to the techniques discussed in more detail in Appendix~\ref{app:spin-fields-correlators}.

\paragraph{$\mathcal{N}=1$ superconformal algebra.} This algebra consists of a stress-tensor $T$ and a supercurrent $G$. The (anti-)commutators in our conventions are
\begin{subequations} \label{eq:n=1-conventions}
	\begin{equation} \label{eq:stress-tensor-commutator}
		[L_n,L_m] = \frac{c}{12} n(n^2-1) \delta_{n+m,0} + (n-m) L_{n+m} \ ,
	\end{equation}
	\begin{equation}
		[L_n,G_r] = \left(\frac{n}{2}-r\right) G_{n+r} \ ,
	\end{equation}
	\begin{equation}
		\{ G_r,G_s \} = \frac{c}{3} \left(r^2-\frac{1}{4}\right) \delta_{r+s,0} + 2 L_{r+s} \ ,
	\end{equation}
\end{subequations}
where $L_n$ are the modes of $T$. It is straightforward to check that in the NS-sectors, the fields given in \eqref{eq:ads3-n=1} form an $\mathcal{N}=1$ superconformal with the central charge \eqref{eq:ads3-central-charge}. Similarly, the fields given in \eqref{eq:s3-n=1} form an $\mathcal{N}=1$ superconformal with the central charge \eqref{eq:ads3-central-charge} where $k^\prime=k$.

In the R-sector, one should be careful about the normal-ordering of the zero modes of fermions. We adopt the following normal-ordering conventions \cite{Fuchs:1992nq,Ferreira:2017pgt}
\begin{equation} \label{eq:RR-normal-ordering-conventions}
	(\psi^a\psi^b)_n = \frac{1}{2} [\psi^a_0,\psi^b_n] + \sum_{m\leq -1} \psi^a_m \psi^b_{n-m} - \sum_{m\geq 1} \psi^b_{n-m} \psi^a_m \ , \quad n\in\mathbb{Z} \ .
\end{equation}
Having this, if we define
\begin{equation} \label{eq:ramond-L0-weight-shift}
	L^{\text{R}}_n = L_n + \frac{3}{16} \delta_{n,0} \ , \quad G^{\text{R}}_r = G_r \ ,
\end{equation}
where $L_n$ and $G_r$ are the modes of the fields given in \eqref{eq:ads3-n=1} and \eqref{eq:s3-n=1} following \eqref{eq:RR-normal-ordering-conventions}, then $L^{\text{R}}_n$ and $G^{\text{R}}_r$ form an $\mathcal{N}=1$ superconformal algebra given in \eqref{eq:n=1-conventions}.

\paragraph{Diffeomorphism ghosts.} We briefly fix our conventions on the (super)diffeomorphism ghosts that enter the definition of superstring theory. We will only discuss the left-moving ghosts, since the right-moving fields are defined similarly. The diffeomorphism ghosts $(b,c)$ have weights $(2,-1)$ and satisfy
\begin{equation}
	b(z) c(w) \sim \frac{1}{(z-w)} \ ,
\end{equation}
while the rest of the OPEs are trivial. The stress-tensor is given by
\begin{equation}
	T_{bc} = (\partial b) c - 2 \partial(bc) \ .
\end{equation}
It is straightforward to check that the central charge is $(-26)$, see \eqref{eq:stress-tensor-commutator}.

The superdiffeomorphism ghosts $(\beta,\gamma)$ have weights $(\frac{3}{2},-\frac{1}{2})$ and satisfy
\begin{equation}
	\beta(z)\gamma(w)\sim \frac{-1}{z-w} \ ,
\end{equation}
where the other OPEs are trivial. The stress-tensor is
\begin{equation}
	T_{\beta\gamma} = (\partial \beta) \gamma - \frac{3}{2} \partial(\beta\gamma) \ .
\end{equation}
The central charge of $T_{\beta\gamma}$ is $(+11)$. In the main text, we need to bosonize the superdiffeomorphism ghosts:
\begin{equation} \label{eq:beta-gamma-bosonization}
	\beta = e^{\phi}\partial \xi \ , \quad \gamma = e^{-\phi}\eta \ .
\end{equation}
$\phi$ is a chiral boson that satisfies
\begin{equation}
	\phi(z)\phi(w)\sim -\ln(z-w) \ ,
\end{equation}
and has the background charge $\Lambda_\phi=2$, i.e.\ its stress-tensor is
\begin{equation}
	T_\phi = -\frac{1}{2} (\partial \phi)^2 + \partial^2 \phi \ .
\end{equation}
$(\eta,\xi)$ is a $bc$-system with weights $(1,0)$ that
\begin{equation} \label{eq:xi-eta-ope}
	\eta(z) \xi(w) \sim \frac{1}{(z-w)} \ ,
\end{equation}
while the rest of the OPEs are trivial. The correlation functions involving these ghosts can be computed using contractions. Putting the left- and right-moving fields together, we have
\begin{subequations} \label{eq:ghost-correlation-functions}
\begin{equation}
	\langle \left(c \bar{c}\right)(z_1,\bar{z}_1) \left(c \bar{c}\right)(z_2,\bar{z}_2) \left(c \bar{c}\right)(z_3,\bar{z}_3) \rangle = |z_1-z_2|^2 |z_1-z_3|^2 |z_2-z_3|^2 \ ,
\end{equation}
and
\begin{align}
	\langle e^{\phi+\bar{\phi}}(z_1,\bar{z}_1) e^{\frac{\phi+\bar{\phi}}{2}}(z_2,\bar{z}_2) &e^{\phi+\bar{\phi}}(z_3,\bar{z}_3) e^{-\frac{\phi+\bar{\phi}}{2}}(z_4,\bar{z}_4) \rangle \\&= |z_1-z_2|^{-1} |z_1-z_3|^{-2} |z_1-z_4| |z_2-z_3|^{-1} |z_2-z_4|^{\frac{1}{2}} |z_3-z_4| \ , \nonumber
\end{align}
\begin{align}
    \langle e^{\frac{\phi+\bar{\phi}}{2}}(z_1,\bar{z}_1) e^{\frac{\phi+\bar{\phi}}{2}}(z_2,\bar{z}_2) &e^{\frac{3\phi+3\bar{\phi}}{2}}(z_3,\bar{z}_3) e^{-\frac{\phi+\bar{\phi}}{2}}(z_4,\bar{z}_4) \rangle \\
    &= |z_1-z_2|^{-\frac{1}{2}} |z_1-z_3|^{-\frac{3}{2}} |z_1-z_4|^{\frac{1}{2}} |z_2-z_3|^{-\frac{3}{2}} |z_2-z_4|^{\frac{1}{2}} |z_3-z_4|^{\frac{3}{2}} \ . \nonumber
\end{align}
\end{subequations}

\paragraph{Correlation functions.} Having $n$ physical fields $V_j(z_j,\bar z_j)$ that satisfy \eqref{eq:brst-physical}, the world-sheet tree-level $n$-pt function is defined as in \eqref{eq:n-pt-function}. We will now explain a few technicalities regarding the definition of this correlation function following \cite{Blumenhagen:2013fgp}. Given a physical field $V$, the corresponding integrated vertex operator is defined via
\begin{equation}
	\int_{\Sigma_W} \text{d}^2 z \, b_{-1} \bar{b}_{-1} V \ ,
\end{equation}
where $b$ and $\bar b$ are left- and right-moving diffeomorphism ghosts, respectively, and $\Sigma_W$ is the world-sheet. In \eqref{eq:n-pt-function}, $(n-3)$ of the physical fields are integrated while $3$ of them are unintegrated. As explained in detail in \cite{Blumenhagen:2013fgp}, this is a consequence of the fact that using the global conformal transformations of sphere, i.e.\ M\"{o}bius group, one can always fix three points to be $(0,1,\infty)$. Given a physical field $V$, let us define $\widetilde{V}$ as the pre-image of $\eta_0$:\footnote{The $\eta_0$ cohomology is trivial, see \eqref{eq:eta-cohomology}. Therefore, it is always possible to find a solution to \eqref{eq:tilded-operator-def} provided that $V$ lies in the small Hilbert space, see \eqref{eq:eta0-simple-solution}.}
\begin{equation} \label{eq:tilded-operator-def}
	\eta_0 \widetilde{V} = V \ .
\end{equation}
Since
\begin{equation} \label{eq:eta-cohomology}
	\{ \eta_n, \xi_m \} = \delta_{n+m,0} \ , \quad \eta_0 V = 0 \ ,
\end{equation}
see eqs.~\eqref{eq:xi-eta-ope} and \eqref{eq:small-hilbert-space}, a solution to \eqref{eq:tilded-operator-def} is obtained via setting
\begin{equation} \label{eq:eta0-simple-solution}
	\widetilde{V} = \xi_0 V \ .
\end{equation}
In \eqref{eq:n-pt-function}, we have already replaced $\widetilde{V}_1 \mapsto \xi_0 \bar{\xi}_0 V_1$ for simplicity. In fact, since all the physical fields $V_j$ lie in the small Hilbert space, see \eqref{eq:small-hilbert-space}, it does not matter which of the physical fields $V_j$ is tilded or equivalently where $\xi_0\bar{\xi}_0$ is inserted. We note that for the case of $2$-pt function, one inserts an additional $0$-mode of the $c$-ghost, see e.g.\ \cite{Witten:2012bh}, and considers
\begin{equation} \label{eq:2-pt-function-rule}
	\langle [c_0 \bar{c}_0 \xi_0 \bar{\xi}_0 V_1](z_1,z_2) V_2(z_2,z_2) \rangle \ .
\end{equation}
In order for the correlation function \eqref{eq:n-pt-function} to be non-zero, the sum of the picture numbers $p_j$ defined in \eqref{eq:picture-number-def} of the physical fields $V_j$ must satisfy
\begin{equation} \label{eq:picture-sum-condition}
	\sum_{j=1}^n p_j =\sum_{j=1}^n \bar{p}_j = -2 \ .
\end{equation}
If this condition is not met (and the sum of the pictures is less than $-2$), one can use the picture-raising operator $P_+$ iteratively, see \eqref{eq:picture-raising}, until \eqref{eq:picture-sum-condition} holds. Provided that the condition \eqref{eq:picture-sum-condition} is satisfied, it can be shown that it does not matter how one distributes the picture number between the physical fields, see \cite{Blumenhagen:2013fgp}.

\section{The dual CFT multiplets} \label{app:n4}
In this appendix, we briefly review the space-time small $\mathcal{N}=(4,4)$ superconformal algebras with $c=6kp$ and their representations. For brevity, we focus on the left-moving fields. The generators of a small $\mathcal{N}=4$ superconformal algebra are stress-tensor $T$, R-symmetry generators $J$, $J^{\pm\pm}$, and supercurrents $G^\pm$ and $\widetilde{G}^\pm$. The non-zero (anti-)commutation relations are\footnote{$L_n$ are the modes of the stress-tensor $T$.}
\begin{equation} \label{eq:n=4-commutators}
	\begin{split}
		[L_n,L_m] &= \frac{c}{12} n(n^2-1) \delta_{n+m,0} + (n-m) L_{n+m} \ , \\
		[L_n,J_m] &= - m J_{n+m} \ , \\
		[L_n,J^{\pm\pm}_m] &= - m J^{\pm\pm}_{n+m} \ , \\
		[L_n,G^{\pm}_r] &= \left( \frac{n}{2} - r \right) G^{\pm}_{n+r} \ , \\
		[L_n,\widetilde{G}^{\pm}_r] &= \left( \frac{n}{2} - r \right) \widetilde{G}^{\pm}_{n+r} \ , \\
		[J_n,J^{\pm\pm}_m] &= \pm J^{\pm\pm}_{n+m} \ , \\
		[J_n,J_m] &= \frac{c}{12} n \delta_{n+m,0} \ ,\\
		[J^{++}_n,J^{--}_m] &= \frac{c}{6} n \delta_{n+m,0} + 2 J_{n+m} \ , \\
		[J_n,G^\pm_r]&=\pm \frac{1}{2} G^\pm_{r+n} \\
		[J_n,\widetilde{G}^\pm_r]&=\pm \frac{1}{2} \widetilde{G}^\pm_{r+n} \\
		[J^{\pm\pm}_n,G^{\mp}_r] &= \pm \widetilde{G}^\pm_{n+r} \ ,\\
		[J^{\pm\pm}_n,\widetilde{G}^{\mp}_r] &= \mp G^\pm_{n+r} \ , \\
		\{ G^+_r,G^-_s \} &= \frac{c}{6} \left( r^2-\frac{1}{4}\right) \delta_{r+s,0} + (r-s) J_{r+s} + L_{r+s} \ , \\
		\{ \widetilde{G}^+_r,\widetilde{G}^-_s \} &= \frac{c}{6} \left( r^2-\frac{1}{4}\right) \delta_{r+s,0} + (r-s) J_{r+s} + L_{r+s} \ , \\
		\{ G^\pm_r,\widetilde{G}^\pm_s \} &= \mp (r-s) J^{\pm\pm}_{r+s} \ .
	\end{split}
\end{equation}
Having this, we can now discuss the structure of multiplets by noticing that $[L_0,J_0]=[L_0,J^{\pm\pm}_0]=0$ and that they form the maximal subalgebra of the zero modes of the bosonic fields. Therefore, we can label the states that transform under this small $\mathcal{N}=4$ algebra with $L_0$ weight $h$ and $\mathfrak{su}(2)$ quantum number, which we denote as $\textbf{j}^\prime$: $(h,\textbf{j}^\prime)$. A superprimary field $\ket{h,\textbf{j}^\prime}$ furthermore satisfies
\begin{equation}
	L_n \ket{h,\textbf{j}^\prime} = J_n \ket{h,\textbf{j}^\prime} = J^{\pm\pm}_n \ket{h,\textbf{j}^\prime} = G^\pm_r \ket{h,\textbf{j}^\prime} = \widetilde{G}^\pm_r \ket{h,\textbf{j}^\prime} = 0 \ , \quad (n\geq 1) \ , \quad (r\geq \tfrac{1}{2}) \ .
\end{equation}
Starting with $\ket{h,\textbf{j}^\prime}$, via applying the $(r=-\tfrac{1}{2})$ modes of the supercurrents, which we collectively show as $\underline{Q}$ with $\underline{Q}\in\{G^+_{-\frac{1}{2}},G^-_{-\frac{1}{2}},\widetilde{G}^+_{-\frac{1}{2}},\widetilde{G}^-_{-\frac{1}{2}}\}$, one (generically) obtains a long multiplet of the small $\mathcal{N}=4$ superconformal algebra:
\begin{equation} \label{eq:n=4-multiplets}
\begin{gathered}
\underline{Q}\underline{Q}\underline{Q}\underline{Q}\ket{h,\mathbf{j}^\prime} \\
\hphantom{\underline{Q}}\underline{Q}\underline{Q}\underline{Q}\ket{h,\mathbf{j}^\prime}\hphantom{\underline{Q}} \\
\hphantom{\underline{Q}\underline{Q}}\underline{Q}\underline{Q}\ket{h,\mathbf{j}^\prime}\hphantom{\underline{Q}\underline{Q}} \\
\hphantom{\underline{Q}\underline{Q}\underline{Q}}\underline{Q}\ket{h,\mathbf{j}^\prime}\hphantom{\underline{Q}\underline{Q}\underline{Q}} \\
\hphantom{\underline{Q}\underline{Q}\underline{Q}\underline{Q}}\ket{h,\mathbf{j}^\prime}\hphantom{\underline{Q}\underline{Q}\underline{Q}\underline{Q}}
\end{gathered}
\end{equation}
In such a multiplet, $\ket{h,\textbf{j}^\prime}$ is called the bottom component while $\underline{Q}\underline{Q}\underline{Q}\underline{Q}\ket{h,\mathbf{j}^\prime}$ is called the top component. These multiplets are shortened provided that the bottom component is $1/4$-BPS, i.e.\footnote{There are further shortenings for small values of $\textbf{j}^\prime$. Moreover, a $1/4$-BPS state is automatically $1/2$-BPS in the small $\mathcal{N}=4$ superconformal algebra. See e.g.\ Appendix~C of \cite{Ferreira:2017pgt} for more details.}
\begin{equation}
	h = |\textbf{j}^\prime| \ .
\end{equation}
At the level of Fock space, different orderings of products of supercurrents in \eqref{eq:n=4-multiplets} give rise to different states. However, this ambiguity can be fixed by requiring that the components are quasi-primary. Using \eqref{eq:n=4-commutators}, one can check that the $(r=-\frac{1}{2})$ modes of the supercurrents anti-commute between one another except
\begin{equation}
	\{ G^+_{-\frac{1}{2}} , G^-_{-\frac{1}{2}} \} = \{ \widetilde{G}^+_{-\frac{1}{2}},\widetilde{G}^-_{-\frac{1}{2}}\} = L_{-1} \ .
\end{equation}
Therefore, different orderings of the supercurrents for components in \eqref{eq:n=4-multiplets} are equal up to fermionic signs and derivatives of lower components. In order to see this more explicitly, let us focus on the top component and require that
\begin{equation} \label{eq:top-component-unique-condition}
	L_1 \ket{\text{top}} = 0 \ ,
\end{equation}
where $\ket{\text{top}}$ is, a yet to be determined, linear combination of states of the form $\underline{Q}\underline{Q}\underline{Q}\underline{Q}\ket{h,\mathbf{j}^\prime}$ and derivatives of lower components. Let us for simplicity set $\textbf{j}^\prime=0$. In this case, using the (anti-)commutation relations in \eqref{eq:n=4-commutators}, one can check that
\begin{equation} \label{eq:top-explicit}
	\ket{\text{top}} = \frac{1}{4} \epsilon^{\alpha\beta} \epsilon^{\mu\rho} G^\alpha_{-\frac{1}{2}} G^\beta_{-\frac{1}{2}} \widetilde{G}^\mu_{-\frac{1}{2}} \widetilde{G}^\rho_{-\frac{1}{2}} \ket{h,0} + \frac{1}{4(2h+1)} (L_{-1})^2 \ket{h,0} \ ,
\end{equation}
satisfies \eqref{eq:top-component-unique-condition}, where $\alpha,\beta,\mu,\rho\in\{+,-\}$ and the summation over them is understood.

\section{The string field theory computation} \label{app:4pt}
In this appendix, we will discuss the elements that enter the calculation of $\delta \Delta$ in \eqref{eq:SFT-deltaE-deltaDelta}. Throughout this appendix, we will follow the conventions spelled out in Section~\ref{sec:NSNS} and Appendix~\ref{app:ws}.

\subsection{The picture-raised RR vertex operator} \label{app:SFT-picture-raised}
Let us begin by discussing the picture-raised vertex operator of \eqref{eq:RR-vertex-operator}. Recall that the RR vertex operator \eqref{eq:RR-vertex-operator} can be written as
\begin{equation}
	\mathcal{O}_{\text{RR}}(z,\bar{z}) = c \, \bar{c} \, e^{\phi/2+\bar{\phi}/2} \, f_{-\frac{1}{2}}(z,\bar{z}) \ ,
\end{equation}
where
\begin{equation}
	f_{-\frac{1}{2}}(z,\bar{z}) = \frac{1}{4\pi} \epsilon^{\alpha m} \epsilon^{\beta p} \epsilon^{\bar \alpha \bar m} \epsilon^{\bar \beta \bar p} \epsilon^{IJ} \Theta^{I+} \bar{\Theta}^{J+} V\Big(\ket{m,\alpha;p,\beta;+}\overline{\ket{\bar m,\bar \alpha;\bar p,\bar \beta;+}};z,\bar z\Big) \ ,
\end{equation}
see eq.~\eqref{eq:spin-fields-notation} and Appendix~\ref{app:ws}. Having this, let us consider the action of $P_+$ on the RR vertex operator, see \eqref{eq:picture-raising}. This can be written as
\begin{align} \label{eq:RR-picture-raised-temp-left}
	P_+ \mathcal{O}_{\text{RR}}(z,\bar{z}) = c \, \bar{c} \, e^{-\phi/2+\bar{\phi}/2} \, f^{\text{L}}_{\frac{1}{2}}(z,\bar{z}) + c \, \bar{c} \, e^{-\phi/2+\bar{\phi}/2} \, f^{\mathbb{T}^4,\text{L}}_{\frac{1}{2}}(z,\bar{z}) + Q^{(2)} \xi_0 \mathcal{O}_{\text{RR}}(z,\bar{z}) \ ,
\end{align}
where
\begin{equation}
	f^{\text{L}}_{\frac{1}{2}}(z,\bar{z}) = (G^{\text{AdS}_3}_{-1}+G^{\text{S}^3}_{-1}) f_{-\frac{1}{2}}(z,\bar{z}) \ ,
\end{equation}
and
\begin{equation}
	f^{\mathbb{T}^4,\text{L}}_{\frac{1}{2}}(z,\bar{z}) =  G^{\mathbb{T}^4}_{-1} f_{-\frac{1}{2}}(z,\bar{z}) \ ,
\end{equation}
see eq.~\eqref{eq:tot-t-g}. In particular, $f^{\text{L}}_{1/2}$ is the contribution of $Q^{(1)}$ in \eqref{eq:brst-q1} restricted to $\text{AdS}_3\times \text{S}^3$. Similarly, $f^{\mathbb{T}^4,\text{L}}_{1/2}$ is the $Q^{(1)}$ contribution associated with $\mathbb{T}^4$. The $Q^{(0)}$ contribution is zero while the last term in \eqref{eq:RR-picture-raised-temp-left} denotes the contribution from $Q^{(2)}$ in \eqref{eq:brst-q2}. In fact, this latter term does \textit{not} contribute to the $4$-pt function \eqref{eq:SFT-4-pt-function} due to the background charge conservation of $(b,c)$, see Appendix~\ref{app:ws}. Moreover, $f^{\mathbb{T}^4,\text{L}}_{\frac{1}{2}}$ that comes from $G^{\mathbb{T}^4}_{-1}$ inside $G^{\text{tot}}_{-1}$, see eq.~\eqref{eq:tot-t-g}, also does not contribute to the 4-pt functions that we have considered in this paper, since 1-pt functions of bosons $\partial X^j$ and $\partial \underline{X}^j$ vanish, see eqs.~\eqref{eq:t4-bosons} and \eqref{eq:t4-n=2}. Therefore, for the calculations at hand, it is enough to only consider $f^{\text{L}}_{\frac{1}{2}}$, i.e.\ to write
\begin{align} \label{eq:RR-picture-raised}
	P_+ \bar{P}_+ \mathcal{O}_{\text{RR}}(z,\bar{z}) = c \, \bar{c} \, e^{-\phi/2-\bar{\phi}/2} \, f_{\frac{1}{2}}(z,\bar{z}) \ ,
\end{align}
and
\begin{equation} \label{eq:RR-P=+1/2}
	f_{\frac{1}{2}}(z,\bar{z}) = \frac{1}{4\pi} \epsilon^{IJ} \Theta^{I+} \bar{\Theta}^{J+} V(\ket{\mathcal{O}^{P=\frac{1}{2}}} \overline{\ket{\mathcal{O}^{P=\frac{1}{2}}}};z,\bar{z}) \ .
\end{equation}
where
\begin{equation} \label{eq:RR-ket-picture-raised}
\begin{split}
\ket{\mathcal{O}^{P=\frac{1}{2}}} = \frac{1}{k}\Big(&
- \mathcal J^3_{-1}\ket{-1,1;-1,1;-1}
+ \mathcal J^3_{-1}\ket{-1,1;1,-1;-1}
- \mathcal J^3_{-1}\ket{1,-1;-1,1;-1} \\
&
+ \mathcal J^3_{-1}\ket{1,-1;1,-1;-1}
- \mathcal J^-_{-1}\ket{1,1;-1,1;-1}
+ \mathcal J^-_{-1}\ket{1,1;1,-1;-1} \\
&
- \mathcal J^+_{-1}\ket{-1,-1;-1,1;-1}
+ \mathcal J^+_{-1}\ket{-1,-1;1,-1;-1}
+ \mathcal K^3_{-1}\ket{-1,1;-1,1;-1} \\
&
+ \mathcal K^3_{-1}\ket{-1,1;1,-1;-1}
- \mathcal K^3_{-1}\ket{1,-1;-1,1;-1}
- \mathcal K^3_{-1}\ket{1,-1;1,-1;-1} \\
&
- \mathcal K^-_{-1}\ket{-1,1;1,1;-1}
+ \mathcal K^-_{-1}\ket{1,-1;1,1;-1}
+ \mathcal K^+_{-1}\ket{-1,1;-1,-1;-1} \\
&
- \mathcal K^+_{-1}\ket{1,-1;-1,-1;-1}
\Big) \ ,
\end{split}
\end{equation}
and similarly for the right-moving sector.

\subsection{Definition of \texorpdfstring{$\mathcal{V}^\dagger$}{Vdagger}} \label{app:dual-pair}
Having fixed the picture-raised RR vertex operator, we now need to be precise about the definition of $\mathcal{V}^\dagger$ in \eqref{eq:SFT-4-pt-function} where $\mathcal{V}$ is one of the physical states discussed in Section~\ref{sec:NSNS}. Following \cite{Belavin:1984vu,Zwiebach:1992ie,Witten:2012bh} and because of the no-ghost theorems \cite{Goddard:1972iy,Evans:1998wq}, the tree-level 2-pt function gives a non-degenerate pairing between physical fields up to BRST exact states in superstring theory, see in particular Section~5.5 in \cite{Witten:2012bh}. More precisely, one defines $\mathcal{V}^\dagger$ such that
\begin{equation} \label{eq:conjugate-2pt}
	\langle [\xi_0 \bar{\xi}_0 c_0 \bar{c}_0\mathcal{V}^\dagger](\infty,\infty) \mathcal{W}(0,0) \rangle = \braket{\mathcal{V}|\mathcal{W}} \ ,
\end{equation}
holds for any physical field $\mathcal{W}$. Here $\mathcal{V}^\dagger$ is obtained by the Hermitian conjugation followed by inverse BPZ conjugation, conventionally called star conjugation \cite{Gaberdiel:1997ia,Sen:2016bwe}. As discussed in \cite{Witten:2012bh}, the definition of $\mathcal{V}^\dagger$ depends on whether $\mathcal{V}$ is in NS-sector or R-sector. We are interested in an explicit realization of $\mathcal{V}^\dagger$ at the level of states for the WZW model considered in Section~\ref{sec:NSNS}.

For a physical field $\mathcal{V}^{P=-1}$ in picture $P=-1$ in the NS-sector, we set\footnote{The coefficient $(-1)^{j^\prime-m^\prime}$ is fixed as follows: by writing $V(\ket{j^\prime,m^\prime},z)^\dagger = \eta_{j^\prime,m^\prime}V(\ket{j^\prime,-m^\prime},z)$, imposing compatibility of $\dagger$ with the $\mathcal{K}^{\pm}_0$ action implies that $\eta_{j^\prime,m^\prime-1}=-\eta_{j^\prime,m^\prime}$. Setting $\eta_{j^\prime,j^\prime}=1$ then yields $\eta_{j^\prime,m^\prime}=(-1)^{j^\prime-m^\prime}$.\label{footnote-app-su2-factors}}
\begin{equation} \label{eq:ns-dual-pair}
	V^{\dagger}(\ket{j,m} \ket{j^\prime,m^\prime},z) = (-1)^{j^\prime-m^\prime} V(\ket{j,-m} \ket{j^\prime,-m^\prime},z) \ ,
\end{equation}
extended anti-linearly, and map all the currents and fermions to their corresponding conjugate fields, i.e.\
\begin{subequations} \label{eq:field-conjugation}
\begin{equation} \label{eq:field-conjugation-ads3}
	\begin{split}
		(\mathcal{J}^{\pm}_m)^\dagger = (-1)^{m+1} \mathcal{J}^{\mp}_m \ , \quad (\mathcal{J}^3_m)^{\dagger} = (-1)^{m+1} \mathcal{J}^3_m \ ,\\ (\psi^{\pm}_r)^\dagger = (-1)^{r+\frac{1}{2}} \psi^{\mp}_r \ , \quad (\psi^3_r)^\dagger = (-1)^{r+\frac{1}{2}} \psi^3_r \ ,
	\end{split}
\end{equation}
and similarly for the $\mathfrak{su}(2)$ currents:
\begin{equation} \label{eq:field-conjugation-su2}
	\begin{split}
		(\mathcal{K}^{\pm}_m)^\dagger = (-1)^{m+1} \mathcal{K}^{\mp}_m \ , \quad (\mathcal{K}^3_m)^{\dagger} = (-1)^{m+1} \mathcal{K}^3_m \ ,\\ (\chi^{\pm}_r)^\dagger = (-1)^{r+\frac{1}{2}} \chi^{\mp}_r \ , \quad (\chi^3_r)^\dagger = (-1)^{r+\frac{1}{2}} \chi^3_r \ ,
	\end{split}
\end{equation}
\end{subequations}
where $m\in\mathbb{Z}$ and $r\in\mathbb{Z}+\frac{1}{2}$. \eqref{eq:ns-dual-pair} together with \eqref{eq:field-conjugation} map the lowest weight discrete representation $\mathcal{D}^+_j$ to the highest weight representation $\mathcal{D}^-_j$. Note that in particular $\dagger$ defined including the signs in eqs.~\eqref{eq:field-conjugation} does \textit{not} change the order of product between operators.\footnote{More specifically, in the NS-sector we have that $(A B)^\dagger=(-1)^{|A||B|} A^\dagger B^\dagger$ where $|A|=0$ if $A$ is bosonic and $|A|=1$ if fermionic.}

For a physical field $\mathcal{V}^{P=-\frac{1}{2}}$ in picture $P=-\frac{1}{2}$ in the R-sector, the procedure is a bit more involved \cite{Witten:2012bh}. The reason is that, as discussed in eq.~\eqref{eq:picture-sum-condition}, the sum of the picture numbers in \eqref{eq:conjugate-2pt} must be $P=-2$. Therefore, the dual pairing must map $\mathcal{V}^{P=-\frac{1}{2}}$ to a physical field in picture $P=-\frac{3}{2}$. This representative is of course not unique, however, the difference is BRST exact. In order to do so, we first find a representative $\mathcal{V}^{P=-\frac{3}{2}}$ at $P=-\frac{3}{2}$, i.e.\ $P_+ \mathcal{V}^{P=-\frac{3}{2}} = \mathcal{V}^{P=-\frac{1}{2}}$. We then apply the conjugation \eqref{eq:field-conjugation} for the bosonic fields, together with
\begin{subequations}
\begin{equation}
	(\psi^{\pm}_m)^\dagger = (-1)^{m} \psi^{\mp}_m \ , \quad (\psi^3_m)^\dagger = (-1)^{m} \psi^3_m \ ,
\end{equation}
\begin{equation}
	(\chi^{\pm}_m)^\dagger = (-1)^{m} \chi^{\mp}_m \ , \quad (\chi^3_m)^\dagger = (-1)^{m} \chi^3_m \ ,
\end{equation}
\end{subequations}
where $m\in\mathbb{Z}$, and define the dual pairing anti-linearly using\footnote{$\eta_{\alpha\beta\gamma}$, similar to footnote~\ref{footnote-app-su2-factors}, are fixed by imposing compatibility with the zero mode actions of the fermions $\psi^a$ and $\chi^a$ defined in eqs.~\eqref{eq:psi-spinfield-0-mode} and \eqref{eq:chi-spinfield-0-mode}. In fact, as we have explicitly checked, the normalized 4-pt function \eqref{eq:SFT-4-pt-function-app} is independent of the exact numerical values of $\eta_{\alpha\beta\gamma}$, since they cancel between the numerator and the denominator.}
\begin{equation}
	V^{\dagger}(\ket{j,m,\alpha;j^\prime,m^\prime,\beta;\gamma},z) = \eta_{\alpha\beta\gamma} (-1)^{j^\prime-m^\prime} V(\ket{j,-m,-\alpha;j^\prime,-m^\prime,-\beta;\gamma},z) \ ,
\end{equation}
with
\begin{equation}
	\eta_{\alpha\beta\gamma} = \beta\gamma \ .
\end{equation}
Note that $\dagger$ acts on $\mathcal{V}^{P=-\frac{3}{2}}$ and in particular, $\gamma$ does not change sign. This defines the field $\mathcal{V}^\dagger$ at picture $P=-\frac{3}{2}$.

\subsection{Ward identities} \label{app:ward}
Following the basic discussions in the previous subsections, we now want to calculate the following 4-pt function:
\begin{equation} \label{eq:SFT-4-pt-function-app}
	\mathcal{I}(z,\bar{z}) = \frac{\langle [\xi_0 \bar{\xi}_0 \mathcal{V}](0,0) \mathcal{O}_{\text{RR}}(1,1) \mathcal{V}^\dagger(\infty,\infty) [b_{-1}\bar{b}_{-1} P_+ \bar{P}_+ \mathcal{O}_{\text{RR}}](z,\bar z) \rangle}{\langle \mathcal{V}(0,0) [\xi_0 \bar{\xi}_0 c_0 \bar{c}_0 \mathcal{V}^\dagger](\infty,\infty) \rangle} \ .
\end{equation}
The correlation functions of ghosts and $\mathbb{T}^4$ fields are given in eqs.~\eqref{eq:ghost-correlation-functions}, \eqref{eq:m4-spin-fields-normalization} and \eqref{eq:t4-spin-field-corr}. As discussed in Appendix~\ref{app:SFT-picture-raised} and Section~\ref{sec:NSNS}, see eq.~\eqref{eq:RR-P=+1/2} and e.g.\ eqs.~\eqref{eq:xi-states}, \eqref{eq:leading-regge-trajectory-states} and \eqref{eq:qsc-state-result}, the vertex operators in \eqref{eq:SFT-4-pt-function-app} include actions of currents and fermions. In particular, $\mathcal{V}^\dagger$ and $P_+ \bar{P}_+ \mathcal{O}_{\text{RR}}$ include only currents acting on primary states, see the states in Section~\ref{sec:NSNS}.\footnote{In the NS-sector, there are $\psi^a$ on $\mathcal{V}^\dagger$ which are primary fields.} Therefore, using conformal Ward identities for currents, it is straightforward to reduce \eqref{eq:SFT-4-pt-function-app} to the highest weight correlators with at most one $\psi^a_{-r}$ and at most one $\bar{\psi}^b_{-\bar{r}}$ acting on $\mathcal{V}(0,0)$, where $r=\bar{r}=\frac{1}{2}$ for the NS-sector computations and $r=\bar{r}=1$ for the R-sector calculations. In order to show this in an example, we put $\mathcal{V}^\dagger(v)$ at a finite point $v$, and consider only the left-moving correlators:
\begin{equation}
	\mathcal{I}^{\text{ex}} = \langle [\xi_0 \mathcal{V}](0) \mathcal{O}_{\text{RR}}(1) \mathcal{V}^\dagger(v) V(\mathcal{J}^+_{-1} \ket{-1,-1;-1,1;-1},z) \rangle \ ,
\end{equation}
see eq.~\eqref{eq:RR-P=+1/2}. We write
\begin{equation}
	\mathcal{I}^{\text{ex}} = \oint_{\mathcal{C}_z} \frac{\text{d}t}{(t-z)} \langle \mathcal{J}^+(t) [\xi_0 \mathcal{V}](0) \mathcal{O}_{\text{RR}}(1) \mathcal{V}^\dagger(v) V(\ket{-1,-1;-1,1;-1},z) \rangle \ ,
\end{equation}
where $\mathcal{C}_z$ is a small counter-clockwise contour around $z$. The integrand falls of at infinity as $1/t^3$, and therefore, one can deform the contour on sphere and put around other insertions. So one gets
\begin{equation}
	\mathcal{I}^{\text{ex}} = -\sum_{s\in\{0,1,v\} } \oint_{\mathcal{C}_s} \frac{\text{d}t}{(t-z)} \langle \mathcal{J}^+(t) [\xi_0 \mathcal{V}](0) \mathcal{O}_{\text{RR}}(1) \mathcal{V}^\dagger(v) V(\ket{-1,-1;-1,1;-1},z) \rangle \ .
\end{equation}
Near each insertion, one can use the OPE expansion,
\begin{equation}
	\mathcal{J}^+(t) V(\ket{\psi},s) \sim \sum_{n\geq 0} (t-s)^{-n-1} V(\mathcal{J}^+_n \ket{\psi},s) \ ,
\end{equation}
where `$\sim$' denotes the singular part of the OPE, see e.g.\ \cite{Gaberdiel:1998fs}. By doing this iteratively, we see that the correlation function will reduce to a sum of correlation functions of highest weight states with at most $\psi^a_{-r}$ or $\bar{\psi}^b_{-\bar{r}}$ actions on $\mathcal{V}(0,0)$.

\subsection{Correlation functions involving fermions and spin fields} \label{app:spin-fields-correlators}
As discussed in the previous section and in eqs.~\eqref{eq:basic-corr-after-ward-identity}, the correlation function of interest can be reduced to the highest weight correlators. In this subsection, we discuss the relevant correlation functions involving spin fields and fermions. This discussion depends on whether $\mathcal{V}$ is in the NS-sector or in the R-sector. We discuss the relevant correlation functions in each case separately. In fact, as it is clear from the states presented in Section~\ref{sec:NSNS}, one needs to only consider correlation functions involving the fermions $\psi^a$, and not $\chi^a$.

In the NS-sector, we need to compute
\begin{equation} \label{eq:ns-sector-spin-field-corr}
	\langle \psi^a(0) S^{\alpha\beta}_{\gamma}(1) \psi^b(\infty) S^{\eta\nu}_\rho(z)\rangle \ ,
\end{equation}
see eq.~\eqref{eq:spin-fields-notation-appendix}. In order to do so, we bosonize the fermions $\psi^a$ as follows:
\begin{equation}
	\psi^\pm = \sqrt{k} \, c_\psi^{\pm 1} \, e^{\pm i H^1} \ , \quad \chi^\pm = \sqrt{k} \, c_\chi^{\pm 1} \, e^{\pm i H^2} \ ,
\end{equation}
and
\begin{equation}
	\psi^3 = \frac{\sqrt{k}}{2} \left( -c_3 \, e^{iH^3} + c_3^{-1}\, e^{-i H^3} \right) \ , \quad \chi^3 = \frac{\sqrt{k}}{2} \left( c_3\, e^{iH^3} + c_3^{-1} \, e^{-i H^3} \right) \ ,
\end{equation}
where $c_\psi$, $c_\chi$ and $c_3$ are cocycles anti-commuting with one another \cite{Frenkel:1980rn,Goddard:1986ts}. We have assumed that
\begin{equation} \label{eq:h-opes-spin-field}
	H^j(z) H^\ell(w) \sim -\delta^{j\ell} \ln(z-w) \ ,
\end{equation}
and that the stress-tensor is
\begin{equation}
	T_{\text{bosonized}} = -\frac{1}{2} \sum_{j=1}^3 (\partial H^j)^2 \ .
\end{equation}
Having this, we write
\begin{equation}
	S^{\alpha\beta}_\mu = c_{\alpha\beta\mu} \, k^{-\frac{\mu}{4}} \, \exp{\left[\frac{i\alpha H^1}{2}+\frac{i \beta H^2}{2}+\frac{i \alpha\beta\mu H^3}{2}\right]} \ ,
\end{equation}
and fix the cocycles $c_{\alpha\beta\mu}$ by using the zero mode actions of the fermions defined in eqs.~\eqref{eq:psi-spinfield-0-mode} and \eqref{eq:chi-spinfield-0-mode}. In fact, the compatibility with those actions implies
\begin{subequations}
\begin{equation}
	c_\psi^{-\alpha} c_{\alpha\beta\mu} = - \alpha \mu c_{(-\alpha)\beta(-\mu)} \ ,
\end{equation}
\begin{equation}
	c_\chi^{-\beta} c_{\alpha\beta\mu} = c_{\alpha(-\beta)(-\mu)} \ ,
\end{equation}
\begin{equation}
	c_3^{-\alpha\beta\mu} c_{\alpha\beta\mu} = \beta c_{\alpha\beta(-\mu)} \ ,
\end{equation}
\end{subequations}
for $\alpha,\beta,\mu\in\{+,-\}$. Let us set $c_{+++}=r$. Solving these relations gives
\begin{subequations}
\begin{equation}
	c_{+++}=r \ , \quad c_{-+-}=-c_\psi^{-1}r \ , \quad c_{+--}=c_\chi^{-1}r \ , \quad c_{++-}=c_3^{-1}r \ .
\end{equation}
\begin{equation}
	c_{---}=-c_\psi^{-1} c_\chi^{-1}c_3^{-1}r \ , \quad c_{+-+}=c_\chi^{-1}c_3^{-1}r \ , \quad c_{-++}=c_\psi^{-1}c_3^{-1}r \ , \quad c_{--+}=c_\psi^{-1} c_\chi^{-1}r \ .
\end{equation}
\end{subequations}
We can also write a closed formula for the cocycles:
\begin{equation} \label{eq:cocyle-closed-formula}
	c_{\alpha\beta\gamma} = (-1)^{\hat{\alpha}\hat{\gamma}} c_\psi^{-\hat{\alpha}} c_\chi^{-\hat{\beta}} c_3^{-(\hat{\alpha\beta\gamma})} r \ ,
\end{equation}
where for a symbol $\theta\in\{+,-\}$, we have defined
\begin{equation} \label{eq:hat-definition}
	\hat{\theta}=\frac{1-\theta}{2} \ .
\end{equation}
We can also obtain products of $c_{\alpha\beta\mu}$'s and the fermionic cocycles from the right. In order to do so, let us impose
\begin{equation}
	c_{\alpha\beta\mu} \, c_x = \lambda_x(\alpha,\beta,\mu) \, c_x \, c_{\alpha\beta\mu} \ , \quad x \in \{3,\psi,\chi\} \ ,
\end{equation}
and determine $\lambda_x$'s using associativity of the cocycle algebra. Indeed, it can be shown that
\begin{equation} \label{eq:braiding-equations-cocycle}
	\lambda_\psi(\alpha,\beta,\mu) = -i \alpha \mu \ , \quad \lambda_\chi(\alpha,\beta,\mu) = -i \beta \mu \ , \quad \lambda_3(\alpha,\beta,\mu) = -i \alpha \beta \ .
\end{equation}
We note that for each $\lambda_x$, the associativity does not fix the overall factor. We have set the latter for each $\lambda_x$ to $(-i)$ using the fact that there is a common square-root branch cut between fermions and spin fields.\footnote{We have checked that using $i$ instead of $(-i)$ in eq.~\eqref{eq:braiding-equations-cocycle} only changes the left-moving total physical $4$-pt functions by an overall sign, which can be absorbed in their definitions.}

Using these, now one can directly compute the $4$-pt function in \eqref{eq:ns-sector-spin-field-corr} by keeping track of the factors that one would get from the algebra of cocycles. The final answer is:
\begin{equation}
	\langle \psi^a(0) S^{\alpha\beta}_{\gamma}(1) \psi^b(\infty) S^{\eta\nu}_{\rho}(z)\rangle = (-i)^{\hat{\gamma}} \delta_{\gamma,-\rho} \Big[\frac{(1-z)^{\frac{1}{4}}}{\sqrt{z}} (t^a_\gamma)^\alpha_{\kappa} (t^b_\rho)^{\kappa}_{\tau} \epsilon^{\tau\eta} \epsilon^{\beta\nu} - \frac{\epsilon^{\beta\nu} \epsilon^{\alpha\eta} q^{ab}}{(1-z)^{\frac{3}{4}}\sqrt{z}} \Big] \ ,
\end{equation}
where
\begin{equation}
	q^{+-}=q^{-+}=k \ , \quad q^{33}=-\frac{k}{2} \ ,
\end{equation}
while the rest vanish. We have fixed the overall normalization via imposing
\begin{equation}
	\langle S^{\alpha_1\beta_1}_{+}(1) S^{\alpha_2\beta_2}_{-}(0) \rangle = \epsilon^{\alpha_1 \alpha_2} \epsilon^{\beta_1 \beta_2} \ .
\end{equation}
In the R-sector, we need to compute
\begin{equation} \label{eq:r-sector-spin-field-corr}
	\langle S^{\alpha_1\beta_1}_{\gamma_1}(0) S^{\alpha_2\beta_2}_{\gamma_2}(1) S^{\alpha_3\beta_3}_{\gamma_3}(v) S^{\alpha_4\beta_4}_{\gamma_4}(z) \rangle \ .
\end{equation}
This can be computed using the bosonization discussed above and the algebra between the cocycles. In order to present the final answer compactly, let us define
\begin{equation}
	s_j := \alpha_j\beta_j\gamma_j \ , \quad \lambda_{ij} := \frac{1}{4}\left(\alpha_i\alpha_j+\beta_i\beta_j+s_i s_j\right) \ .
\end{equation}
As the fermions are bosonized and the $H^j$ bosons in eq.~\eqref{eq:h-opes-spin-field} have no background charge, the charge conservation is implemented by
\begin{equation} \label{eq:charge-spin-fields}
	\delta_{\mathrm{charge}}(\{\alpha_j,\beta_j,\gamma_j\}):=\delta_{\sum_{j=1}^{4}\alpha_j,0}\,\delta_{\sum_{j=1}^{4}\beta_j,0}\,\delta_{\sum_{j=1}^{4}s_j,0} \ .
\end{equation}
Using the cocycle algebra, the complete cocycle factor can be computed and one gets
\begin{equation}
	(-1)^{\Xi(\{\alpha_j,\beta_j,\gamma_j\})} \ ,
\end{equation}
where\footnote{We use $(-1)^{s}=e^{i\pi s}$ for a real number $s$. In particular, then we have $(-1)^{\frac{1}{2}}=i$.}
\begin{align}
	\Xi(\{\alpha_j,\beta_j,&\gamma_j\})\\=&1+\sum_{j=1}^{4}\widehat{\alpha_j}\widehat{\gamma_j}+\sum_{1\leq i<j\leq4}\left[\left(\widehat{\beta_i}+\widehat{\alpha_i\beta_i\gamma_i}\right)\widehat{\alpha_j}+\widehat{\alpha_i\beta_i\gamma_i}\widehat{\beta_j}+\frac{1}{2}\left(\widehat{\alpha_j}+\widehat{\beta_j}+\widehat{\alpha_j\beta_j\gamma_j}\right)\right] \ . \nonumber
\end{align}
The $4$-pt function of spin fields is then
\begin{align} \label{eq:r-sector-spin-field-corr-result}
	\langle S^{\alpha_1\beta_1}_{\gamma_1}(0)&S^{\alpha_2\beta_2}_{\gamma_2}(1)S^{\alpha_3\beta_3}_{\gamma_3}(v)S^{\alpha_4\beta_4}_{\gamma_4}(z)\rangle=\delta_{\mathrm{charge}}(\{\alpha_j,\beta_j,\gamma_j\}) \, (-1)^{\Xi(\{\alpha_j,\beta_j,\gamma_j\})} \\ & \times k^{-\frac{1}{4}\sum_{j=1}^{4}\gamma_j}(-1)^{\lambda_{12}+\lambda_{13}+\lambda_{14}+\lambda_{23}}v^{\lambda_{13}}(v-1)^{\lambda_{23}}(v-z)^{\lambda_{34}}z^{\lambda_{14}}(1-z)^{\lambda_{24}} \ . \nonumber
\end{align}
A spin field has conformal weight $(\frac{3}{8})$. Therefore, we can read the $v=\infty$ result by multiplying the correlation function with $v^{\frac{3}{4}}$ and sending $v\to \infty$. We note that the charge conservation, see eq.~\eqref{eq:charge-spin-fields}, implies that the power of $v$ is indeed $-\frac{3}{4}$ at large $v$ (whenever the correlation function is non-zero), i.e.\
\begin{equation}
	\lambda_{13}+\lambda_{23}+\lambda_{34}=-\frac{3}{4} \ .
\end{equation}
Using this, we have
\begin{align}
	\langle S^{\alpha_1\beta_1}_{\gamma_1}(0)&S^{\alpha_2\beta_2}_{\gamma_2}(1)S^{\alpha_3\beta_3}_{\gamma_3}(\infty)S^{\alpha_4\beta_4}_{\gamma_4}(z)\rangle\\&=\delta_{\mathrm{charge}}(\{\alpha_j,\beta_j,\gamma_j\}) \, (-1)^{\Xi(\{\alpha_j,\beta_j,\gamma_j\})}\,k^{-\frac{1}{4}\sum_{j=1}^{4}\gamma_j}(-1)^{\lambda_{12}+\lambda_{13}+\lambda_{14}+\lambda_{23}}z^{\lambda_{14}}(1-z)^{\lambda_{24}} \ . \nonumber
\end{align}
Similarly, the correlation functions of one $\text{AdS}_3$ fermion inserted at the first vertex operator can be computed:
\begin{subequations}
\begin{align}
	\langle [\psi^\pm_{-1}S^{\alpha_1\beta_1}_{\gamma_1}](0) &S^{\alpha_2\beta_2}_{\gamma_2}(1) S^{\alpha_3\beta_3}_{\gamma_3}(\infty) S^{\alpha_4\beta_4}_{\gamma_4}(z)\rangle \\ &= k^{\frac{1}{2}-\frac{1}{4}\sum_{j=1}^{4}\gamma_j} \delta_{\pm 2 + \sum_{j=1}^{4}\alpha_j,0} \delta_{\sum_{j=1}^{4}\beta_j,0} \delta_{\sum_{j=1}^{4}s_j,0} \nonumber \\ &\quad\times (-1)^{\Xi(\{\alpha_j,\beta_j,\gamma_j\})+\lambda_{12}+\lambda_{13}+\lambda_{14}+\lambda_{23}\pm\frac{1}{2}(\alpha_2+\alpha_3+\alpha_4)} z^{\lambda_{14}\pm\frac{\alpha_4}{2}}(1-z)^{\lambda_{24}} \nonumber \\ &\quad\times \left[\delta_{\pm \alpha_1,1}\mp\frac{1}{2}\left(\alpha_2+\frac{\alpha_4}{z}\right)\delta_{\pm\alpha_1,-1}\right] \ , \nonumber
\end{align}
and
\begin{align}
	\langle [\psi^3_{-1}S^{\alpha_1\beta_1}_{\gamma_1}](0) &S^{\alpha_2\beta_2}_{\gamma_2}(1) S^{\alpha_3\beta_3}_{\gamma_3}(\infty) S^{\alpha_4\beta_4}_{\gamma_4}(z)\rangle \\&= \frac{k^{\frac{1}{2}-\frac{1}{4}\sum_{j=1}^{4}\gamma_j}}{2}\, \delta_{\sum_{j=1}^{4}\alpha_j,0}\delta_{\sum_{j=1}^{4}\beta_j,0} (-1)^{\Xi(\{\alpha_j,\beta_j,\gamma_j\})+\lambda_{12}+\lambda_{13}+\lambda_{14}+\lambda_{23}}z^{\lambda_{14}}(1-z)^{\lambda_{24}} \nonumber \\ &\quad\times \sum_{\tau=\pm1}(-\tau)\delta_{2\tau+\sum_{j=1}^{4}s_j,0}(-1)^{\frac{\tau}{2}(s_2+s_3+s_4)}z^{\frac{\tau s_4}{2}} \left[\delta_{\tau s_1,1}-\frac{\tau}{2}\left(s_2+\frac{s_4}{z}\right)\delta_{\tau s_1,-1}\right] \ . \nonumber
\end{align}
\end{subequations}

\subsection{Correlation functions of \texorpdfstring{$\text{AdS}_3$}{AdS3} and \texorpdfstring{$\text{S}^3$}{S3} primaries} \label{app:kz-equations}
This appendix is based on \cite{Teschner:1997ft,Teschner:1999ug,Cho:2018nfn}, and we include the details of some of the derivations for completeness of the paper. We will calculate the following correlation functions
\begin{subequations}
\begin{equation} \label{eq:ads3-primary-4pt-app}
	\langle V(\ket{j,m_1};z_1,\bar{z}_1) V(\ket{-\tfrac{1}{2},m_2};z_2,\bar{z}_2) V(\ket{j,m_3};z_3,\bar{z}_3) V(\ket{-\tfrac{1}{2},m_4};z_4,\bar{z}_4) \rangle \ ,
\end{equation}
\begin{equation} \label{eq:su2-primary-4pt-app}
	\langle V(\ket{j^\prime,n_1};z_1,\bar{z}_1) V(\ket{\tfrac{1}{2},n_2};z_2,\bar{z}_2) V(\ket{j^\prime,n_3};z_3,\bar{z}_3) V(\ket{\tfrac{1}{2},n_4};z_4,\bar{z}_4) \rangle \ ,
\end{equation}
\end{subequations}
in the $\mathfrak{sl}(2,\mathbb{R})_{k+2}$ and $\mathfrak{su}(2)_{k-2}$ WZW models, respectively. See Section~\ref{sec:NSNS} and Appendix~\ref{app:ws} for our conventions. Throughout this appendix, we use the following definitions
\begin{subequations} \label{eq:def-f-ws-st}
\begin{equation}
	f_{\text{WS}}(\{z_i\}) = z_{12}^{-\Delta_1-\Delta_2+\Delta_3-\Delta_4} z_{13}^{-\Delta_1+\Delta_2-\Delta_3+\Delta_4} z_{23}^{\Delta_1-\Delta_2-\Delta_3+\Delta_4} z_{34}^{-2\Delta_4} \ ,
\end{equation}
and
\begin{equation}
	f_{\text{ST}}(\{x_i\}) = x_{12}^{-h_1-h_2+h_3-h_4} x_{13}^{-h_1+h_2-h_3+h_4} x_{23}^{h_1-h_2-h_3+h_4} x_{34}^{-2h_4} \ ,
\end{equation}
\end{subequations}
where
\begin{equation}
	z_{i\ell}:=z_i-z_\ell \ , \quad x_{i\ell} = x_i - x_\ell \ .
\end{equation}
$\Delta_\ell$ are world-sheet weights of the primaries inserted, and it will depend on whether one considers the $\text{AdS}_3$ or the $\text{S}^3$ theory. $h_\ell$ are related to the eigenvalue under the Cartan. In order to define these numbers compactly, we set
\begin{equation}
	j_1=j_3=j \ , \quad j_2=j_4=-\frac{1}{2} \ , 
\end{equation}
and
\begin{equation}
	j^\prime_1=j^\prime_3=j^\prime \ , \quad j^\prime_2=j^\prime_4=\frac{1}{2} \ .
\end{equation}
In the case of $\mathfrak{sl}(2,\mathbb{R})_{k+2}$ we have
\begin{equation} \label{eq:sl2-delta-h}
	\Delta_\ell = -\frac{j_\ell(j_\ell-1)}{k} \ , \quad h_\ell = j_\ell \ ,
\end{equation}
and for $\mathfrak{su}(2)_{k-2}$,
\begin{equation} \label{eq:su2-delta-h}
	\Delta_\ell = \frac{j^\prime_\ell(j^\prime_\ell+1)}{k} \ , \quad h_\ell = -j^\prime_\ell \ .
\end{equation}
The basic idea for calculating the $4$-pt function is to use Knizhnik-Zamolodchikov (KZ) equation \cite{Knizhnik:1984nr,Teschner:1997ft,Teschner:1999ug}, and the fact that there is a finite-dimensional representation in both cases. As the details are different for $\mathfrak{sl}(2,\mathbb{R})_{k+2}$ and $\mathfrak{su}(2)_{k-2}$, we will present the calculation for both cases separately below.

\subsubsection{\texorpdfstring{$\text{AdS}_3$}{AdS3}} \label{app:app-kz-ads3}
First, we define the $x$-basis via\footnote{We note that for the primary states, one could equally use $\mathcal{J}^+_0$ in this definition.}
\begin{equation}
	V(\ket{\psi};x,\bar{x};z,\bar{z}) = e^{x J^+_0+\bar{x} \bar{J}^+_0} e^{z L_{-1} + \bar{z} \bar{L}_{-1}} V(\ket{\psi};0,0;0,0) e^{-z L_{-1}- \bar{z} \bar{L}_{-1}} e^{-x J^+_0-\bar{x} \bar{J}^+_0} \ .
\end{equation}
We note that since $[L_{-1},J^+_0]=0$, this definition is unambiguous. The correlation function of interest is
\begin{align} \label{eq:ads3-primary-4pt-app-x-basis}
	\mathcal{I}^{\text{AdS}_3} = \langle V(\ket{j,j};x_1&,\bar{x}_1;z_1,\bar{z}_1) V(\ket{-\tfrac{1}{2},-\tfrac{1}{2}};x_2,\bar{x}_2;z_2,\bar{z}_2) \\ &V(\ket{j,j};x_3,\bar{x}_3;z_3,\bar{z}_3) V(\ket{-\tfrac{1}{2},-\tfrac{1}{2}};x_4,\bar{x}_4;z_4,\bar{z}_4) \rangle \ , \nonumber
\end{align}
where for brevity we have suppressed the dependency on $\{x_\ell\}$ and $\{z_\ell\}$ in the LHS. Using standard M\"{o}bius symmetry arguments, the reduced $4$-pt function depends only on the cross-ratios. More specifically, we define
\begin{equation} \label{eq:ads3-cross-ratio}
	u = \frac{x_{23} x_{14}}{x_{21} x_{34}} \ , \quad v = \frac{z_{23} z_{14}}{z_{21} z_{34}} \ ,
\end{equation}
and note that by setting
\begin{equation} \label{eq:mobius-ads3-choice}
	z_1=0 \ , \quad z_2=1 \ , \quad z_3=\infty \ , \quad x_1=0 \ , \quad x_2=1 \ , \quad x_3=\infty \ , 
\end{equation}
we have
\begin{equation}
	v = z_4 \ , \quad u = x_4 \ .
\end{equation}
Therefore, we can write
\begin{equation} \label{eq:ads3-4pt-kz-cross-ratio}
	\mathcal{I}^{\text{AdS}_3} = G_{j_1,j_2}(u,v;\bar{u},\bar{v}) |f_{\text{ST}}(\{x_\ell\})|^2 |f_{\text{WS}}(\{z_\ell\})|^2 \ ,
\end{equation}
where $G_{j_1,j_2}$ is a function of cross-ratios and the quantum numbers, also see eqs.~\eqref{eq:def-f-ws-st} and \eqref{eq:sl2-delta-h}. For later purposes, we explicitly specify the spin of the vertex operators inserted at positions $1$ and $2$. The KZ equation for $\mathfrak{sl}(2,\mathbb{R})_{k+2}$ is:
\begin{equation} \label{eq:sl2-kz-equation}
	k \, \partial_{z_4} \mathcal{I}^{\text{AdS}_3}  = \sum_{\ell\neq 4} \frac{\mathcal{Q}_{\ell,4} \, \mathcal{I}^{\text{AdS}_3}}{z_{4}-z_\ell} \ ,
\end{equation}
where
\begin{equation}
	\mathcal{Q}_{\ell,4} = (x_\ell-x_4)^2 \partial_{x_\ell} \partial_{x_4} + 2(x_\ell-x_4) \Big[ j_\ell \partial_{x_4} - j_4 \partial_{x_\ell} \Big] -  2 j_\ell j_4 \ .
\end{equation}
Substituting \eqref{eq:ads3-4pt-kz-cross-ratio} into \eqref{eq:sl2-kz-equation} gives \cite{Knizhnik:1984nr,Teschner:1997ft,Teschner:1999ug,Maldacena:2001km,Dei:2021yom}
\begin{equation}
	\Big[ k \partial_v - \frac{\mathcal{P}}{v} - \frac{\mathcal{Q}}{v-1} \Big] G_{j_1,j_2}(u,v;\bar{u},\bar{v}) = 0 \ ,
\end{equation}
where we have defined
\begin{equation}
	\mathcal{P} = u^2 (u-1) \partial_u^2 - \left( (\kappa-1) u^2+2j_1 u -2j_4 u(u-1) \right) \partial_u - 2 \kappa j_4 u - 2j_1 j_4 \ ,
\end{equation}
\begin{equation}
	\mathcal{Q} = -(1-u)^2 u \partial_u^2 + (1-u) \left( (\kappa-1)(1-u) + 2j_2+2j_4 u \right) \partial_u - 2 \kappa j_4 (1-u) - 2j_2 j_4 \ ,
\end{equation}
and
\begin{equation}
	\kappa = j_3 - j_1 -j_2-j_4 \ .
\end{equation}
Solving the KZ equation in general is difficult. However, in the case where the representation at $z_4=v$ is two-dimensional, we have that $(J^+_0)^2\ket{-\tfrac12,-\tfrac12}=0$, or equivalently
\begin{equation}
	\partial_u^2 G_{j_1,j_2}(u,v;\bar{u},\bar{v}) = \partial_{\bar{u}}^2 G_{j_1,j_2}(u,v;\bar{u},\bar{v}) = 0 \ .
\end{equation}
Therefore, the dependency on $u$ and $\bar{u}$ is linear. As the KZ equation is a linear differential equation, let us first look for a solution that is purely left-moving, i.e.\
\begin{equation} \label{eq:kz-ads3-g0-g1}
	\Big[ k \partial_v - \frac{\mathcal{P}}{v} - \frac{\mathcal{Q}}{v-1} \Big] g_{j_1,j_2}(u,v) = 0 \ , \quad g_{j_1,j_2}(u,v) = g_{j_1,j_2}^{(0)}(v) + u g_{j_1,j_2}^{(1)}(v) \ .
\end{equation}
This will give two coupled first order linear differential equations for $g_{j_1,j_2}^{(0)}$ and $g_{j_1,j_2}^{(1)}$, which generically possess two independent solutions. In order to find those solutions, let us define
\begin{subequations} \label{eq:g0-g1-ads3}
\begin{equation}
	g_{j_1,j_2}^{(0)}(v) = (1-v)^{\frac{1-j_2}{k}} v^{\frac{j_1}{k}} p(v) \ ,
\end{equation}
\begin{equation}
	g_{j_1,j_2}^{(1)}(v) = \frac{1}{C} \, (1-v)^{A} v^B \partial_v[(1-v)^{C} p(v)] \ ,
\end{equation}
\end{subequations}
where
\begin{equation}
	A = \frac{1-2j_1+2j_3+2k}{2k} \ , \quad B = \frac{j_1}{k} \ , \quad C = \frac{1+2j_1-2j_2-2j_3}{2k} \ .
\end{equation}
Substituting \eqref{eq:g0-g1-ads3} into the KZ equation \eqref{eq:kz-ads3-g0-g1} we get
\begin{equation}
	v(1-v) \, \partial_v^2 p(v)+\big[ c-(a+b+1)v \big] \partial_v p(v) - a b\,  p(v) = 0 \ ,
\end{equation}
with
\begin{equation}
	a = C = \frac{1+2j_1-2j_2-2j_3}{2k} \ , \quad b = \frac{-1+2j_1-2j_2+2j_3}{2k} \ , \quad c = \frac{2j_1-1}{k} \ .
\end{equation}
So we have the following two solutions for the function $p$:
\begin{equation}
	p^+(v) = {}_2F_1(a,b;c;v) \ ,
\end{equation}
\begin{equation}
	p^-(v) = v^{1-c} {}_2F_1(1+a-c,1+b-c;2-c;v) \ .
\end{equation}
Thus, we have found the following two solutions to the KZ equation:
\begin{equation}
	g_{j_1,j_2}^{\pm}(u,v) = (1-v)^{\frac{1-j_2}{k}} v^{\frac{j_1}{k}} p^{\pm}(v) + u \, \frac{1}{C} \, (1-v)^{A} v^B \partial_v[(1-v)^{C} p^\pm(v)] \ .
\end{equation}
It is more convenient for us to use the following two identities, see e.g.\ eqs.~(15.2.7) and (15.2.8) in \cite{Abramowitz:1964} for an $n\geq 0$,
\begin{subequations}
\begin{equation} \label{eq:handbook-1}
	\partial_v^n \big[(1-v)^{a+n-1} {}_2F_1(a,b;c;v) \big] = \frac{(-1)^n (a)_n (c-b)_n}{(c)_n} (1-v)^{a-1} {}_2F_1(a+n,b;c+n;v) \ ,
\end{equation}
\begin{align} \label{eq:handbook-2}
	\partial_v^n \Big[v^{C-1}(1-v)^{A-C+n}&{}_2F_1(A,B;C;v) \Big] \\ &= (C-n)_n v^{C-n-1} (1-v)^{A-C} {}_2F_1(A,B-n;C-n;v) \ , \nonumber
\end{align}
where
\begin{equation}
	(x)_n = \frac{\Gamma(x+n)}{\Gamma(x)} \ ,
\end{equation}
to simplify $g_{j_1,j_2}^{\pm}$. Note that in eq.~\eqref{eq:handbook-2} we have also used the identity (15.1.1) in \cite{Abramowitz:1964}, i.e.\
\begin{equation}
	{}_2F_1(a,b;c;v) = {}_2F_1(b,a;c;v) \ .
\end{equation}
\end{subequations}
Using these identities with $n=1$, it is straightforward to see that we can actually write
\begin{equation}
	g_{j_1,j_2}^{+}(u,v) = (1-v)^{\frac{1-j_2}{k}} v^{\frac{j_1}{k}} {}_2F_1(a,b;c;v) - u \, (1-v)^{A+a-1} v^B \frac{(c-b)}{c} {}_2F_1(a+1,b;c+1;v) \ ,
\end{equation}
and
\begin{align}
	g_{j_1,j_2}^{-}(u,v) = (1-v)^{\frac{1-j_2}{k}} & v^{1+\frac{j_1}{k}-c} {}_2F_1(1+a-c,1+b-c;2-c;v) \\&+ u \frac{(1-c)}{a} \, (1-v)^{A+a-1} v^{B-c} {}_2F_1(1+a-c,b-c;1-c;v) \ . \nonumber
\end{align}
Having found the left-moving solutions, we write
\begin{equation} \label{eq:ads3-4pt-function-block-expansion}
	G_{j_1,j_2}(u,v;\bar{u},\bar{v}) = \sum_{\sigma=\pm} h^\sigma_{j_1,j_2} g_{j_1,j_2}^{\sigma}(u,v) g_{j_1,j_2}^{\sigma}(\bar{u},\bar{v}) \ ,
\end{equation}
and impose the crossing symmetry and factorization into $2$-pt functions to determine the unknowns $h^\sigma_{j_1,j_2}$. By exchanging the vertex operators inserted at positions $1$ and $2$ in \eqref{eq:ads3-primary-4pt-app-x-basis}, the correlation function must remain the same. By exchanging the fields at $1$ and $2$, we switch $j_1$ with $j_2$, and also $x_1$ with $x_2$. This implies that the cross-ratios $u$ and $v$ defined in \eqref{eq:ads3-cross-ratio} are changed as follows:
\begin{equation}
	u \mapsto 1-u \ , \quad v \mapsto 1-v \ .
\end{equation}
Note $|f_{\text{WS}}|^2$ and $|f_{\text{ST}}|^2$ do not change under exchanging $1$ and $2$. Therefore, we must have
\begin{equation} \label{eq:cross-equation}
	G_{j_1,j_2}(u,v;\bar{u},\bar{v}) = G_{j_2,j_1}(1-u,1-v;1-\bar{u},1-\bar{v}) \ .
\end{equation}
In order to impose \eqref{eq:cross-equation}, we specialize to $j_2=-\tfrac{1}{2}$ and first write the blocks $g^{\pm}_{j,-\frac{1}{2}}$ and $g^{\pm}_{-\frac{1}{2},j}$ explicitly:
\begin{subequations} \label{eq:ads3-explicit-blocks}
\begin{align}
	g_{j,-\frac{1}{2}}^{+}(u,v) = (1-v)^{\frac{3}{2k}} &v^{\frac{j}{k}} {}_2F_1\left(\frac{1}{k},\frac{2j}{k};\frac{2j-1}{k};v\right) \\
&+ \frac{u}{2j-1} \, (1-v)^{\frac{3}{2k}} v^{\frac{j}{k}} {}_2F_1\left(1+\frac{1}{k},\frac{2j}{k};1+\frac{2j-1}{k};v\right) \ , \nonumber
\end{align}
and
\begin{align}
	g_{j,-\frac{1}{2}}^{-}(u,v) = (1-v)^{\frac{3}{2k}} &v^{1+\frac{1-j}{k}} {}_2F_1\left(1+\frac{2-2j}{k},1+\frac{1}{k};2-\frac{2j-1}{k};v\right) \\
&+ u (k+1-2j) \, (1-v)^{\frac{3}{2k}} v^{\frac{1-j}{k}} {}_2F_1\left(1+\frac{2-2j}{k},\frac{1}{k};1-\frac{2j-1}{k};v\right) \ , \nonumber
\end{align}
\end{subequations}
and
\begin{subequations}
\begin{align}
	g_{-\frac{1}{2},j}^{+}(u,v) = (1-v)^{\frac{1-j}{k}} &v^{-\frac{1}{2k}} {}_2F_1\left(-\frac{2j}{k},-\frac1k;-\frac2k;v\right)  \\ &-\frac{u}{2} \, (1-v)^{\frac{1-j}{k}} v^{-\frac{1}{2k}} {}_2F_1\left(1-\frac{2j}{k},-\frac{1}{k};1-\frac{2}{k};v\right) \ ,\nonumber
\end{align}
\begin{align}
	g_{-\frac{1}{2},j}^{-}(u,v) = (1-v)^{\frac{1-j}{k}} &v^{1+\frac{3}{2k}} {}_2F_1\left(1+\frac{2-2j}{k},1+\frac1k;2+\frac2k;v\right) \\&- u \frac{k+2}{2j} \, (1-v)^{\frac{1-j}{k}} v^{\frac{3}{2k}} {}_2F_1\left(1+\frac{2-2j}{k},\frac{1}{k};1+\frac{2}{k};v\right) \ . \nonumber
\end{align}
\end{subequations}
We now write
\begin{equation} \label{eq:crossing-blocks-ads3}
	g^{\sigma}_{j,-\frac{1}{2}}(u,v) = \sum_{\tau=\pm} d^\sigma_\tau \, g^{\tau}_{-\frac{1}{2},j}(1-u,1-v) \ ,
\end{equation}
and solve for $d^\sigma_\tau$. In fact, using the following identities (see eqs.~(15.3.6) and (15.3.3) in \cite{Abramowitz:1964}),
\begin{subequations} \label{eq:crossing-handbook-3}
\begin{align}
	{}_2F_1(a,b;c;z) &= \frac{\Gamma(c) \Gamma(c-a-b)}{\Gamma(c-a) \Gamma(c-b)} {}_2F_1(a,b;a+b-c+1;1-z) \\ &+ (1-z)^{c-a-b} \frac{\Gamma(c) \Gamma(a+b-c)}{\Gamma(a) \Gamma(b)} {}_2F_1(c-a,c-b;c-a-b+1;1-z) \ , \nonumber
\end{align}
\begin{equation}
	{}_2F_1(a,b;c;z) = (1-z)^{c-a-b} {}_2F_1(c-a,c-b;c;z) \ ,
\end{equation}
\end{subequations}
one can show that
\begin{subequations} \label{eq:d-values}
\begin{equation}
	d^{+}_{+} = \frac{2 \Gamma \left( \frac{2j-1}{k} \right) \Gamma \left( \frac{2}{k} \right)}{\Gamma \left( \frac{1}{k} \right) \Gamma \left( \frac{2j}{k} \right)} \ ,
\end{equation}
\begin{equation}
	d^{+}_{-} = -\frac{2j}{k+2} \frac{\Gamma \left( \frac{2j-1}{k} \right) \Gamma \left( -\frac{2}{k} \right)}{\Gamma \left( \frac{2j-2}{k} \right) \Gamma \left( -\frac{1}{k} \right)} \ ,
\end{equation}
\begin{equation}
	d^{-}_{+} = \frac{2 \Gamma \left( 2-\frac{2j-1}{k} \right) \Gamma \left( \frac{2}{k} \right)}{\Gamma \left( 1+\frac{2-2j}{k} \right) \Gamma \left( 1+\frac{1}{k} \right)} \ ,
\end{equation}
\begin{equation}
	d^{-}_{-} = -\frac{2j}{k+2} \frac{\Gamma \left( 2-\frac{2j-1}{k} \right) \Gamma \left( -\frac{2}{k} \right)}{\Gamma \left( 1-\frac{1}{k} \right) \Gamma \left( 1-\frac{2j}{k} \right)} \ .
\end{equation}
\end{subequations}
In the limit where $v \to 0$, one expects to find the identity in the OPE expansion of $G_{-\frac{1}{2},j}$. In this limit, the blocks $g^{\pm}_{-\frac{1}{2},j}(u,v)$ behave as
\begin{equation}
	g^{+}_{-\frac{1}{2},j}(u,v) \sim v^{-\frac{1}{2k}} \left( 1 - \frac{u}{2} \right) \ , \quad g^{-}_{-\frac{1}{2},j}(u,v) \sim v^{\frac{3}{2k}} \left( -u \frac{k+2}{2j} \right) \ .
\end{equation}
The world-sheet weights of the fields with $j_1=j_4=-\frac{1}{2}$ are equal to $(\frac{-3}{4k})$, and therefore, the identity appears in the block $g^{-}_{-\frac{1}{2},j}$. Requiring that the coefficient of the identity operator is $1$ implies
\begin{equation} \label{eq:ads3-identity-normalization}
	h^-_{-\frac{1}{2},j} = \frac{4j^2}{(k+2)^2}\ .
\end{equation}
Substituting \eqref{eq:crossing-blocks-ads3} into \eqref{eq:ads3-4pt-function-block-expansion} and imposing \eqref{eq:cross-equation} implies
\begin{subequations}
\begin{equation}
	h^{+}_{j,-\frac{1}{2}} d^{+}_{+} d^{+}_{-} + h^{-}_{j,-\frac{1}{2}} d^{-}_{+} d^{-}_{-} = 0 \ ,
\end{equation}
\begin{equation}
	h^{+}_{j,-\frac{1}{2}} \left( d^{+}_{+} \right)^2 + h^{-}_{j,-\frac{1}{2}} \left( d^{-}_{+} \right)^2 = h^{+}_{-\frac{1}{2},j} \ ,
\end{equation}
\begin{equation}
	h^{+}_{j,-\frac{1}{2}} \left( d^{+}_{-} \right)^2 + h^{-}_{j,-\frac{1}{2}} \left( d^{-}_{-} \right)^2 = h^{-}_{-\frac{1}{2},j} \ .
\end{equation}
\end{subequations}
Solving these equations together with eq.~\eqref{eq:ads3-identity-normalization} gives
\begin{equation}
	h^{-}_{j,-\frac{1}{2}} = \frac{4j^2}{(k+2)^2} \left[ \left( d^{-}_{-} \right)^2 - \frac{d^{-}_{+} d^{-}_{-} d^{+}_{-}}{d^{+}_{+}} \right]^{-1} \ , \quad h^{+}_{j,-\frac{1}{2}} = -h^{-}_{j,-\frac{1}{2}} \frac{d^{-}_{+} d^{-}_{-}}{d^{+}_{+} d^{+}_{-}} \ .
\end{equation}
Substituting $d^\sigma_\tau$ given in eqs.~\eqref{eq:d-values} we get
\begin{subequations}
\begin{equation}
	h^{-}_{j,-\frac{1}{2}} = \frac{4 \, \Gamma \left( \frac{2}{k} \right) \Gamma \left( 1-\frac{1}{k} \right) \Gamma \left( \frac{2j-1}{k} \right) \Gamma \left( 1-\frac{2j}{k} \right)}{(k+1-2j)^2 \Gamma \left( 1-\frac{2}{k} \right) \Gamma \left( \frac{1}{k} \right) \Gamma \left( 1-\frac{2j-1}{k} \right) \Gamma \left( \frac{2j}{k} \right)} \ ,
\end{equation}
\begin{equation}
	h^{+}_{j,-\frac{1}{2}} = \frac{4\, \Gamma \left( \frac{2}{k} \right) \Gamma \left( 1-\frac{1}{k} \right) \Gamma \left( \frac{2j-2}{k} \right) \Gamma \left( 1-\frac{2j-1}{k} \right)}{\Gamma \left( 1-\frac{2}{k} \right) \Gamma \left( \frac{1}{k} \right) \Gamma \left( 1-\frac{2j-2}{k} \right) \Gamma \left( \frac{2j-1}{k} \right)} \ .
\end{equation}
\end{subequations}
Using these, we eventually write
\begin{equation}
	G(u,v;\bar{u},\bar{v}) = \sum_{\sigma,\bar{\sigma}=\pm} u^{\frac{1-\sigma}{2}} \, \bar{u}^{\frac{1-\bar{\sigma}}{2}} \, g_{\sigma\bar{\sigma};(-\sigma)(-\bar{\sigma})}(v,\bar{v}) \ ,
\end{equation}
where
\begin{subequations} \label{eq:ads3-final-result-4pt}
	\begin{equation}
		g_{++;--}(v,\bar{v}) = \sum_{\tau=\pm} h^\tau_{j,-\frac{1}{2}} \, g^\tau_{j,-\frac{1}{2}}(0,v) \, g^\tau_{j,-\frac{1}{2}}(0,\bar{v}) \ ,
	\end{equation}
	\begin{equation}
		g_{+-;-+}(v,\bar{v}) = \sum_{\tau=\pm} h^\tau_{j,-\frac{1}{2}} \, g^\tau_{j,-\frac{1}{2}}(0,v) \, \Big( \partial_{\bar{u}} g^\tau_{j,-\frac{1}{2}}(\bar{u},\bar{v}) \Big) \ ,
	\end{equation}
	\begin{equation}
		g_{-+;+-}(v,\bar{v}) = \sum_{\tau=\pm} h^\tau_{j,-\frac{1}{2}} \, \Big( \partial_u g^\tau_{j,-\frac{1}{2}}(u,v) \Big) \, g^\tau_{j,-\frac{1}{2}}(0,\bar{v}) \ ,
	\end{equation}
	\begin{equation}
		g_{--;++}(v,\bar{v}) = \sum_{\tau=\pm} h^\tau_{j,-\frac{1}{2}} \, \Big( \partial_u g^\tau_{j,-\frac{1}{2}}(u,v) \Big) \, \Big( \partial_{\bar{u}} g^\tau_{j,-\frac{1}{2}}(\bar{u},\bar{v}) \Big) \ ,
	\end{equation}
\end{subequations}
see eqs.~\eqref{eq:ads3-explicit-blocks}.

\subsubsection{\texorpdfstring{$\text{S}^3$}{S3}} \label{app:app-kz-s3}
The discussion for $\mathfrak{su}(2)_{k-2}$ is very similar to Appendix~\ref{app:app-kz-ads3}. For this reason, we will mainly focus on stating our conventions and the final result for the $4$-pt function. To begin with, we define the $y$-basis via
\begin{equation}
	V(\ket{\psi};y,\bar{y};z,\bar{z}) = e^{y K^-_0+\bar{y} \bar{K}^-_0} e^{z L_{-1} + \bar{z} \bar{L}_{-1}} V(\ket{\psi};0,0;0,0) e^{-z L_{-1}- \bar{z} \bar{L}_{-1}} e^{-y K^-_0-\bar{y} \bar{K}^-_0} \ .
\end{equation}
The $4$-pt function of interest is
\begin{align} \label{eq:s3-primary-4pt-app-y-basis}
	\mathcal{I}^{\text{S}^3} = \langle V(\ket{j^\prime,j^\prime};y_1&,\bar{y}_1;z_1,\bar{z}_1) V(\ket{\tfrac{1}{2},\tfrac{1}{2}};y_2,\bar{y}_2;z_2,\bar{z}_2) \\ &V(\ket{j^\prime,j^\prime};y_3,\bar{y}_3;z_3,\bar{z}_3) V(\ket{\tfrac{1}{2},\tfrac{1}{2}};y_4,\bar{y}_4;z_4,\bar{z}_4) \rangle \ , \nonumber
\end{align}
which depends on the cross-ratios,
\begin{equation} \label{eq:s3-cross-ratio}
	w = \frac{y_{23} y_{14}}{y_{21} y_{34}} \ , \quad v = \frac{z_{23} z_{14}}{z_{21} z_{34}} \ .
\end{equation}
In other words, we can write
\begin{equation} \label{eq:s3-4pt-kz-cross-ratio}
	\mathcal{I}^{\text{S}^3} = L_{j_1^\prime,j_2^\prime}(w,v;\bar{w},\bar{v}) |f_{\text{ST}}(\{y_\ell\})|^2 |f_{\text{WS}}(\{z_\ell\})|^2 \ ,
\end{equation}
see eqs.~\eqref{eq:def-f-ws-st} and \eqref{eq:su2-delta-h}. The KZ equation for $\mathfrak{su}(2)_{k-2}$ is:
\begin{equation} \label{eq:s3-kz-equation}
	k \, \partial_{z_4} \mathcal{I}^{\text{S}^3}  + \sum_{\ell\neq 4} \frac{\mathcal{Q}^\prime_{\ell,4} \, \mathcal{I}^{\text{S}^3}}{z_{4}-z_\ell} = 0 \ ,
\end{equation}
where
\begin{equation}
	\mathcal{Q}^\prime_{\ell,4} = (y_\ell-y_4)^2 \partial_{y_\ell} \partial_{y_4} - 2(y_\ell-y_4) \Big[ j_\ell^\prime \partial_{y_4} - j_4^\prime \partial_{y_\ell} \Big] -  2 j_\ell^\prime j_4^\prime \ .
\end{equation}
Plugging eq.~\eqref{eq:s3-4pt-kz-cross-ratio} into \eqref{eq:s3-kz-equation}, we get
\begin{equation}
    \Big[ k \, \partial_v - \frac{\mathcal{P}^\prime}{v} - \frac{\mathcal{Q}^\prime}{v-1} \Big] L_{j^\prime_1,j^\prime_2}(w,v;\bar{w},\bar{v}) = 0 \ ,
\end{equation}
where we have defined
\begin{equation}
    \mathcal{P}^\prime = -w^2 (w-1) \partial_w^2 - \left( (\kappa^\prime+1) w^2+2j^\prime_1 w-2j^\prime_4 w(w-1) \right) \partial_w + 2\kappa^\prime j^\prime_4 w+2j^\prime_1 j^\prime_4 \ ,
\end{equation}
\begin{equation}
    \mathcal{Q}^\prime = (1-w)^2 w \partial_w^2 + (1-w) \left( (\kappa^\prime+1)(1-w)+2j^\prime_2+2j^\prime_4 w \right) \partial_w + 2\kappa^\prime j^\prime_4 (1-w)+2j^\prime_2 j^\prime_4 \ ,
\end{equation}
where we have defined
\begin{equation}
    \kappa^\prime = j^\prime_3-j^\prime_1-j^\prime_2-j^\prime_4 \ .
\end{equation}
Since $\text{S}^3$ is compact, the irreducible (zero mode) representations are finite dimensional. In the following, we focus on the case where $j^\prime_4=\frac{1}{2}$. Therefore, similar to the $\text{AdS}_3$ case, we have that $(K^-_0)^2 \ket{\frac{1}{2},\frac{1}{2}}=0$, and that the dependency on $w$ and $\bar{w}$ is linear. So we focus on finding the left-moving blocks by writing
\begin{equation} \label{eq:kz-s3-l0-l1}
	\Big[ k \partial_v - \frac{\mathcal{P}^\prime}{v} - \frac{\mathcal{Q}^\prime}{v-1} \Big] \ell_{j_1^\prime,j_2^\prime}(w,v) = 0 \ , \quad \ell_{j^\prime_1,j^\prime_2}(w,v) = \ell_{j_1^\prime,j_2^\prime}^{(0)}(v) + w \ell_{j_1^\prime,j_2^\prime}^{(1)}(v) \ .
\end{equation}
We define
\begin{subequations} \label{eq:l0-l1-s3}
\begin{equation}
	\ell_{j_1^\prime,j_2^\prime}^{(0)}(v) = \frac{1}{C^\prime} \, (1-v)^{A^\prime} v^{B^\prime} \partial_v[(1-v)^{C^\prime} p^\prime(v)] \ ,
\end{equation}
\begin{equation}
	\ell_{j_1^\prime,j_2^\prime}^{(1)}(v) = (1-v)^{-\frac{1+j_2^\prime}{k}} v^{-\frac{1+j_1^\prime}{k}} p^\prime(v) \ ,
\end{equation}
\end{subequations}
where
\begin{equation}
	A^\prime = -\frac{1-2j_1^\prime+2j_3^\prime-2k}{2k} \ , \quad B^\prime = -\frac{1+j_1^\prime-k}{k} \ , \quad C^\prime = -\frac{1+2j_1^\prime+2j_2^\prime-2j_3^\prime}{2k} \ .
\end{equation}
Upon substituting \eqref{eq:l0-l1-s3} into \eqref{eq:kz-s3-l0-l1}, we see that
\begin{equation}
	v(1-v) \, \partial_v^2 p^\prime(v)+\big[ c^\prime-(a^\prime+b^\prime+1)v \big] \partial_v p^\prime(v) - a^\prime b^\prime\,  p^\prime(v) = 0 \ ,
\end{equation}
with
\begin{align}
		a^\prime = C^\prime = -\frac{1+2j^\prime_1+2j^\prime_2-2j^\prime_3}{2k} & \ , \quad b^\prime = 1-\frac{3+2j^\prime_1+2j^\prime_2+2j^\prime_3}{2k} \ , \quad \\ c^\prime &= 1-\frac{1+2j^\prime_1}{k} \ . \nonumber
\end{align}
and thus,
\begin{equation}
    p^{+,\prime}(v) = {}_2F_1 \left(a^\prime,b^\prime;c^\prime;v\right) \ ,
\end{equation}
\begin{equation}
    p^{-,\prime}(v) = v^{1-c^\prime} {}_2F_1 \left(1+a^\prime-c^\prime,1+b^\prime-c^\prime;2-c^\prime;v\right) \ .
\end{equation}
Using the identities in eqs.~\eqref{eq:handbook-1} and \eqref{eq:handbook-2} with $n=1$, we get the following two blocks
\begin{align}
	\ell_{j_1^\prime,j_2^\prime}^{+}(w,v) = -\frac{c^\prime-b^\prime}{c^\prime} (1-v)^{-\frac{1+j_2^\prime}{k}} &v^{1-\frac{1+j_1^\prime}{k}} {}_2F_1 \left(1+a^\prime,b^\prime;1+c^\prime;v\right) \\
&+ w (1-v)^{-\frac{1+j_2^\prime}{k}} v^{-\frac{1+j_1^\prime}{k}} {}_2F_1 \left(a^\prime,b^\prime;c^\prime;v\right) \ , \nonumber
\end{align}
and
\begin{align}
	\ell_{j_1^\prime,j_2^\prime}^{-}(w,v) = \frac{1-c^\prime}{a^\prime} (1-v)^{-\frac{1+j_2^\prime}{k}} &v^{\frac{j_1^\prime}{k}} {}_2F_1 \left(1+a^\prime-c^\prime,b^\prime-c^\prime;1-c^\prime;v\right) \\ &+ w (1-v)^{-\frac{1+j_2^\prime}{k}} v^{\frac{j_1^\prime}{k}} {}_2F_1 \left(1+a^\prime-c^\prime,1+b^\prime-c^\prime;2-c^\prime;v\right) \ . \nonumber
\end{align}
Thus, we write
\begin{equation} \label{eq:s3-4pt-function-block-expansion}
	L_{j_1^\prime,j_2^\prime}(w,v;\bar{w},\bar{v}) = \sum_{\sigma=\pm} H^\sigma_{j_1^\prime,j_2^\prime} \ell_{j_1^\prime,j_2^\prime}^{\sigma}(w,v) \ell_{j_1^\prime,j_2^\prime}^{\sigma}(\bar{w},\bar{v}) \ ,
\end{equation}
and fix $H^\sigma_{j_1^\prime,j_2^\prime}$ using crossing equation,
\begin{equation} \label{eq:cross-equation-s3}
	L_{j_1^\prime,j_2^\prime}(w,v;\bar{w},\bar{v}) = L_{j_2^\prime,j_1^\prime}(1-w,1-v;1-\bar{w},1-\bar{v}) \ .
\end{equation}
From now on we focus on the case where $j^\prime_1=j^\prime_3=j^\prime$ and $j^\prime_2=\frac{1}{2}$, and write
\begin{equation} \label{eq:crossing-blocks-s3}
	\ell^{\sigma}_{j^\prime,\frac{1}{2}}(w,v) = \sum_{\tau=\pm} e^\sigma_\tau \, \ell^{\tau}_{\frac{1}{2},j^\prime}(1-w,1-v) \ .
\end{equation}
We explicitly have
\begin{subequations} \label{eq:s3-explicit-blocks}
\begin{align}
	\ell_{j^\prime,\frac{1}{2}}^{+}(w,v) &= -\frac{1}{k-2j^\prime-1} (1-v)^{-\frac{3}{2k}} v^{1-\frac{1+j^\prime}{k}} {}_2F_1 \left(1-\frac{1}{k},1-\frac{2j^\prime+2}{k};2-\frac{2j^\prime+1}{k};v\right) \\ &+ w (1-v)^{-\frac{3}{2k}} v^{-\frac{1+j^\prime}{k}} {}_2F_1 \left(-\frac{1}{k},1-\frac{2j^\prime+2}{k};1-\frac{2j^\prime+1}{k};v\right) \ , \nonumber
\end{align}
\begin{align}
	\ell_{j^\prime,\frac{1}{2}}^{-}(w,v) = -(2j^\prime+1) \, (1-v)^{-\frac{3}{2k}} &v^{\frac{j^\prime}{k}} {}_2F_1 \left(\frac{2j^\prime}{k},-\frac{1}{k};\frac{2j^\prime+1}{k};v\right) \\ &+ w (1-v)^{-\frac{3}{2k}} v^{\frac{j^\prime}{k}} {}_2F_1 \left(\frac{2j^\prime}{k},1-\frac{1}{k};1+\frac{2j^\prime+1}{k};v\right) \ . \nonumber
\end{align}
\end{subequations}
and
\begin{subequations}
\begin{align}
	\ell_{\frac{1}{2},j^\prime}^{+}(w,v) = -\frac{2j^\prime}{k-2} (1-v)^{-\frac{1+j^\prime}{k}} &v^{1-\frac{3}{2k}} {}_2F_1 \left(1-\frac{1}{k},1-\frac{2j^\prime+2}{k};2-\frac{2}{k};v\right) \\ &+ w (1-v)^{-\frac{1+j^\prime}{k}} v^{-\frac{3}{2k}} {}_2F_1 \left(-\frac{1}{k},1-\frac{2j^\prime+2}{k};1-\frac{2}{k};v\right) \ , \nonumber
\end{align}
\begin{align}
	\ell_{\frac{1}{2},j^\prime}^{-}(w,v) = -2 (1-v)^{-\frac{1+j^\prime}{k}} &v^{\frac{1}{2k}} {}_2F_1 \left(\frac{1}{k},-\frac{2j^\prime}{k};\frac{2}{k};v\right) \\ &+ w (1-v)^{-\frac{1+j^\prime}{k}} v^{\frac{1}{2k}} {}_2F_1 \left(\frac{1}{k},1-\frac{2j^\prime}{k};1+\frac{2}{k};v\right) \ . \nonumber
\end{align}
\end{subequations}
Similar to the $\text{AdS}_3$ case, in particular see eqs.~\eqref{eq:crossing-handbook-3}, one can then show that
\begin{subequations} \label{eq:e-values}
	\begin{equation}
		e^{+}_{+} = -\frac{\Gamma \left( 1-\frac{2j^\prime+1}{k} \right) \Gamma \left( \frac{2}{k} \right)}{\Gamma \left( 1-\frac{2j^\prime}{k} \right) \Gamma \left( \frac{1}{k} \right)} \ ,
	\end{equation}
	\begin{equation}
		e^{+}_{-} = -\frac{\Gamma \left( 1-\frac{2j^\prime+1}{k} \right) \Gamma \left( -\frac{2}{k} \right)}{\Gamma \left( 1-\frac{2j^\prime+2}{k} \right) \Gamma \left( -\frac{1}{k} \right)} \ ,
	\end{equation}
	\begin{equation}
		e^{-}_{+} = -\frac{\Gamma \left( 1+\frac{2j^\prime+1}{k} \right) \Gamma \left( \frac{2}{k} \right)}{\Gamma \left( \frac{2j^\prime+2}{k} \right) \Gamma \left( 1+\frac{1}{k} \right)} \ ,
	\end{equation}
	\begin{equation}
		e^{-}_{-} = -\frac{\Gamma \left( 1+\frac{2j^\prime+1}{k} \right) \Gamma \left( -\frac{2}{k} \right)}{\Gamma \left( \frac{2j^\prime}{k} \right) \Gamma \left( 1-\frac{1}{k} \right)} \ .
	\end{equation}
\end{subequations}
In the limit where $v\to 0$, the blocks $\ell^{\pm}_{\frac{1}{2},j^\prime}(w,v)$ behave as
\begin{equation}
    \ell^{+}_{\frac{1}{2},j^\prime}(w,v) \sim v^{-\frac{3}{2k}} w \ , \quad \ell^{-}_{\frac{1}{2},j^\prime}(w,v) \sim v^{\frac{1}{2k}} (w-2) \ .
\end{equation}
As the world-sheet weights of the fields with $j^\prime_1=j^\prime_4=\frac{1}{2}$ are $(\frac{3}{4k})$, the identity operator appears in the block $\ell^{+}_{\frac{1}{2},j^\prime}$ and normalizing it to $1$ gives
\begin{equation} \label{eq:su2-two-point-normalization}
    H^{+}_{\frac{1}{2},j^\prime} = 1 \ .
\end{equation}
Now we proceed as in the previous section: we substitute \eqref{eq:crossing-blocks-s3} into \eqref{eq:s3-4pt-function-block-expansion} and require \eqref{eq:cross-equation-s3}. This implies the following equations:
\begin{subequations}
\begin{equation}
	H^{+}_{j^\prime,\frac{1}{2}} e^{+}_{+} e^{+}_{-} + H^{-}_{j^\prime,\frac{1}{2}} e^{-}_{+} e^{-}_{-} = 0 \ ,
\end{equation}
\begin{equation}
	H^{+}_{j^\prime,\frac{1}{2}} \left( e^{+}_{+} \right)^2 + H^{-}_{j^\prime,\frac{1}{2}} \left( e^{-}_{+} \right)^2 = H^{+}_{\frac{1}{2},j^\prime} \ ,
\end{equation}
\begin{equation}
	H^{+}_{j^\prime,\frac{1}{2}} \left( e^{+}_{-} \right)^2 + H^{-}_{j^\prime,\frac{1}{2}} \left( e^{-}_{-} \right)^2 = H^{-}_{\frac{1}{2},j^\prime} \ .
\end{equation}
\end{subequations}
Using eq.~\eqref{eq:su2-two-point-normalization}, we get
\begin{equation}
	H^{+}_{j^\prime,\frac{1}{2}} = \left[ \left( e^{+}_{+} \right)^2 - \frac{e^{+}_{+} e^{+}_{-} e^{-}_{+}}{e^{-}_{-}} \right]^{-1} \ , \quad H^{-}_{j^\prime,\frac{1}{2}} = -H^{+}_{j^\prime,\frac{1}{2}} \frac{e^{+}_{+} e^{+}_{-}}{e^{-}_{+} e^{-}_{-}} \ .
\end{equation}
$e^\sigma_\tau$ are given in eqs.~\eqref{eq:e-values}. Hence, we get
\begin{subequations}
\begin{equation}
	H^{+}_{j^\prime,\frac{1}{2}} = \frac{\Gamma \left( \frac{1}{k} \right) \Gamma \left( 1-\frac{2}{k} \right) \Gamma \left( \frac{2j^\prime+1}{k} \right) \Gamma \left( 1-\frac{2j^\prime}{k} \right)}{\Gamma \left( 1-\frac{1}{k} \right) \Gamma \left( \frac{2}{k} \right) \Gamma \left( 1-\frac{2j^\prime+1}{k} \right) \Gamma \left( \frac{2j^\prime}{k} \right)} \ ,
\end{equation}
\begin{equation}
	H^{-}_{j^\prime,\frac{1}{2}} = \frac{\Gamma \left( \frac{1}{k} \right) \Gamma \left( 1-\frac{2}{k} \right) \Gamma \left( \frac{2j^\prime+2}{k} \right) \Gamma \left( 1-\frac{2j^\prime+1}{k} \right)}{(2j^\prime+1)^2 \Gamma \left( 1-\frac{1}{k} \right) \Gamma \left( \frac{2}{k} \right) \Gamma \left( 1-\frac{2j^\prime+2}{k} \right) \Gamma \left( \frac{2j^\prime+1}{k} \right)} \ .
\end{equation}
\end{subequations}
To summarize, we write
\begin{equation}
	L_{j^\prime,\frac{1}{2}}(w,v;\bar{w},\bar{v}) = \sum_{\sigma,\bar{\sigma}=\pm} w^{\frac{1+\sigma}{2}} \, \bar{w}^{\frac{1+\bar{\sigma}}{2}} \, \ell_{\sigma\bar{\sigma};(-\sigma)(-\bar{\sigma})}(v,\bar{v}) \ ,
\end{equation}
where
\begin{subequations} \label{eq:s3-final-result-4pt}
\begin{equation}
	\ell_{++;--}(v,\bar{v}) = \sum_{\tau=\pm} H^\tau_{j^\prime,\frac{1}{2}} \, \Big( \partial_w \ell^\tau_{j^\prime,\frac{1}{2}}(w,v) \Big) \, \Big( \partial_{\bar{w}} \ell^\tau_{j^\prime,\frac{1}{2}}(\bar{w},\bar{v}) \Big) \ ,
\end{equation}
\begin{equation}
	\ell_{+-;-+}(v,\bar{v}) = \sum_{\tau=\pm} H^\tau_{j^\prime,\frac{1}{2}} \, \Big( \partial_w \ell^\tau_{j^\prime,\frac{1}{2}}(w,v) \Big) \, \ell^\tau_{j^\prime,\frac{1}{2}}(0,\bar{v}) \ ,
\end{equation}
\begin{equation}
	\ell_{-+;+-}(v,\bar{v}) = \sum_{\tau=\pm} H^\tau_{j^\prime,\frac{1}{2}} \, \ell^\tau_{j^\prime,\frac{1}{2}}(0,v) \, \Big( \partial_{\bar{w}} \ell^\tau_{j^\prime,\frac{1}{2}}(\bar{w},\bar{v}) \Big) \ ,
\end{equation}
\begin{equation}
	\ell_{--;++}(v,\bar{v}) = \sum_{\tau=\pm} H^\tau_{j^\prime,\frac{1}{2}} \, \ell^\tau_{j^\prime,\frac{1}{2}}(0,v) \, \ell^\tau_{j^\prime,\frac{1}{2}}(0,\bar{v}) \ ,
\end{equation}
\end{subequations}
see eqs.~\eqref{eq:s3-explicit-blocks}.

\subsection{The SFT integrands} \label{app:SFT-integrands}
In this appendix, we report the final results of the $4$-pt function in \eqref{eq:SFT-4-pt-function} for the states discussed in Section~\ref{sec:NSNS}. As the world-sheet is a $2$d CFT, each of the states can be put either on the left- or the right-moving sector. Therefore, in order to report the results in a form that can be easily used for various states put on the left- and right-moving parts, we report the integrands in a matrix form. More specifically, we write the final answer as a linear combination over $g_{\downarrow}$, $g_{\uparrow}$, $\ell_{\downarrow}$ and $\ell_{\uparrow}$ and only in terms of $z$. Having this, in order to combine an integrand put on the left with the coordinate $z$, with another integrand put on the right with the coordinate $\bar{z}$, we multiply them and use
\begin{subequations}
	\begin{equation}
		g_{\uparrow} \bar{g}_{\uparrow} \mapsto g_{++;--}(z,\bar{z}) \ ,
	\end{equation}
	\begin{equation}
		g_{\uparrow} \bar{g}_{\downarrow} \mapsto g_{+-;-+}(z,\bar{z}) \ ,
	\end{equation}
	\begin{equation}
		g_{\downarrow} \bar{g}_{\uparrow} \mapsto g_{-+;+-}(z,\bar{z}) \ ,
	\end{equation}
	\begin{equation}
		g_{\downarrow} \bar{g}_{\downarrow} \mapsto g_{--;++}(z,\bar{z}) \ ,
	\end{equation}
\end{subequations}
and
\begin{subequations}
	\begin{equation}
		\ell_{\uparrow} \bar{\ell}_{\uparrow} \mapsto \ell_{++;--}(z,\bar{z}) \ ,
	\end{equation}
	\begin{equation}
		\ell_{\uparrow} \bar{\ell}_{\downarrow} \mapsto \ell_{+-;-+}(z,\bar{z}) \ ,
	\end{equation}
	\begin{equation}
		\ell_{\downarrow} \bar{\ell}_{\uparrow} \mapsto \ell_{-+;+-}(z,\bar{z}) \ ,
	\end{equation}
	\begin{equation}
		\ell_{\downarrow} \bar{\ell}_{\downarrow} \mapsto \ell_{--;++}(z,\bar{z}) \ ,
	\end{equation}
\end{subequations}
see eqs.~\eqref{eq:ads3-final-result-4pt} and \eqref{eq:s3-final-result-4pt}. In the end, we multiply the obtained expression by an overall factor of $1/(4\pi)^2$ because of the normalization factor of the RR vertex operator in eq.~\eqref{eq:RR-vertex-operator}.

\paragraph{The state $\ket{j-n-1}\ket{j^\prime,j^\prime}$.} This state, see eq.~\eqref{eq:leading-regge-right}, depends on the parameters $n$ and $j^\prime$. As the picture-raised RR vertex operator in \eqref{eq:RR-ket-picture-raised} includes $\mathfrak{su}(2)_{k-2}$ currents, the calculation depends on whether $j^\prime=0$ or $j^\prime\geq \frac{1}{2}$. In the case where $j^\prime=0$, we have
\begin{equation}
	\mathcal{I}^{\ket{j-n-1}}_{j^\prime=0}(z) = \frac{\left[ k(j+n)-2j(j-1) \right] \left( g_{\uparrow}+z g_{\downarrow} \right) \left( \ell_{\uparrow}-\ell_{\downarrow} \right)}{k(k+2-2j)z(1-z)} \ .
\end{equation}
If $j^\prime \geq \frac{1}{2}$, we have
\begin{align}
	\mathcal{I}^{\ket{j-n-1}}_{j^\prime\geq \frac{1}{2}}(z) = \frac{1}{k(k+2-2j)z(1-z)} \Big\{ \ell_{\uparrow} &\Big[ z \left( (j+j^\prime+1)(k+2-2j)+kn \right) g_{\downarrow} \nonumber \\&+\left( (j-j^\prime-1)(k+2-2j)+kn \right) g_{\uparrow} \Big] \\ &-\ell_{\downarrow} \Big[ z \left( (j-j^\prime-1)(k+2-2j)+kn \right) g_{\downarrow}\nonumber\\&+\left( (j+j^\prime+1)(k+2-2j)+kn \right) g_{\uparrow} \Big] \Big\} \ . \nonumber
\end{align}

\paragraph{The state $\ket{j+n+1}\ket{j^\prime,j^\prime}$.}
In this case, see \eqref{eq:leading-regge-left}, the answer again depends on whether $j^\prime=0$ or not. If $j^\prime=0$, we have
\begin{align}
	\mathcal{I}^{\ket{j+n+1}}_{j^\prime=0}(z)& \\&= \frac{\left[ 2j(j-1)+k(j-1-n) \right] \left[ \left( (2j-1)z-1 \right) g_{\uparrow}+(2j-1-z)g_{\downarrow} \right] \left( \ell_{\uparrow}-\ell_{\downarrow} \right)}{2jk(2j+k)z(1-z)} \ . \nonumber
\end{align}
For the case $j^\prime \geq \frac{1}{2}$, we get
\begin{equation}
	\begin{aligned}
		&\mathcal{I}^{\ket{j+n+1}}_{j^\prime\geq \frac{1}{2}}(z) = \frac{1}{2jk(2j+k)z(1-z)} \\[2pt] &\quad \times \Big\{ \ell_{\uparrow} \Big[ \Big( (2j+k) \big( (2j-1)(j+j^\prime)-z(j-j^\prime-2) \big)-kn(2j-1-z) \Big) g_{\downarrow} \\ &\qquad\quad +\Big( (2j+k) \big( (2j-1)z(j-j^\prime-2)-(j+j^\prime) \big)+kn \big( 1-(2j-1)z \big) \Big) g_{\uparrow} \Big] \\[2pt] &\qquad -\ell_{\downarrow} \Big[ \Big( (2j+k) \big( (2j-1)(j-j^\prime-2)-z(j+j^\prime) \big)-kn(2j-1-z) \Big) g_{\downarrow} \\ &\qquad\quad +\Big( (2j+k) \big( (2j-1)z(j+j^\prime)-(j-j^\prime-2) \big)+kn \big( 1-(2j-1)z \big) \Big) g_{\uparrow} \Big] \Big\} \ .
	\end{aligned}
\end{equation}

\paragraph{The state $\ket{j-n-\frac{1}{2};\varepsilon=+}$.} In this case, see eq.~\eqref{eq:qsc-state-right}, we get
\begin{align}
	&\mathcal{I}^{\ket{j-n-\frac{1}{2};\varepsilon=+}}(z) = \frac{-1}{k(j+j^\prime)(k+2-2j)z \, (1-z)} \nonumber \\ &\quad \times \Big\{ \ell_{\downarrow} \Big[ \Big( (k+2-2j)(j+j^\prime)(1-j+j^\prime) +kn \big( k+2-2j-z(k+1-j+j^\prime) \big) \Big) g_{\downarrow} \nonumber \\ &\qquad\quad -\Big( (k+2-2j)(j+j^\prime)^2+(j+j^\prime-1)kn \Big) g_{\uparrow} \Big] \\ &\qquad +\ell_{\uparrow} \Big[ z\Big( (k+2-2j)(j+j^\prime)^2+(j+j^\prime-1)kn \Big) g_{\downarrow} \nonumber \\ &\qquad\quad +\Big( (k+2-2j)(j+j^\prime)(j-j^\prime-1)+(j+j^\prime-1)kn \Big) g_{\uparrow} \Big] \Big\} \ . \nonumber
\end{align}

\paragraph{The state $\ket{j-n-\frac{1}{2};\varepsilon=-}$.} In this case, see eq.~\eqref{eq:qsc-state-right-epsilon-}, we get
\begin{equation}
	\begin{aligned}
		&\mathcal{I}^{\ket{j-n-\frac{1}{2};\varepsilon=-}}(z) = \frac{-1}{2(j-j^\prime-1)j^\prime k(k+2-2j)z(1-z)} \\[2pt] &\quad \times \Big\{ \ell_{\uparrow} \Big[ \Big( (j-j^\prime-1)(k+2-2j) \big( (2j^\prime+1)(j+j^\prime)-(j-j^\prime-1)z \big) \\ &\qquad -kn \big( (2j^\prime+1)(k+2-2j)(1-z)-2j^\prime(j-j^\prime-2)z \big) \Big) g_{\downarrow} \\ &\quad +\Big( (j-j^\prime-1)(k+2-2j) \big( 2j^\prime(j-j^\prime-2)-1 \big)+2j^\prime(j-j^\prime-2)kn \Big) g_{\uparrow} \Big] \\[2pt] &\qquad +\ell_{\downarrow} \Big[ \Big( (j-j^\prime-1)(k+2-2j) \big( j+j^\prime-(2j^\prime+1)(j-j^\prime-1)z \big) \\ &\qquad -kn \big( (k+2-2j)(1-z)+2j^\prime(j-j^\prime-2)z \big) \Big) g_{\downarrow} \\ &\quad -\Big( (j-j^\prime-1)(k+2-2j) \big( 1+2j^\prime(j+j^\prime+1) \big)+2j^\prime(j-j^\prime-2)kn \Big) g_{\uparrow} \Big] \Big\} \ .
	\end{aligned}
\end{equation}

\paragraph{The state $\ket{j+n+\frac{1}{2};\varepsilon=+}$.} For the state given in eq.~\eqref{eq:qsc-state-left-epsilon=+}, we get
\begin{equation}
	\begin{aligned}
		&\mathcal{I}^{\ket{j+n+\frac{1}{2};\varepsilon=+}}(z) = \frac{-1}{2j(j-j^\prime-1)k(2j+k)z(1-z)} \\[3pt] &\quad \times \Big\{ \ell_{\uparrow} \Big[ \Big( (j-j^\prime-1)(2j+k) \big( (2j-1)(j+j^\prime)-(j-j^\prime-1)z \big) \\ &\qquad\qquad -(j-j^\prime)kn(2j-1-z) \Big) g_{\downarrow} \\ &\qquad\quad -\Big( (j-j^\prime-1)(2j+k) \big( j+j^\prime-(2j-1)(j-j^\prime-1)z \big) \\ &\qquad\qquad -(j-j^\prime)kn \big( 1-(2j-1)z \big) \Big) g_{\uparrow} \Big] \\[3pt] &\qquad -\ell_{\downarrow} \Big[ \Big( (j-j^\prime-1)(2j+k) \big( 1+2j(j-j^\prime-2) \big) \\ &\qquad\qquad +kn \big( (2j+k)(1-z)-(j-j^\prime)(2j-1-z) \big) \Big) g_{\downarrow} \\ &\qquad\quad +\Big( (j-j^\prime-1)(2j+k) \big( 1+2j(j+j^\prime-1) \big) \\ &\qquad\qquad -kn \big( (2j-1)(2j+k)(1-z)-(j-j^\prime)(1-(2j-1)z) \big) \Big) g_{\uparrow} \Big] \Big\} \ .
	\end{aligned}
\end{equation}

\paragraph{The state $\ket{j+n+\frac{1}{2};\varepsilon=-}$.} This state is given in eq.~\eqref{eq:qsc-state-left} and the integrand is
\begin{align}
	&\mathcal{I}^{\ket{j+n+\frac{1}{2};\varepsilon=-}}(z)=\frac{-1}{4jj^\prime(j+j^\prime)k(2j+k)z(1-z)} \nonumber \\ &\quad\times\Bigg\{g_{\downarrow}\Bigg[\ell_{\downarrow}\Bigg\{kn\Big[\big(2(1+j^\prime)(j+j^\prime)+k\big)(1-z)+4j^\prime(j-1)(j+j^\prime+1)\Big] \nonumber \\ &\qquad +(j+j^\prime)(2j+k)\Big[2\big(j-j^\prime+2jj^\prime(3-j+j^\prime)\big)-(j+j^\prime)(2j^\prime+1)(1-z)\Big]\Bigg\} \nonumber \\ &\qquad+\ell_{\uparrow}\Bigg\{(j+j^\prime)^2(2j+k)\Big[2(1+2jj^\prime)-(1-z)\Big] \nonumber \\ &\qquad+kn\Big[\big(2(j-j^\prime)(1+j^\prime)+(2j^\prime+1)k\big)(1-z)-4j^\prime(j-1)(j+j^\prime+1)\Big]\Bigg\}\Bigg] \\ &\qquad+g_{\uparrow}\Bigg[\ell_{\downarrow}\Bigg\{(j+j^\prime)^2(2j+k)\Big[(2j-1)(2j^\prime+1)(1-z)-2(1+2jj^\prime)\Big] \nonumber \\ &\qquad+kn\Big[4j^\prime(j-1)(j+j^\prime+1)-(2j-1)\big(2(1+j^\prime)(j+j^\prime)+k\big)(1-z)\Big]\Bigg\} \nonumber \\ &\qquad+\ell_{\uparrow}\Bigg\{(j+j^\prime)(2j+k)\Big[(2j-1)(j+j^\prime)(1-z)-2\big(j-j^\prime+2jj^\prime(3-j+j^\prime)\big)\Big] \nonumber \\ &\qquad-kn\Big[4j^\prime(j-1)(j+j^\prime+1)+(2j-1)\big(2(j-j^\prime)(1+j^\prime)+(2j^\prime+1)k\big)(1-z)\Big]\Bigg\}\Bigg]\Bigg\} \ . \nonumber
\end{align}

\subsection{Large-\texorpdfstring{$k$}{k} expansion of the plane integral}
\label{app:SFT-large-k-integral}
In this appendix, we discuss how we can calculate the SFT integrals \eqref{eq:ws-anomalous-dimension} analytically as a $1/\sqrt{k}$ expansion. More specifically, given the integrands presented in Appendix~\ref{app:SFT-integrands}, the space-time energy is computed as follows:
\begin{equation} \label{eq:SFT-integrand-split}
	\frac{\delta\Delta(k)}{\sqrt{k}} = - \pi \mu^2 \frac{2\sqrt{k}}{2j-1} \text{Reg} \int_{\mathbb C}\text{d}^2 z\, \mathcal{I}(z,\bar{z}) \ ,
\end{equation}
see eq.~\eqref{eq:SFT-deltaE-deltaDelta}, where $\mathcal{I}(z,\bar{z})$ is the integrand in \eqref{eq:SFT-4-pt-function}. In this appendix, we use the prescription presented in \cite{Sen:2019jpm} for regularizing the integrals
\begin{equation} \label{eq:i1234-app}
	\mathcal{I}_{1234} = \int_{\mathbb C}\text{d}^2 z\, \langle [\xi_0 \bar{\xi}_0 \mathcal{V}_1](0,0) \mathcal{V}_2(1,1) \mathcal{V}_3(\infty,\infty) b_{-1}\bar{b}_{-1} \mathcal{V}_4(z,\bar z) \rangle \ ,
\end{equation}
where the sum of the picture numbers satisfy \eqref{eq:picture-sum-condition}. In particular, one expects that if one of the $\mathcal{V}_\ell$ is BRST exact, then $\mathcal{I}_{1234}=0$. In fact, since
\begin{equation}
	\{ Q,b_{-1} \} = T_{-1} \ , \quad \{ \bar{Q},\bar{b}_{-1} \} = \bar{T}_{-1} \ ,
\end{equation}
where $Q$ is defined in \eqref{eq:q-brst-charge} and $T$ is the total world-sheet stress-tensor, a BRST exact state produces a total derivative integrand. Although the integral of a total derivative is zero on a set with no boundary, the integrand can have singularities near $z\in\{0,1,\infty\}$ where the physical fields are inserted. In order to see this more explicitly, let us write the string field propagator in Schwinger parametrization as
\begin{equation} \label{eq:sft-propagator}
	\frac{1}{L_0+\bar{L}_0} = \int_0^\infty \text{d}t \, e^{-t (L_0+\bar{L}_0)} \ .
\end{equation}
However, in string field theory, the LHS is ill-defined on-shell and there are off-shell states with negative $L_0+\bar{L}_0$ eigenvalues, and thus, one needs to regulate this propagator as discussed in \cite{Sen:2019jpm}. To begin with, we choose small neighborhoods around the three insertion points,
\begin{equation}
	D_0=\{|z|<\epsilon_0\} \ , \qquad D_1=\{|1-z|<\epsilon_1\} \ , \qquad D_\infty=\{|z|>R\} \ ,
\end{equation}
where $\epsilon_0$ and $\epsilon_1$ are small positive real numbers, while $R$ is a large positive real number. We also write the bulk region as\footnote{In the notations of \cite{Sen:2019jpm}, $\mathcal{C}^{(0)}$ is $\Omega_{\epsilon_0,\epsilon_1,R}$ while $\mathcal{C}^{(1)}_z$ are $\partial D_z$ for $z\in\{0,1,\infty\}$.}
\begin{equation}
	\Omega_{\epsilon_0,\epsilon_1,R} = \mathbb C \setminus (D_0\cup D_1\cup D_\infty) \ .
\end{equation}
Having this, following \cite{Sen:2019jpm}, we define\footnote{These are equivalent to considering the $\int_{\mathcal{C}^{(1)}_z} \mathcal{I}^{(1)}_z$ contributions in the notation of \cite{Sen:2019jpm}, see eq.~(1.6) there.}
\begin{equation} \label{eq:reg-z-delta-app}
	\text{Reg} \, \int_{|z|<\epsilon} \text{d}^2 z \, |z|^{-2+2\delta} = \frac{\pi\epsilon^{2\delta}}{\delta} \ , \quad \delta\in \mathbb{C}-\{0\} \ , \quad \epsilon \in \mathbb{R}_{>0} \ .
\end{equation}
We then calculate \eqref{eq:SFT-integrand-split} or \eqref{eq:i1234-app} by writing
\begin{equation} \label{eq:SFT-endpoint-region-split}
	\mathcal{I}_{\mathrm{bulk}}(k) +\mathcal{I}_{0}(k) +\mathcal{I}_{1}(k) +\mathcal{I}_{\infty}(k) \ ,
\end{equation}
where each subscript denotes the contribution from the corresponding region, e.g.\ $\mathcal{I}_{\text{bulk}}(k)$ is the integral evaluated on $\Omega_{\epsilon_0,\epsilon_1,R}$. The evaluation of \eqref{eq:SFT-endpoint-region-split} is now done in the following way:
\begin{enumerate}
	\item On the region $\Omega_{\epsilon_0,\epsilon_1,R}$, we first expand the integrand at large $k$ and then integrate term by term.
	\item On each of $D_0$, $D_1$ and $D_\infty$, we keep the exact finite-$k$ integrand, perform the corresponding analytically continued integrals, and only then expand the result at large $k$. Indeed, as $\delta(k)\to0$ as $k\to\infty$, see \eqref{eq:reg-z-delta-app}, expanding the integrand first gives
	\begin{equation}
		|z|^{-2+2\delta(k)}=\frac{1}{|z|^2}\left[1+2\delta(k)\log|z|+\cdots\right] \ .
	\end{equation}
	whose individual terms are not integrable at $z=0$.
	\item We finally sum the four contributions up. The dependence on the auxiliary parameters $\epsilon_0$, $\epsilon_1$, and $R$ must cancel order by order \cite{Sen:2019jpm}, providing a check that the partition has not changed the original integral.
\end{enumerate}
We note that, using this prescription, BRST exact terms decouple since the bulk contribution cancels the contributions from $D_0$, $D_1$ and $D_\infty$. After imposing the relation \eqref{eq:regge-sol-j} and assembling the contributions from the above four regions, the space-time energy \eqref{eq:SFT-integrand-split} is given as a series expansion over $1/\sqrt{k}$:
\begin{equation}
	\frac{\delta\Delta(k)}{\sqrt{k}} =
	\mathcal{A}_0 +\frac{\mathcal{A}_1}{\sqrt{k}} +\frac{\mathcal{A}_2}{k} +\frac{\mathcal{A}_3}{k^{3/2}} + \frac{\mathcal{A}_4}{k^{2}} + \mathcal{O}(1/k^{5/2}) \ .
\end{equation}
The results reported in Section~\ref{sec:SFT-summary} are obtained by performing the above procedure to the integrands attained through Appendix~\ref{app:SFT-integrands} for the relevant states. Before we close this section, let us mention that in \cite{Cho:2018nfn} another regularization prescription was used, where one instead considers the following integral
\begin{equation} \label{eq:SFT-integrand-split-xi}
	\frac{\delta\Delta(k)}{\sqrt{k}} = -\pi \mu^2 \frac{2\sqrt{k}}{2j-1} \int_{\mathbb C}\text{d}^2 z\, \Big[ \mathcal{I}(z,\bar{z}) - \mathcal{C}(z,\bar{z}) \Big] \ .
\end{equation}
Here $\mathcal{C}(z,\bar{z})$ is a counterterm predicted in \cite{Cho:2018nfn},
\begin{equation} \label{eq:counter-term-sft-app}
	\mathcal{C}(z,\bar{z}) = |1-z|^{-2-\frac{4}{k}} \, \lim_{w \to 1} |1-w|^{2+\frac{4}{k}} \, \mathcal{I}(w,\bar{w}) \ ,
\end{equation}
that subtracts the singularity $z\to 1$ as two RR vertex operators collide. In fact, by a direct computation it can be shown that
\begin{equation}
	\text{Reg} \, \int_{\mathbb{C}} \text{d}^2 z \, \mathcal{C}(z,\bar{z}) = 0 \ .
\end{equation}
Therefore, retaining the counterterm $\mathcal{C}(z,\bar{z})$ in \eqref{eq:SFT-integrand-split-xi} does not change the value of the analytically continued integral. However, it cancels the singularity near $z=1$ and it often allows numerical evaluations of \eqref{eq:SFT-integrand-split}, as done in \cite{Cho:2018nfn}.

\bibliographystyle{JHEP}
\bibliography{SFTdraft}

\end{document}